\documentclass[twocolumn]{aastex631}
\usepackage{mhchem}

\newcommand{\Rmnum}[1]{\uppercase\expandafter{\romannumeral #1}}

\newcommand{\teff}{$T_{\rm eff}$}
\newcommand{\logg}{$\log{g}$}
\newcommand{\feh}{[Fe/H]}
\newcommand{\snrr}{SNR\_r}

\newcommand{\specNum}{763,136} 
\newcommand{\starNum}{616,314}
\newcommand{\RefNum}{16}

\newcommand{\TDSA}{176\,K}
\newcommand{\TDDA}{74\,K}
\newcommand{\TGSA}{140\,K}
\newcommand{\TGDA}{58\,K}
\newcommand{\GDSA}{0.16\,dex}
\newcommand{\GDDA}{0.19\,dex}
\newcommand{\GGSA}{0.30\,dex}
\newcommand{\GGDA}{0.32\,dex}

\newcommand{\MDDA}{0.16\,dex}

\newcommand{\MGDA}{0.26\,dex}
\newcommand{\TDSAs}{73\,K}

\newcommand{\TGSAs}{22\,K}

\newcommand{\GDSAs}{0.52\,dex}

\newcommand{\GGSAs}{0.25\,dex}

\newcommand{\TDSR}{88\,K}

\newcommand{\TGSR}{152\,K}
\newcommand{\TGDR}{139\,K}
\newcommand{\GDSR}{0.05\,dex}

\newcommand{\GGSR}{0.14\,dex}

\newcommand{\MDSR}{0.26\,dex}

\newcommand{\MGSR}{0.16\,dex}

\newcommand{\TrandSNR}{45\,K}
\newcommand{\GrandSNR}{0.25\,dex}
\newcommand{\MrandSNR}{0.22\,dex}

\shorttitle{LAMOST M parameters}
\shortauthors{Ding et al.}
\accepted{2022.04.14}
\submitjournal{ApJS}

\graphicspath{./}

\begin{document}

\title{Stellar Atmospheric Parameters of M-type Stars from LAMOST DR8}

\author[0000-0001-6898-7620]{Ming-Yi Ding}
\affiliation{CAS Key Laboratory of Optical Astronomy, National Astronomical Observatories, Chinese Academy of Sciences, Beijing 100101, China}
\affiliation{School of Astronomy and Space Science, University of Chinese Academy of Sciences, Beijing 100049, China}

\author[0000-0002-0349-7839]{Jian-Rong Shi}
\affiliation{CAS Key Laboratory of Optical Astronomy, National Astronomical Observatories, Chinese Academy of Sciences, Beijing 100101, China}
\affiliation{School of Astronomy and Space Science, University of Chinese Academy of Sciences, Beijing 100049, China}
\email{sjr@nao.cas.cn}

\author{Yue Wu}
\affiliation{CAS Key Laboratory of Optical Astronomy, National Astronomical Observatories, Chinese Academy of Sciences, Beijing 100101, China}

\author[0000-0003-0433-3665]{Hugh R.A. Jones}
\affiliation{Centre for Astrophysics Research, University of Hertfordshire, College Lane, AL10 9AB, Hatfield, UK}

\author[0000-0002-8609-3599]{Hong-liang Yan}
\affiliation{CAS Key Laboratory of Optical Astronomy, National Astronomical Observatories, Chinese Academy of Sciences, Beijing 100101, China}
\affiliation{School of Astronomy and Space Science, University of Chinese Academy of Sciences, Beijing 100049, China}
\email{hlyan@nao.cas.cn}

\author[0000-0002-6647-3957]{Chun-Qian Li}
\affiliation{CAS Key Laboratory of Optical Astronomy, National Astronomical Observatories, Chinese Academy of Sciences, Beijing 100101, China}
\affiliation{School of Astronomy and Space Science, University of Chinese Academy of Sciences, Beijing 100049, China}

\author[0000-0003-4972-0677]{Qi Gao}
\affiliation{CAS Key Laboratory of Optical Astronomy, National Astronomical Observatories, Chinese Academy of Sciences, Beijing 100101, China}
\affiliation{School of Astronomy and Space Science, University of Chinese Academy of Sciences, Beijing 100049, China}

\author[0000-0002-6448-8995]{Tian-Yi Chen}
\affiliation{CAS Key Laboratory of Optical Astronomy, National Astronomical Observatories, Chinese Academy of Sciences, Beijing 100101, China}
\affiliation{School of Astronomy and Space Science, University of Chinese Academy of Sciences, Beijing 100049, China}

\author[0000-0002-2510-6931]{Jing-Hua Zhang}
\affiliation{CAS Key Laboratory of Optical Astronomy, National Astronomical Observatories, Chinese Academy of Sciences, Beijing 100101, China}

\author[0000-0001-5193-1727]{Shuai Liu}
\affiliation{CAS Key Laboratory of Optical Astronomy, National Astronomical Observatories, Chinese Academy of Sciences, Beijing 100101, China}
\affiliation{School of Astronomy and Space Science, University of Chinese Academy of Sciences, Beijing 100049, China}

\author[0000-0003-2318-9013]{Tai-Sheng Yan}
\affiliation{CAS Key Laboratory of Optical Astronomy, National Astronomical Observatories, Chinese Academy of Sciences, Beijing 100101, China}
\affiliation{School of Astronomy and Space Science, University of Chinese Academy of Sciences, Beijing 100049, China}

\author{Xiao-Jin Xie}
\affiliation{CAS Key Laboratory of Optical Astronomy, National Astronomical Observatories, Chinese Academy of Sciences, Beijing 100101, China}
\affiliation{Tibet University, Lhasa 850000, China}

\begin{abstract}

The Large Sky Area Multi-Object Fiber Spectroscopic Telescope (LAMOST) Low Resolution Spectroscopic Survey (LRS) provides massive spectroscopic data of M-type stars, and the derived stellar parameters could bring vital help to various studies.
We adopt the ULySS package to perform $\chi^2$ minimization with model spectra generated from the MILES interpolator, and determine the stellar atmospheric parameters for the M-type stars from LAMOST LRS Data Release (DR) 8.
Comparison with the stellar parameters from APOGEE Stellar Parameter and Chemical Abundance Pipeline (ASPCAP) suggests that most of our results have good consistency.
For M dwarfs, we achieve dispersions better than \TDDA, \GDDA\ and \MDDA\ for \teff, \logg\ and \feh, while for M giants, the internal uncertainties are \TGDA, \GGDA\ and \MGDA, respectively.
Compared to ASPCAP we also find a systematic underestimation of $\Delta {T_{\rm eff}} =$ $-$\TDSA\ for M dwarfs, and a systematic overestimation of $\Delta {\log{g}} =$ \GGSA\ for M giants.
However, such differences are less significant when we make comparison with common stars from other literature, which indicates that systematic biases exist in the difference of ASPCAP and other measurements.
A catalog of \specNum\ spectra corresponding to \starNum\ M-type stars with derived stellar parameters is presented.
We determine the stellar parameters for stars with \teff\ higher than 2,900\,K, with \logg\ from -0.24\,dex to 5.9\,dex.
The typical precisions are \TrandSNR, \GrandSNR\ and \MrandSNR, for \teff, \logg\ and \feh, respectively, which are estimated from the duplicate observations of the same stars.

\end{abstract}


\section{Introduction} \label{sect:intro}

Of the adopted astronomical spectral types, the M spectral type shows the most wide-ranging properties. On the Hertzsprung-Russell (H-R) diagram, the M dwarfs (dMs) and the M giants (gMs) showed the most extreme differences in luminosity and radius \citep[e.g.,][]{Gray.2009}.
The dMs are the faintest of the core hydrogen burners \citep[e.g.,][]{Kirkpatrick.1992}, while the gMs might have the largest brightness variations.
Studies of low-mass main-sequence dMs show that dMs are the most common stars in the Galaxy, which comprise 70\% of all stars \citep[e.g.,][]{laughlin.1997}.
The dMs are very important to determine the initial mass function which constrains theoretical star formation studies, and they are helpful to trace the chemical and dynamical history of the Galaxy.

The main-sequence types---O, B, A, F, G and K stars are considered to be completely hydrogen-burning stars, while the later types (giant -- S and C types, dwarf -- L, T and Y type) stars appear more likely to be carbon stars or brown dwarfs \citep{kirkpatrick.1999}.
The turning point for these later spectral types is the M-type stars, and they are a milestone in the study of the chemical and dynamical evolution of the Galaxy.

M-type stars are iconic for their crowded molecular absorption bands \citep{Gray.2009}, making their continuum difficult to define in the optical region. 
Over the years, methods based on the measurement of the atomic and molecular features in the optical and near-infrared region (6,300--9,000\,\AA), such as the use of \ce{TiO}, \ce{CaH} and \ce{Ca}\Rmnum{2} triplet, are highlighted in stellar atmospheric parameter determination for M-type stars \citep[etc.]{Bessell.1991,Kirkpatrick.1992,reid.1995,Cenarro.2001}. 
In a work of M-subdwarf (sdM) classification, \citet{Gizis.1997} showed the metallicity dependency on the \ce{TiO} and \ce{CaH} features in the region of 6,200 to 7,400\,\AA.

Likewise, an empirical spectroscopic index has been suggested \citep{Lepine.2007} for sdM classification, which is defined as the relative strength of \ce{TiO}5 and \ce{CaH} molecular absorption bands.
\citet{Mann.2012} used six different gravity-sensitive molecular and atomic indices to determine the luminosity class for late-type \textit{Kepler} exoplanet candidates. The spectral index of \ce{TiO}5 and \ce{CaH}2+\ce{CaH}3 was used to separate dMs from gMs and sdMs in Large Sky Area Multi-Object Fiber Spectroscopic Telescope (LAMOST) Data Release (DR) 1 \citep{Luo.2015}, and assemble these stars into a set of M-type spectral template library \citep{Zhong.2015,Zhong.2015b}. 
By employing this template, \citet{zhang.2019} calibrated the spectroscopic index and applied a new separator to identify 2,791 new sdMs from LAMOST DR4.

Currently, thanks to several ongoing large-scale spectroscopic surveys, large numbers of spectra along with stellar parameters are available now. 
For example, the Sloan Digital Sky Survey (SDSS) uses the Sloan Foundation 2.5\,m telescope at Apache Point Observatory \citep{gunn.2006} to determine spectroscopic abundances on a large scale. 
The Apache Point Observatory Galactic Evolution Experiment \citep[APOGEE,][]{Majewski.2017} is one of subprojects of SDSS-\Rmnum{3} \citep{eisenstein.2011}, which provides high resolution (R $\sim$ 22,500), high signal-to-noise ratio (SNR $>$ 100), near-infrared spectra. 
During the second generation of the APOGEE project (APOGEE-2) with the SDSS-\Rmnum{4}, the latest DR17 presents spectra for about 650,000 stars. 
Meanwhile, a set of stellar parameters and chemical abundances of up to 26 elements are also provided though the APOGEE Stellar Parameter and Chemical Abundances Pipeline (ASPCAP) \citep{GarciaPerez.2016}, which fits the observed spectra via comparison to the synthetic spectra generated from the MARCS stellar atmospheric model. 
\citet{Jonsson.2020} updated ASPCAP with the precalculated grids \citep{gustafsson.2008}, which improved the model performance under the effective temperature of 3,500\,K, and made the stellar parameters for cool stars more accurate compared to the previous version. 
Other spectroscopic surveys such as the Large Sky Area Multi-Object Fiber Spectroscopic Telescope \citep[LAMOST,][]{Cui.2012}, the Radial Velocity Experiment \citep[RAVE,][]{steinmetz.2006}, the GALactic Archaeology with HERMES \citep[GALAH,][]{desilva.2015}, Gaia-ESO \citep{brown.2018} and the Calar Alto high-Resolution search for M-dwarfs with Exo-earths with Near-infrared and optical Echelle Spectrographs \citep[CARMENES,][]{reiners.2018, quirrenbach.2020}, also provide relevant large-scale spectroscopic data of M-type stars.

In stellar astrophysics research, stellar atmospheric parameters (effective temperature \teff, surface gravity \logg, metallicity \feh) derived from high resolution and high signal-to-noise ratio (SNR) spectra are important indicators. 
During the past few years, different methods have been developed for determining the stellar parameters from high quality spectra. 
Generally, these approaches could be simply divided into two categories \citep{Wu.2011b}, the first one is the minimum distance method (MDM), including the measurement of equivalent width (EWs) or the synthetic spectra based on absorption lines \citep{Jofre.2019}.
For example, the $\chi^2$ minimization is a widely used MDM method, which searches the minimum $\chi^2$ between the observed spectra and the templates or spectra generated from stellar atmospheric models. 
The other one is so-called machine-learning approaches such as the Artificial Neural Network (ANN) and non-linear regression methods.

\citet{Mann.2015} determined the precise stellar parameters for 183 nearby dMs via MDM method and presented empirical relations between \teff, absolute magnitude, radius, mass and bolometric flux. 
In a search for exoplanets around dMs, \citet{Passegger.2018} determined the stellar atmospheric parameters for over 300 dMs by fitting the spectra generated from the PHOENIX-ACES \citep{Husser.2013} models. 
\citet{Rajpurohit.2018} determined the stellar parameters for 292 early to late M-type stars by comparing high resolution spectra with the synthetic spectra obtained from BT-Settl \citep{allard.2011,allard.2012,Allard.2013}. 
This model is also used to derive the stellar atmospheric parameters and kinematics for sdMs from the LAMOST survey by \citet{zhang.2021}.

Similarly, \citet{Kesseli.2019} obtained \teff\ for 88 sdMs using the BT-Settl grid and \feh\ values estimated from color empirical relation \citep{Newton.2014}. 
\citet{Hejazi.2020} presented a catalog of 1,544 nearby dMs with stellar parameters determined from the latest version of BT-Settl model. 
\citet{sarmento.2021} derived the stellar parameters for 313 dMs through $\chi^2$ minimization using APOGEE H-band spectra. 
The LAMOST Stellar Parameter Pipeline at Peking University (LSP3) adopts a cross-correlation algorithm to determine stellar radial velocities (RV) and uses a template matching method for stellar atmospheric parameter determination \citep{Xiang.2015}.

The machine-learning methods make stellar parameter predictions based on data-driven models learned from large data set. 
The advantage of these data-driven models is their flexibility of learning patterns from spectra and transforming them into various stellar parameters.
\citet{maldonado.2020a} have presented a catalog of \teff\ and \feh\ via Principal Component Analysis (PCA) and sparse Bayesian methods for 204 dMs from HARPS GTO M-dwarf survey \citep{bonfils.2013}.
Based on \textit{The Cannon} \citep{Ness.2015,Casey.2016,Ho.2017}, \citet{Birky.2020} trained a data-driven model with high resolution spectra from APOGEE and derived stellar parameters (\teff\ and \feh) for 5,875 M-type stars.  
\citet{passegger.2020} trained a Convolutional Neural Network (CNN) based on synthetic spectra generated from PHOENIX-ACES model to estimate the stellar parameters for CARMENES dMs.

The Stellar Parameters and Chemical Abundances Network \citep[SPCANet,][]{Wang.2020} have constructed a CNN with three convolutional layers trained on spectra from the LAMOST Medium Resolution Spectroscopic survey \citep[MRS,][]{Liu.2020} with the APOGEE \textit{Payne} \citep{Ting.2019} stellar labels. 
Based on the Support Vector Regression \citep[SVR,][]{Smola.2004} method, \citet{Zhang.2020} developed a data-driven approach called the Stellar LAbel Machine (SLAM), which performs a nonlinear regression on stellar labels and the spectrum itself, and estimated the stellar labels for stars from LAMOST DR5 over a wide range of spectral types. 
\citet{li.2021} have adapted the SLAM program by introducing the APOGEE stellar labels and synthetic spectra from BT-Settl model as training sets, and obtained the stellar parameters for dMs from LAMOST DR6.

The empirical spectral library is also commonly used for determining the stellar parameters in large scale spectroscopic surveys, which directly or indirectly compare the observed spectra with templates listed in the spectral library. 
\citet{Yee.2017} presented a high resolution, high SNR empirical spectral library of 76 dMs, along with a parameterizing tool ``Empirical SpecMatch'', which estimates the stellar parameters for FGKM stars by comparing them against this spectral library. 
Other empirical spectral libraries, including STELIB \citep{LeBorgne.2003}, ELODIE \citep{Prugniel.2001,Prugniel.2007}, CFLIB (Indo-US library) \citep{Valdes.2004} and MILES \citep{SanchezBlazquez.2006,FalconBarroso.2011}, are composed of high quality spectra with good coverage of stellar atmospheric parameter space. 
With the help of empirical libraries of homogeneous M-type star templates, we are able to derive precise stellar atmospheric parameters from matching our spectra to these templates.

In this study, we employ the ULySS \citep{Koleva.2009} package to perform $\chi^2$ minimization between observed and model spectrum, and use the MILES interpolator \citep{Sharma.2016} to estimate the stellar atmospheric parameters for M-type stars from the LAMOST Low Resolution Spectroscopic survey (LRS) DR8.

The structure of this paper is as follows. 
Section~\ref{sect:data} introduces the preparation of our spectra from the LAMOST survey, and Section~\ref{sect:method} gives a brief description of the interpolation model and the ULySS program. 
In Section~\ref{sect:valid}, we present the summary of our results and compare the stellar parameters derived from our method with those in the literature. 
The deviation and precision are discussed in Section~\ref{sect:error}. 
Finally, in Section~\ref{sect:conclusion} we show our conclusions. 

\section{Observational Data} \label{sect:data}

The LAMOST spectral survey provides us with a massive number of medium (R $\sim$ 7,500) and low (R $\sim$ 1,800) resolution spectra, which are collected by the innovative active reflecting Schmidt telescope located in Xinglong Observatory, China \citep{Cui.2012,Zhao.2006,Deng.2012}.
The Schmidt optics system has a large aperture (effective aperture of 3.6\,m $\sim$ 4.9\,m) with a wide field of view (FOV $\sim 5^{\circ}$). 
The spectroscopic system contains 16 spectrographs with 32 integrated CCD cameras (4K $\times$ 4K). A total of 4,000 spectra can be obtained simultaneously via 4,000 fibers plugged into the spectrographs.

\citet{yi.2014} utilized the spectra from LAMOST pilot survey to determine the spectral subtype for 67,082 dMs by matching the relative strength of atomic and molecular features in the spectral region from 6,000\,\AA\ to 9,000\,\AA.
\citet{Zhong.2015} classified 8,639 gMs and 101,690 dMs/sdMs in LAMOST DR1 using spectral templates.
\citet{guo.2015} presented a catalog of 110,321 spectra for 93,619 dMs from LAMOST DR1, and separated gMs from dMs by spectral features and 2MASS near-infrared photometry.

After the eighth year of the regular low-resolution survey, a total of 11,214,076 low resolution spectra is available in LAMOST DR8\footnote{http://DR8.lamost.org/}. 
This catalog contains 10,388,423 stellar spectra with a resolution of R $\sim$ 1,800 at the 5,500\,\AA, among them 773,721 spectra are from M-type stars, 520,934 of them have r-band SNR higher than 10. 
These spectra cover a wavelength range from 3,690--9,100\,\AA, including the blue channel which is optimized for 3,690--5,900\,\AA, and the red channel for 5,700--9,100\,\AA. 
In this study, we adopt the identified M-type stars from LAMOST DR8 to derive the stellar parameters by applying the $\chi^2$ minimization performed by the ULySS program.

\section{Methodology} \label{sect:method}

\subsection{MILES Interpolator}
\label{subsect:interpolator}
Spectra libraries collect a set of widely-used templates and corresponding stellar parameters to classify stars and synthesize stellar populations. 
For example, the ELODIE library \citep{Prugniel.2001,Prugniel.2007} contains 1,962 spectra of 1,070 stars obtained from the ELODIE spectrograph with a resolution of R $\sim$ 42,000. 
\citet{Soubiran.2008} have determined the stellar parameters of ELODIE stars using the TGMET code \citep{Katz.1998}. The MILES library \citep{SanchezBlazquez.2006,FalconBarroso.2011} consists of 985 stars at optical region with a resolution of R $\sim$ 2,200, which were obtained from the 2.5\,m Issac Newton Telescope (INT). 
For stars in the MILES library, \citet{Cenarro.2007} compiled and calibrated stellar atmospheric parameters from the literature. 

Thus, the MILES library is considered to be an ideal empirical library for stellar atmospheric parameter determination. 
\citet{Prugniel.2011} redetermined a set of homogeneous stellar parameters for MILES stars, and built an interpolator function (version 1, hereafter V1) based on MILES spectra which can generate model spectra based on a function of stellar parameters. 
The reliability of V1 has been proofed as it shows better performance for both hot evolved and cool stars. 

Based on V1, \citet{Sharma.2016} extended the parameter space with several cool stars and refined V1 interpolator with effective temperature downwards to 2,900\,K. 
This new interpolator\footnote{https://cdsarc.cds.unistra.fr/ftp/J/A+A/585/A64/miles\_tgm2.fits.gz} (version 2, hereafter V2) recalculated the stellar parameters of V1 in the well-populated regions of the parameter space, while in the sparse border regions, it included external spectra for improvement. 
The V2 interpolator also extended the validity of M-type stars, which used a fine-tuned 26 terms polynomial to improve the modelling and decrease the biases.
The valid spectral range of V2 interpolator is 3,536\,\AA--7,410\,\AA, which is limited by the wavelength coverage of the MILES library.

\subsection{ULySS: Université de Lyon Spectroscopic analysis Software}

ULySS\footnote{http://ulyss.univ-lyon1.fr/} \citep{Koleva.2009} is a full-spectrum fitting package based on IDL/GDL language, which has been used in various types of tasks.
In this work, we employ ULySS to derive the stellar atmospheric parameters by fitting a spectrum with a linear function of non-linear models:

\begin{equation}
	\textit{F}_{\text{obs}} = \text{P}_n(\lambda) \times {\text{TGM}}(T_{\rm eff}, \log{g}, \text{[Fe/H]}, \lambda) \otimes {\text{G}}(v_{\text{sys}}, \sigma) 
	\label{eq:ULySSEqu}
\end{equation}

Where $\text{P}_n$ is an $\textit{n}$-th order polynomial, $\text{TGM}$ represents an interpolator function modeling the spectrum with the variables denoting effective temperature, surface gravity, metallicity and wavelength respectively.
$\text{G}(v_{\text{sys}}, \sigma)$ describes the Gaussian broadening caused by the systemic velocity $v_{\text{sys}}$, and the velocity dispersion $\sigma$. The interpolator TGM is a function of the stellar atmospheric parameters (\teff, \logg\ and \feh) and wavelength $\lambda$, providing a deduction of stellar flux distribution and $\chi^2$ minimization.

For M-type stars, the spectra are typically occupied by dense molecular absorption bands, which make the continuum hard to determine.
Because of this, we do not calculate the pseudo-continuum for normalization.
Instead, we adopt a multiplicative polynomial $\text{P}_n$ as the scaling factor making the theoretical spectrum comparable to the observed ones.
This procedure achieves the same effect as the ordinary continuum normalization.
Due to the existence of molecular absorption lines, a polynomial of lower orders may not fit the features correctly, while a higher order of the multiplicative polynomial can easily over-fit.

To find the best degree of polynomial $\text{P}_n$, we use a series of $\text{P}_n$ with different orders in the fit procedure, calculate the loss value, then evaluate the multiple sets of results obtained.
The loss value is defined as:

\begin{equation}
	\textit{loss} = \sqrt{\sum_{i=0}^{n} (x_i - x_{\text{ref}})^2}
	\label{eq:lossvalue}
\end{equation}

The $x_i$ represents each derived stellar parameter, and the $x_{\text{ref}}$ is the corresponding parameter provided by ASPCAP.

To estimate the overall performance of these stellar parameters,
 we calculate the total loss value by performing a weighted average for the loss values of effective temperature ($\textit{loss}_T$), surface gravity ($\textit{loss}_G$) and metallicity ($\textit{loss}_M$): 

\begin{equation}
	\textit{loss}_{\text{tot}} = \frac{w_T \times \textit{loss}_T + w_G \times \textit{loss}_G + w_M \times \textit{loss}_M}{w_T + w_G + w_M}
	\label{eq:loss_total}
\end{equation}

In general, the \rm{Fe} absorption lines significantly influence the determination of \feh, therefore we decide to empirically give higher weights to \feh, specifically, $w_T = 1$, $w_G = 1$ and $w_M = 1.5$.

\begin{figure}[h]
	\centering
	\includegraphics[width=0.47\textwidth]{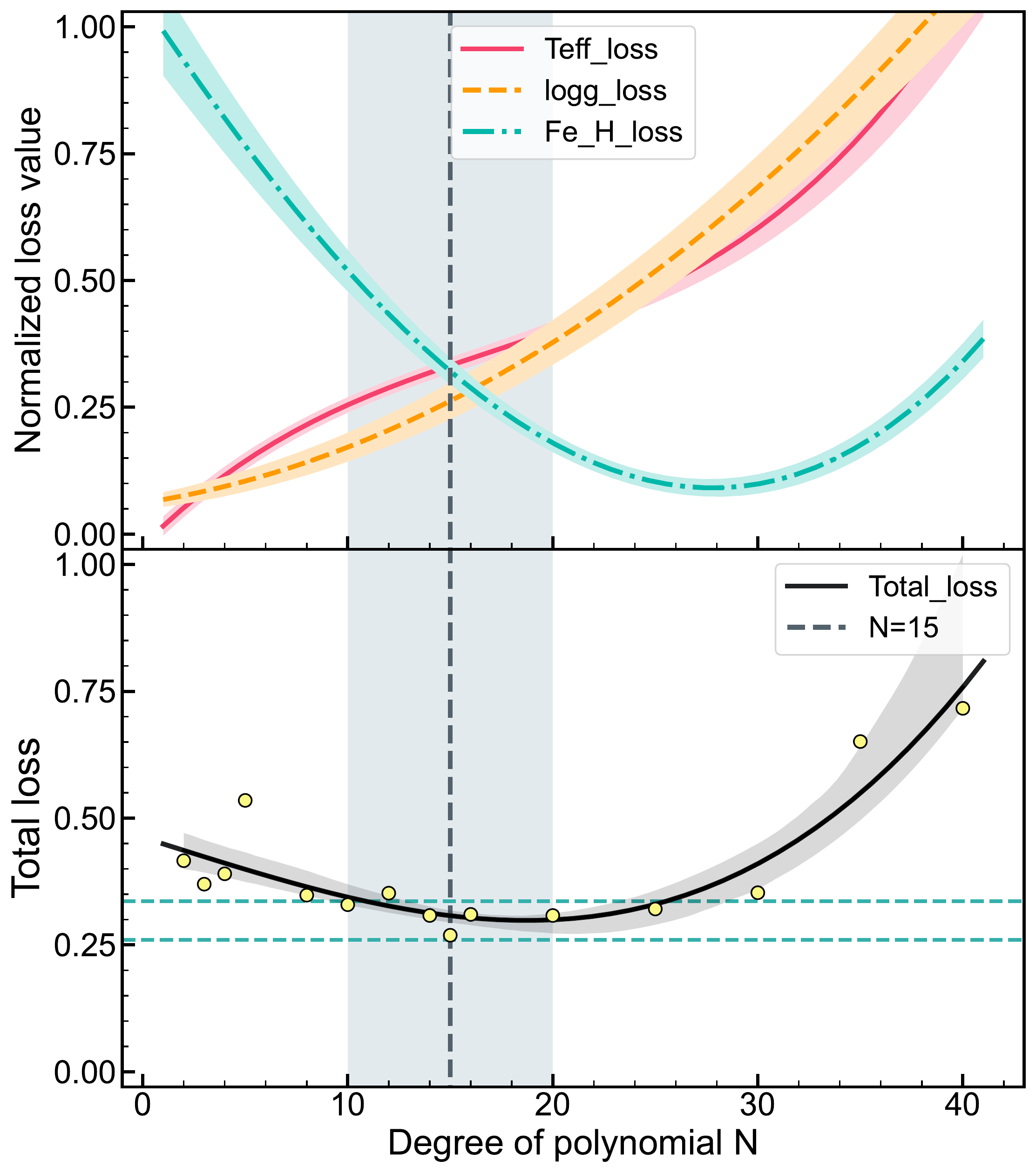}
	\caption{The variations of loss value as a function of the degree of $\text{P}_n$. Top panel : the variation trends of loss value for stellar parameters : \teff\ (red), \logg\ (orange) and \feh\ (blue) by changing the degree of the polynomial $\text{P}_n$. Bottom panel : the total loss value (weighted average of three loss values above). To make the loss values of different stellar parameters comparable, each loss value is normalized to [0, 1].}
	\label{fig:n_degrees_loss}
\end{figure}

Figure~\ref{fig:n_degrees_loss} shows the loss function of each stellar parameter, as well as the total loss value. 
The loss values of effective temperature and surface gravity show positive correlations with the degree of polynomial $\text{P}_n$, while the loss value of metallicity declines first, and then rises as the order N increases.

Eventually, we decide to adopt a maximum degree $N = 15$ for the polynomial $\text{P}_n$, which ensures the spectra fitted properly, and does not hit the marginal effect wall. 
Also, we apply ``\text{/CLEAN}'' option in ULySS program to eliminate the spikes (emission lines, bad pixels) during the fitting procedure.

After scaling the spectrum, we adopt a $\chi^2$ minimization approach to deduce the stellar parameters.
The $\chi^2$ is defined as:

\begin{equation}
	\chi^2 = \sum_{i = 1}^{n_{pxl}} (\frac{O_i - S_i}{\sigma_i})^2
	\label{eq:chi-sqr}
\end{equation}

Here, $n_{pxl}$ stands for the number of pixels of the spectrum, while $O_i$ and $S_i$ represent the flux at the $\textit{i}$-th pixel of the observed spectrum and the synthetic spectrum. The $\sigma_i$ is the standard deviation of flux measurement at the $\textit{i}$-th pixel of the observed spectrum. 

More specifically, we iteratively change the input parameters until the reduced $\chi^2$ reaches a local minimum. 
The reduced $\chi^2$ is defined as :

\begin{equation}
	\chi^2_{\nu} = \frac{\chi^2}{\nu}
	\label{eq:reduced_chi-sqr}
\end{equation}

Here, $\nu$ is the degree of freedom (DOF), $\nu = n_{pxl} - n_{para}$, which equals the difference between the number of pixels ($n_{pxl}$) and the number of free parameters ($n_{para}$).

\begin{figure}[h]
	\centering
	\includegraphics[width=0.5\textwidth]{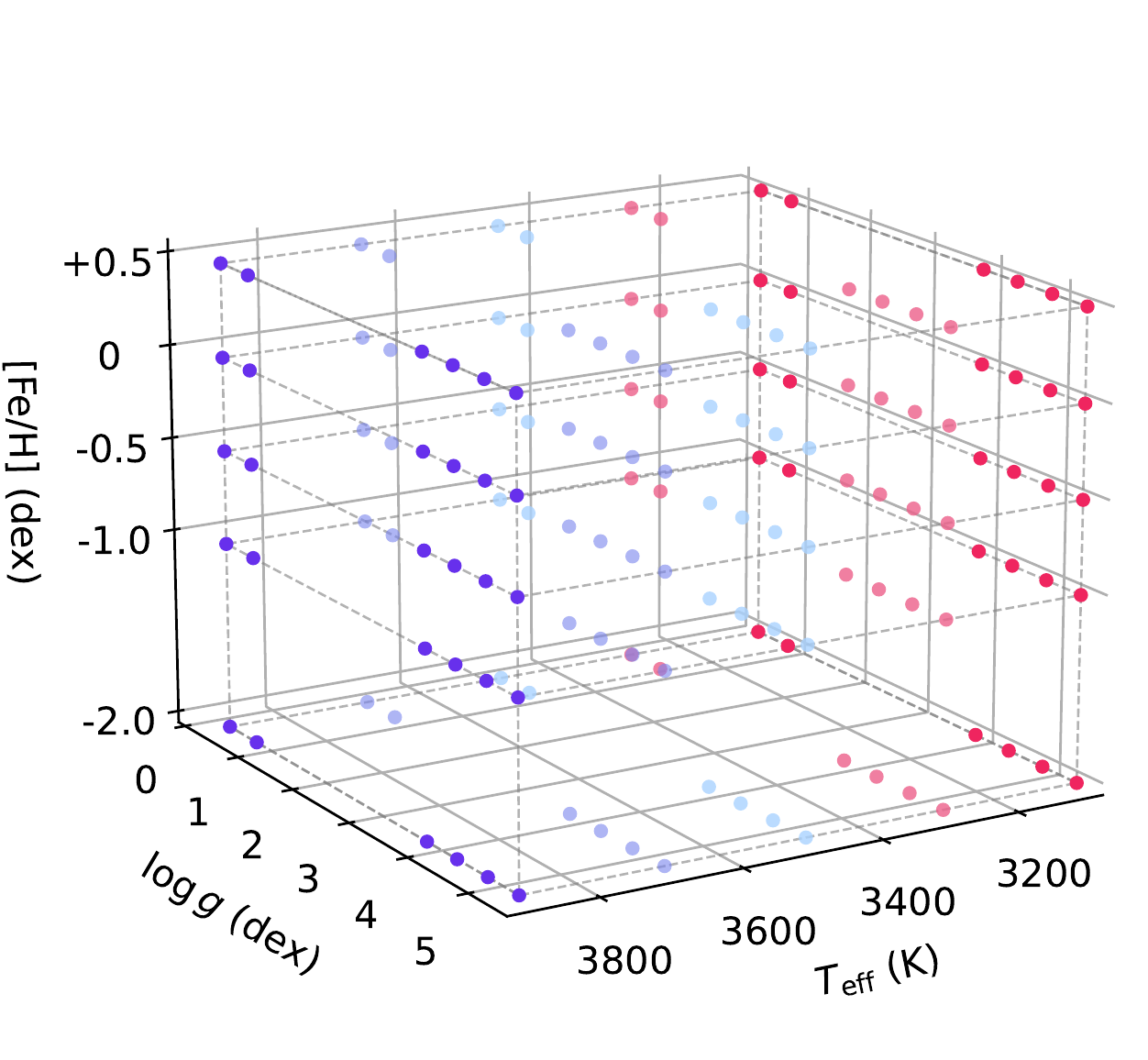}
	\caption{The initial guess values of \teff, \logg\ and \feh. Dots are color coded in different effective temperatures. Detailed information is listed in Table~\ref{Tab:guess_grid}.}
	\label{fig:initial_guess}
\end{figure}

Table~\ref{Tab:guess_grid} lists the initial guess grid of stellar parameters adopted in this work.
It should be noted that, the initial guess values provide our program with a grid of start points, and the final results are determined by the $\chi^2$ minimization algorithm which may exceed the grid region (see Figure~\ref{fig:initial_guess}).

\begin{deluxetable}{lrrrrrr}
\centering
\tablecaption{The initial guess grid of stellar parameters.}
\label{Tab:guess_grid}
\tablehead{
\colhead{Variable} & \multicolumn{6}{c}{Initial guesses}}
\startdata
	\teff\ (K)                & 3100   & 3300   & 3500   & 3700   & 3900   &     \\
	\logg\ (dex)  & 0.50    & 1.00    & 4.00    & 4.50    & 5.00    & 5.50 \\
	\feh\ (dex)       & $-2.00$   & $-1.00$   & $-0.50$   & 0.00    & 0.50   &     \\
\enddata
\end{deluxetable}

In order to apply the V2 interpolator to LAMOST spectra, we have separately estimated the stellar parameters for a group of spectra by fitting the blue (3,690--5,700\,\AA) and the red (5,900--9,100\,\AA) segments of the spectra, as well as the assembled (3,690--9,100\,\AA) spectra. 
After a series of experiments, we found that rather than using the blue regions or the assembled spectra, fitting the red regions usually shows more robust dispersion and lower systematic error. 
The red segment generally has better SNR and the \logg\ sensitive molecular bands such as \ce{TiO} and \ce{CaH} bands locate in this spectral region \citep{Gizis.1997}. 
Although some spectral signatures may be lost, we have to manually set the upper limit of our spectra as 7,400\,\AA\ to avoid unreliable extrapolation. 
Therefore, we decide to use the overlap parts (5,700 $\sim$ 7,400\,\AA) of V2 spectral range and the red region of LAMOST spectrum during the fitting. 

\begin{figure}[h]
	\centering
		\includegraphics[width=0.5\textwidth]{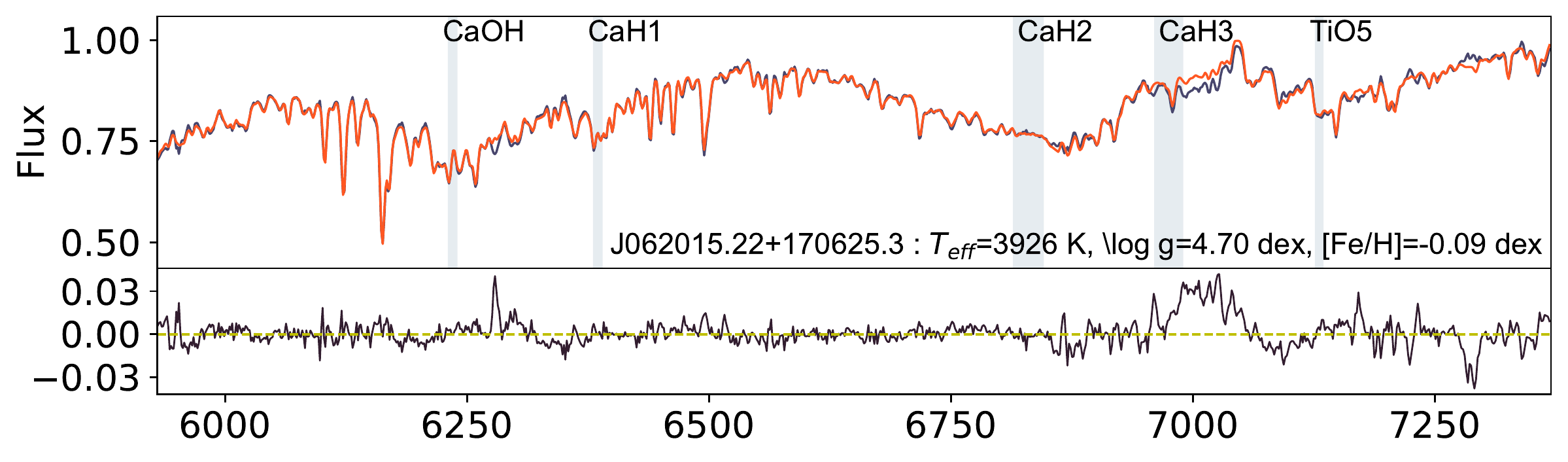}
		\includegraphics[width=0.5\textwidth]{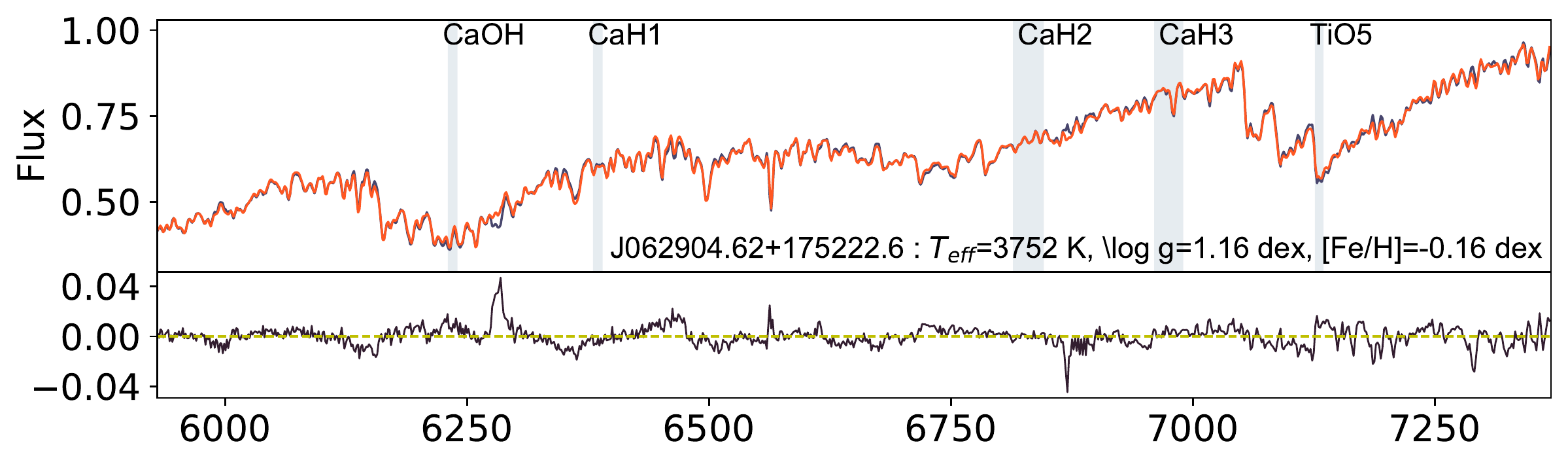}
		\includegraphics[width=0.5\textwidth]{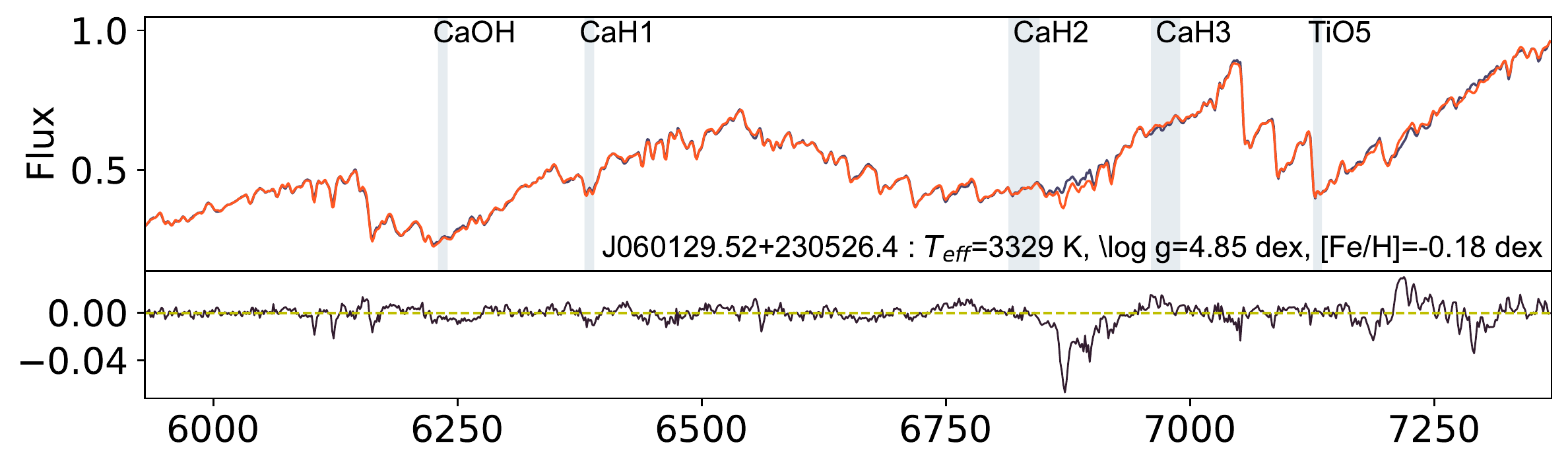}
		\includegraphics[width=0.5\textwidth]{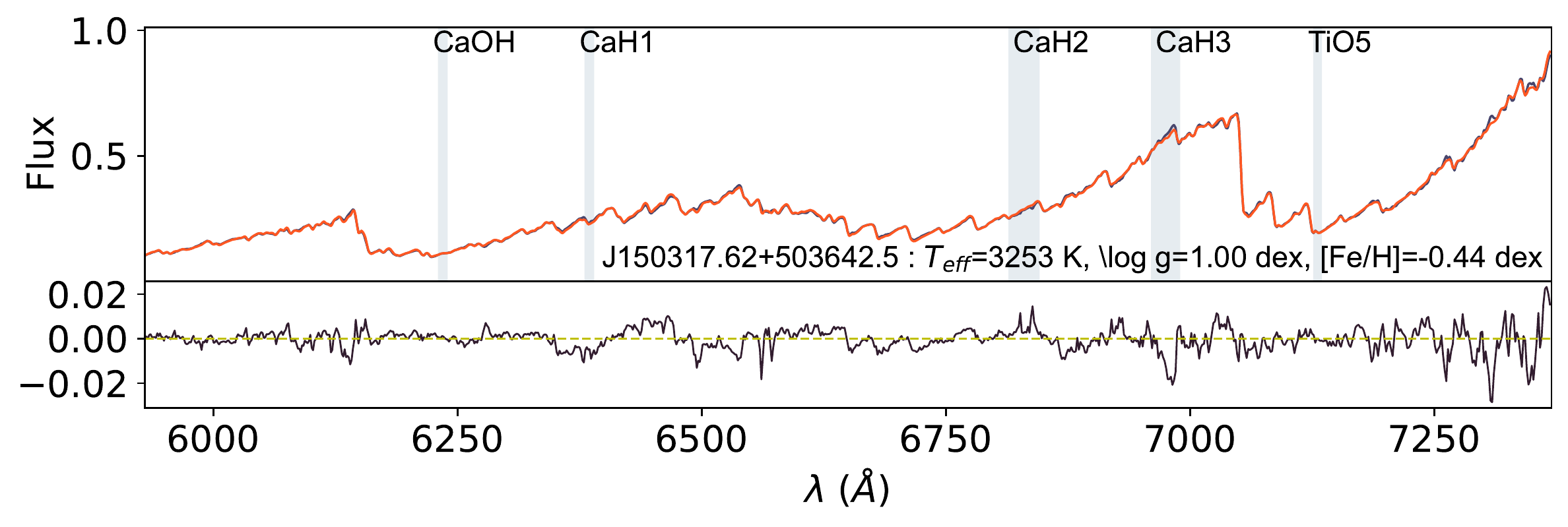}
	\caption{Samples of our fitting results with different spectral subtypes (from top to bottom: dM0, gM2, dM4 and gM6). 
	Black and red lines are observed and model spectra, respectively. The designations and corresponding stellar parameters are listed, and the residuals between model and observed spectra are shown at the bottom.}
	\label{fig:spectra_samples}
\end{figure}

Finally, we construct a template matching routine for loading the observational spectrum and finding the best solution, by using ULySS 1.3.1 combined with the V2 interpolator. 
We display four samples of our fitting results with different spectral subtypes (dM0, gM2, dM4 and gM6) in Figure~\ref{fig:spectra_samples}, and the residuals exhibit good consistencies in our fitting region of 5,700--7,400\,\AA. 
However, the fitting is not as good as the others in the spectral region near the red end, which indicates the instability of model spectrum generation restricted by the interpolator.

\section{Validation of Our Method} \label{sect:valid}
Before applying our method to the entire sample set, it is necessary to validate this method using a spectrum subset equipped with reliable external stellar parameters.
Therefore, we cross-match with APOGEE DR17 \citep{ahumada.2020,Jonsson.2020,abdurrouf.2022} and \RefNum\ suitable papers from the literature, then compare the stellar parameters for the common stars.

\subsection{Comparison with APOGEE DR17}
\label{subsec:APOGG_valid}

We cross-match our sample set with APOGEE DR17 allStar-file\footnote{https://dev.sdss.org/dr17/irspec/spectro\_data/}, following the criteria below:

\begin{enumerate}
	\item{} $2900 < T_{\text{eff}} < 4200$ K;
	\item{} The spectra tpye property in the LAMOST DR8 catalog is classified as M-type by LASP;
	\item{} The signal-to-noise ratio of r-band (\snrr) for the observed spectra should larger than 20;
	\item{} The SNR of u, g, r, i and z-band for the observed spectra should not equal to $-9999$.
\end{enumerate}

\snrr\ is the SNR value at SDSS r-band filter \citep[5550--6900\,\AA,][]{gunn.1998} and is defined as the median value of each pixel in this band.
Note that unless specified, all SNR presented hereafter refer to the \snrr\ value. 
The aim of this constraint is to ensure that our sample has good observational quality in the r-band region where covers the distinctive spectral features. 
The criterion of SNR $>$ 20 can exclude most of the misfits caused by low SNR, which will be discussed in Section~\ref{sect:error}. 
To avoid the influence of poor data quality, we discard the spectra whose SNR value have been artificially set as $-9999$ in corresponding spectral region. 
In this way, we have selected 19,592 spectra of 14,532 unique stars after cross-matching. 
For targets with repeated observations, we only keep the spectrum with highest SNR in order to avoid the impact of low spectral quality. 
Thus, we obtain a total of 14,532 common stars.

\begin{figure}[!htbp]
	\centering
    \fig{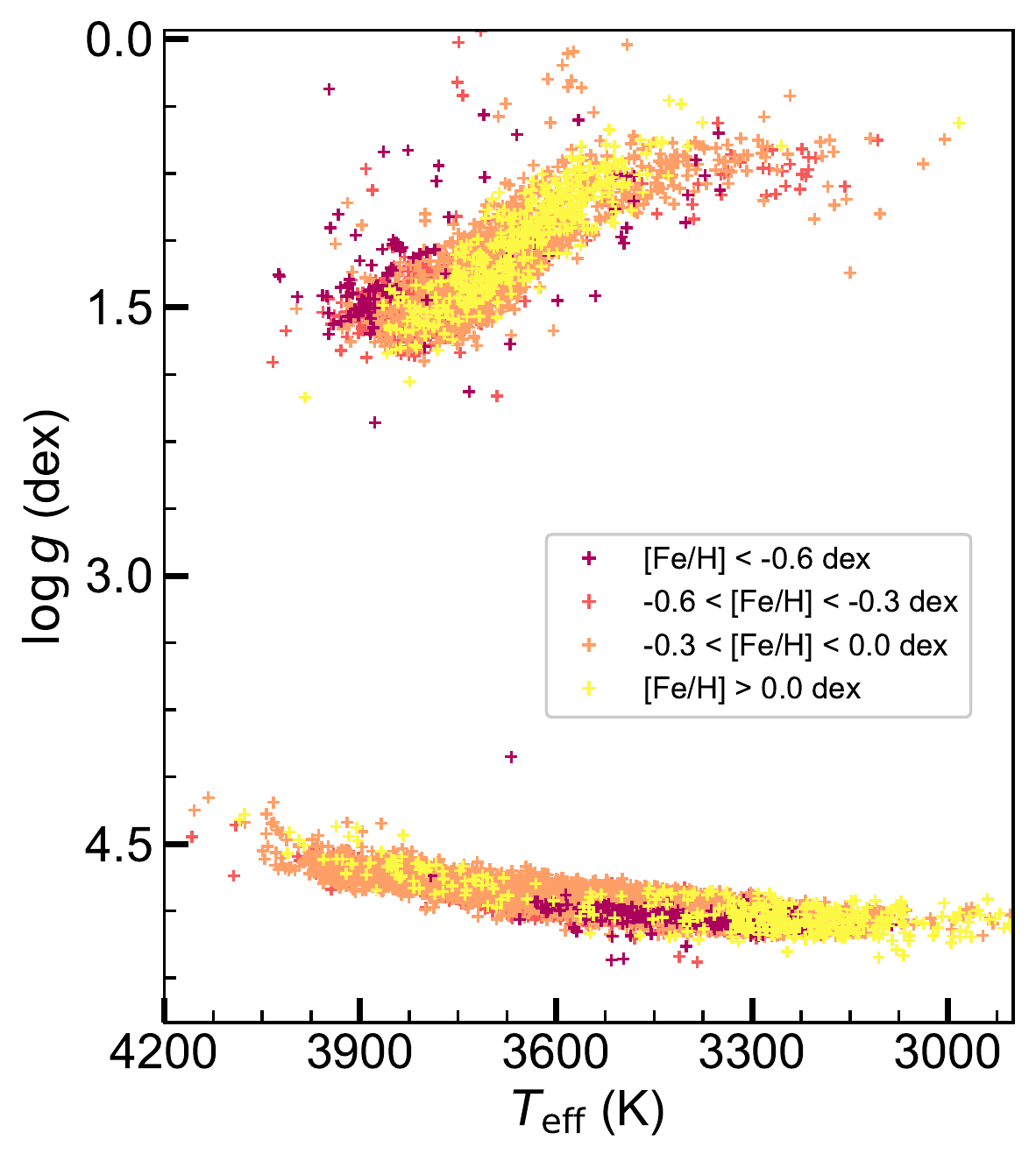}{0.4\textwidth}{}
	\caption{The distribution of \teff\ versus \logg\ for common stars from APOGEE DR17. Stars are color coded in different metallicity groups.}
	\label{fig:APG_para_dist}
\end{figure}

Furthermore, we exclude the binaries by cross-matching with the binary catalogs including the binaries of APOGEE \citep{el-badry.2018}, the Washington Double Star Catalog \citep[WDS,][]{mason.2001}, the third revision of Kepler Eclipsing Binary Catalog \citep[KEB\Rmnum{3},][]{kirk.2016} and the Gaia-ESO binary candidates \citep{merle.2017}. 
Similarly, we exclude the variables by cross-matching with the General catalog of Variable Stars \citep[GCVS,][]{samus.2017}, the International Variable Star Index \citep[VSX,][]{watson.2006} and the ASAS-SN catalog of variable stars \citep{jayasinghe.2021}.
After these cuts, there are 13,846 common M-type stars with APOGEE DR17. 
In Figure~\ref{fig:APG_para_dist}, we present the parameter distribution (\teff\ versus \logg) of these stars.

\begin{figure*}[!ht]
	\centering
    \gridline{
    \fig{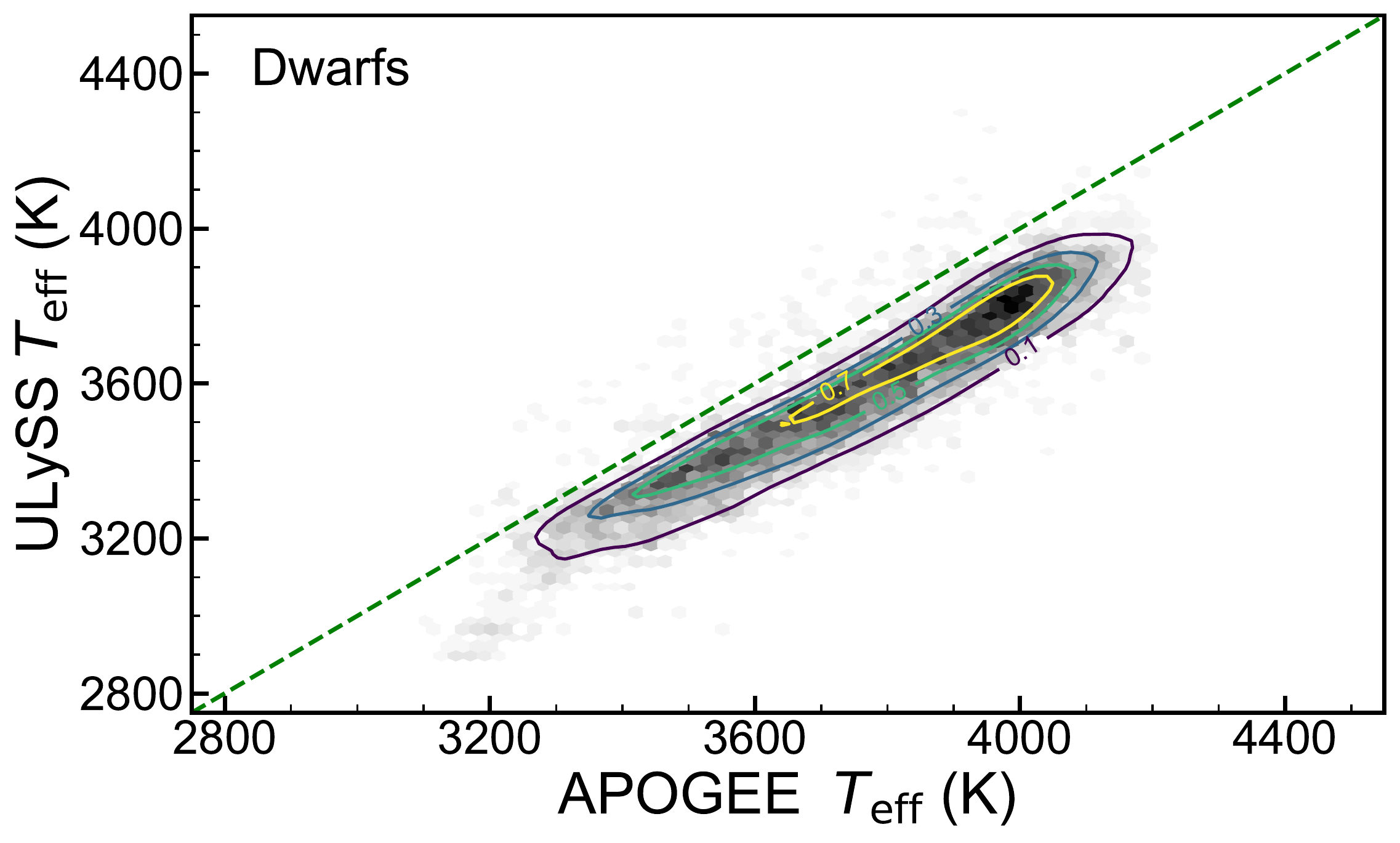}{0.45\textwidth}{}
    \fig{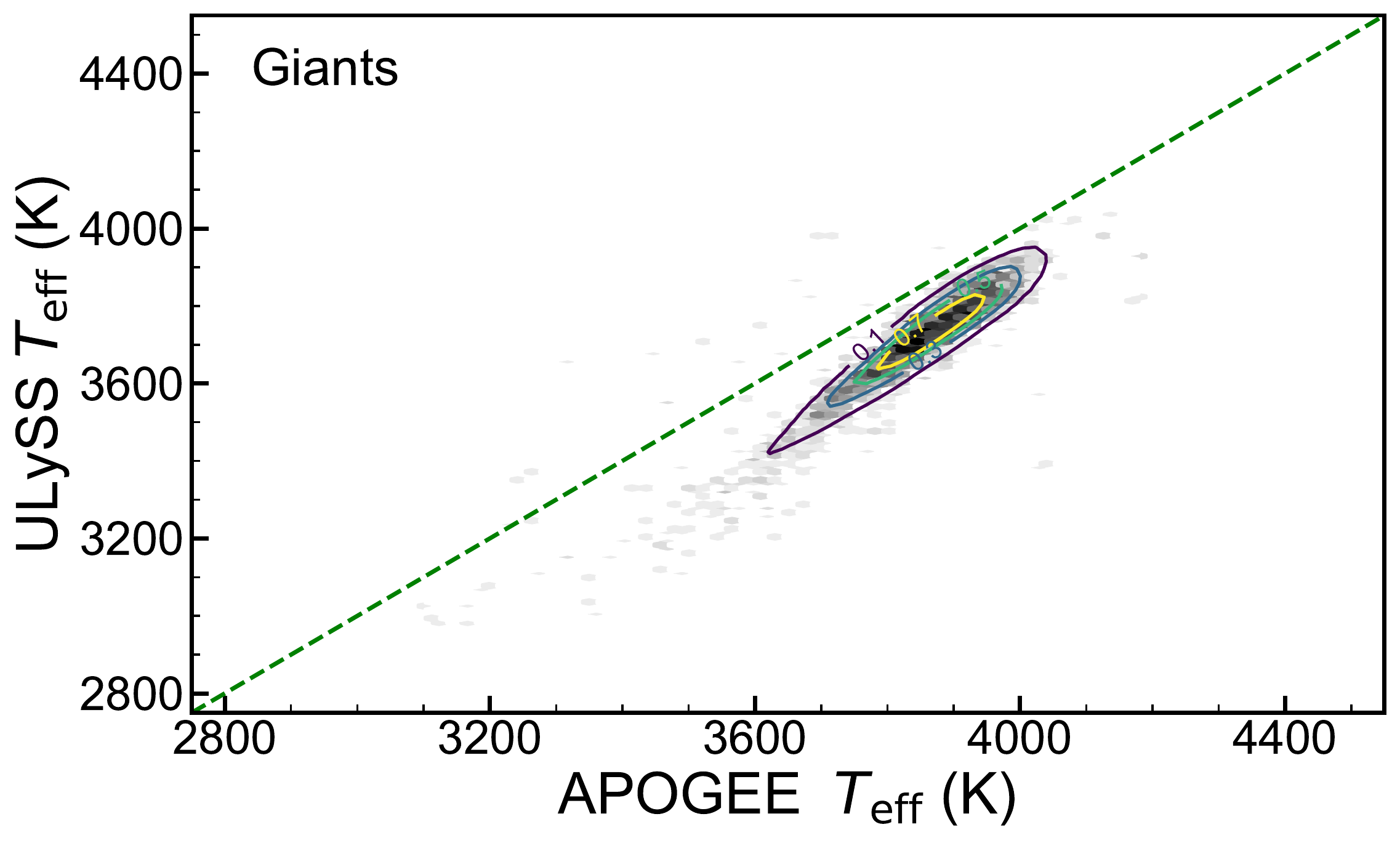}{0.45\textwidth}{}
	}  \vspace{-7mm}
        \gridline{
    \fig{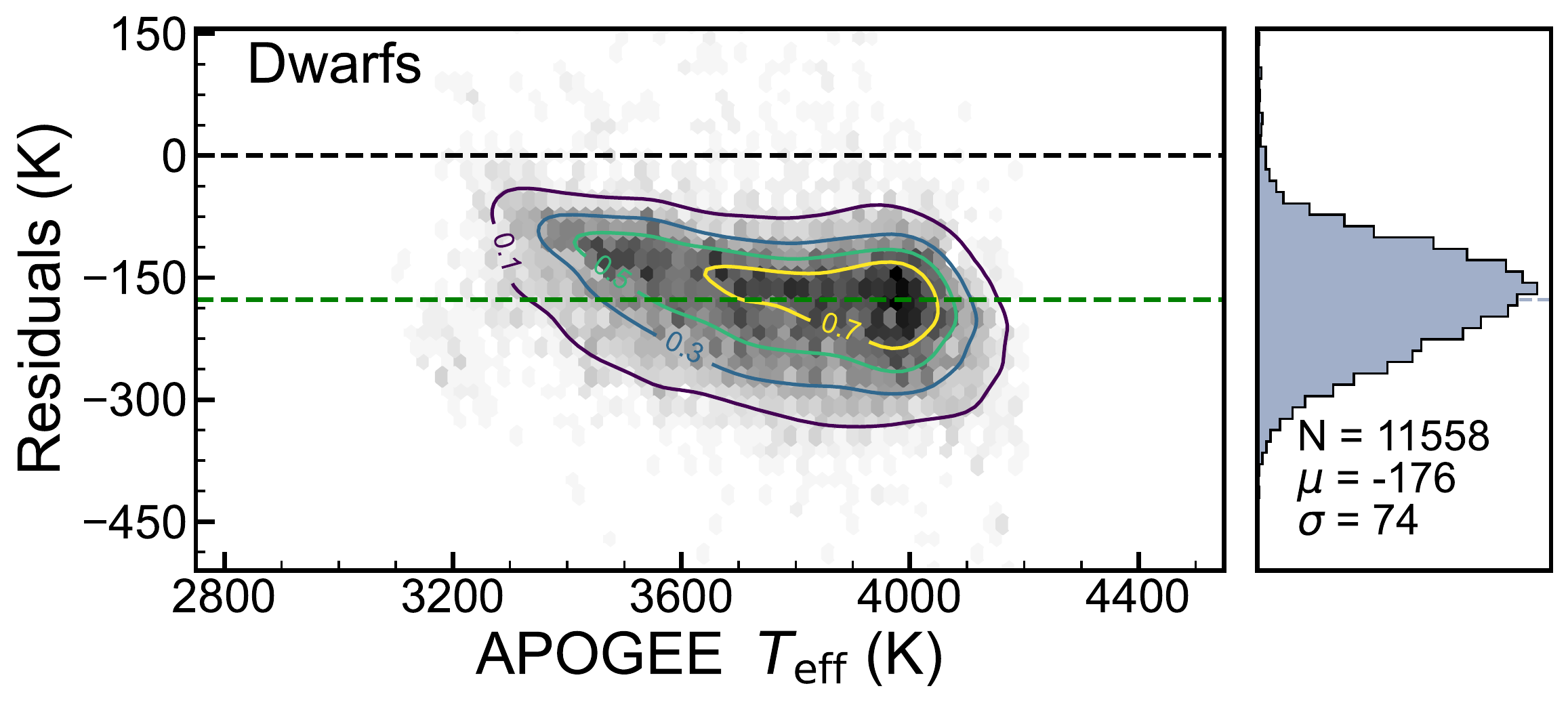}{0.45\textwidth}{}
    \fig{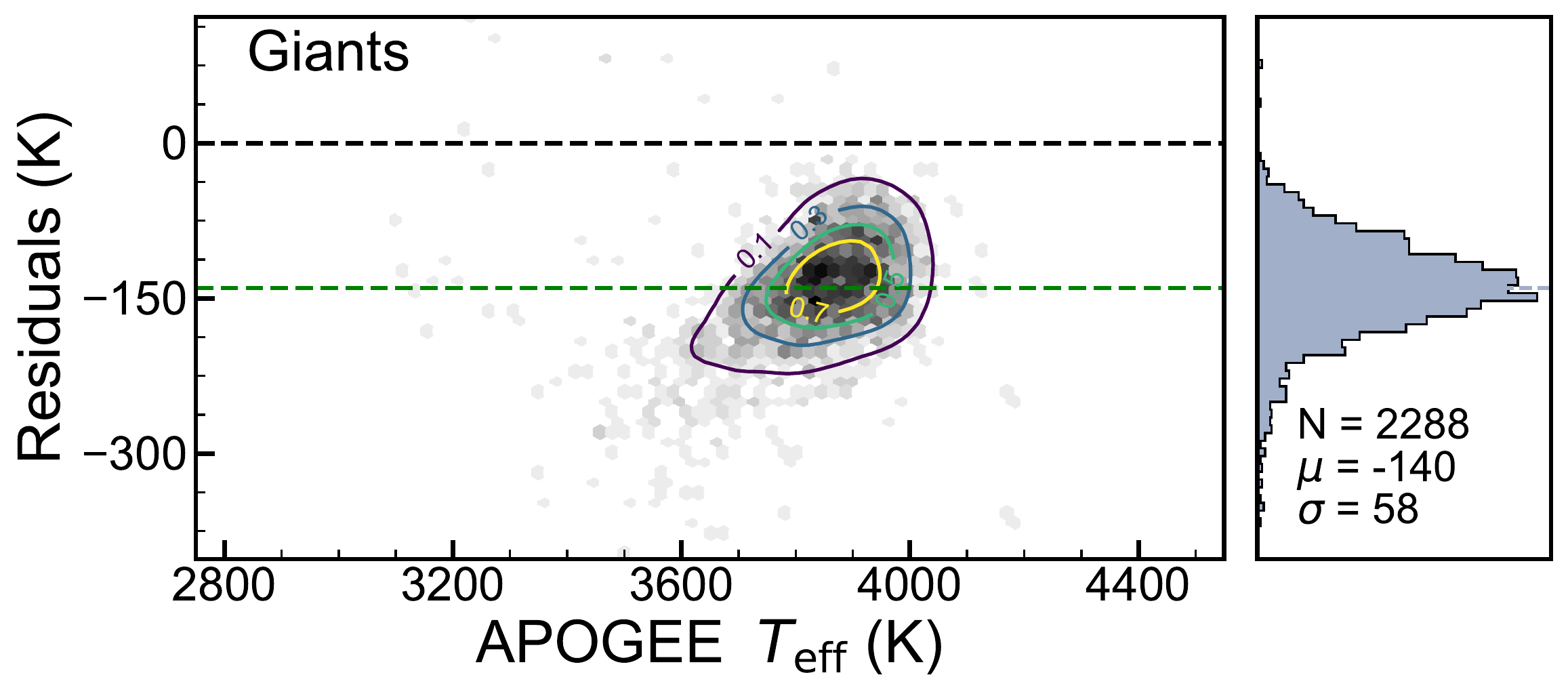}{0.45\textwidth}{}
    }  \vspace{-7mm}    
    
	\caption{The comparison of the effective temperatures derived from our method with the APOGEE DR17 calibrated values. 
	The top-left panel shows the \teff\ comparison for dMs, while the top-right panel for gMs. 
	The bottom-left and bottom-right panels show the residuals' distributions for dMs and gMs, respectively. 
	The color scale as well as the contour lines indicate the number density of each region.
	}
	\label{fig:TvsAPOGEE}
\end{figure*}

\begin{figure*}[!ht]
	\centering
    \gridline{
    \fig{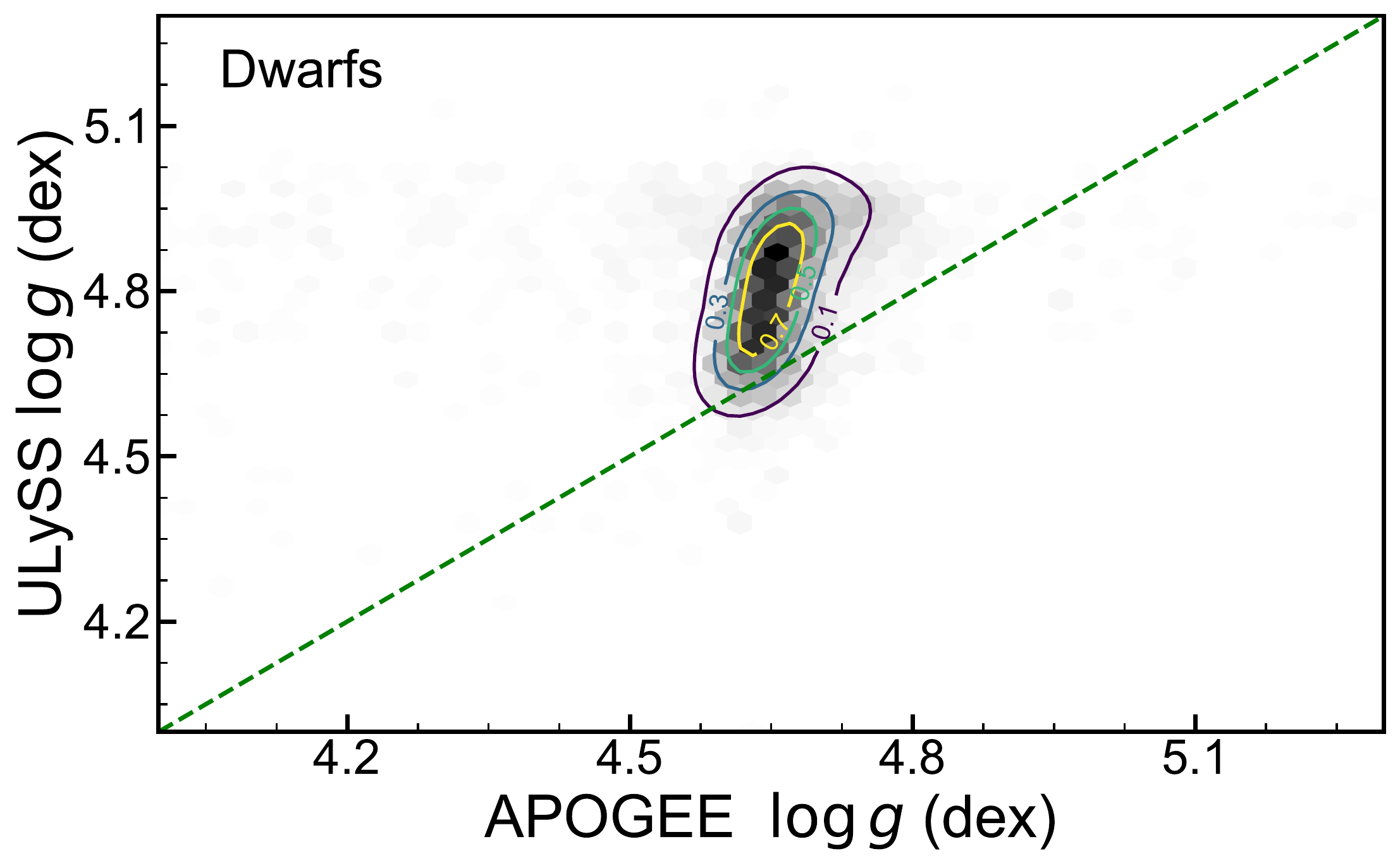}{0.45\textwidth}{}
    \fig{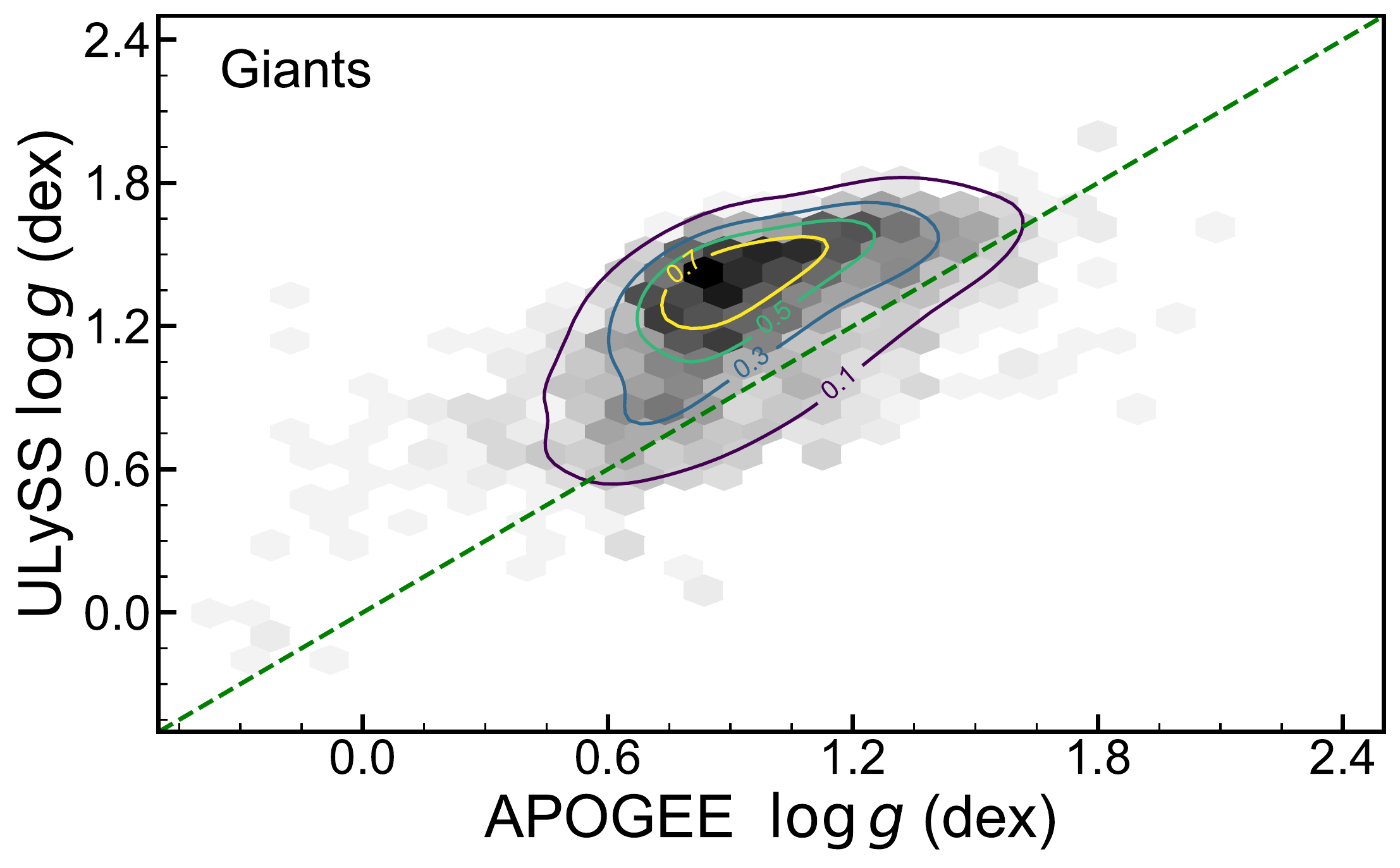}{0.45\textwidth}{}
	}  \vspace{-7mm}
        \gridline{
    \fig{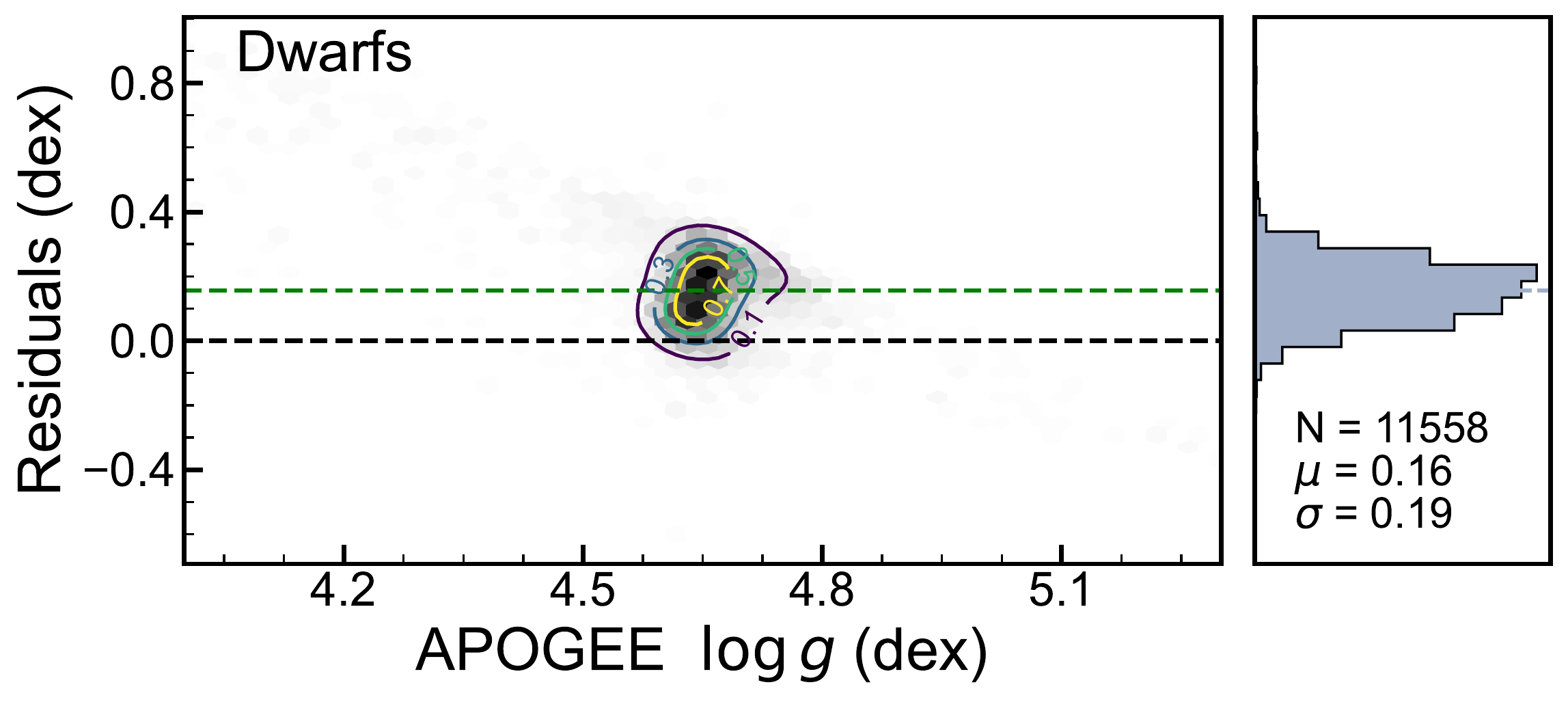}{0.45\textwidth}{}
    \fig{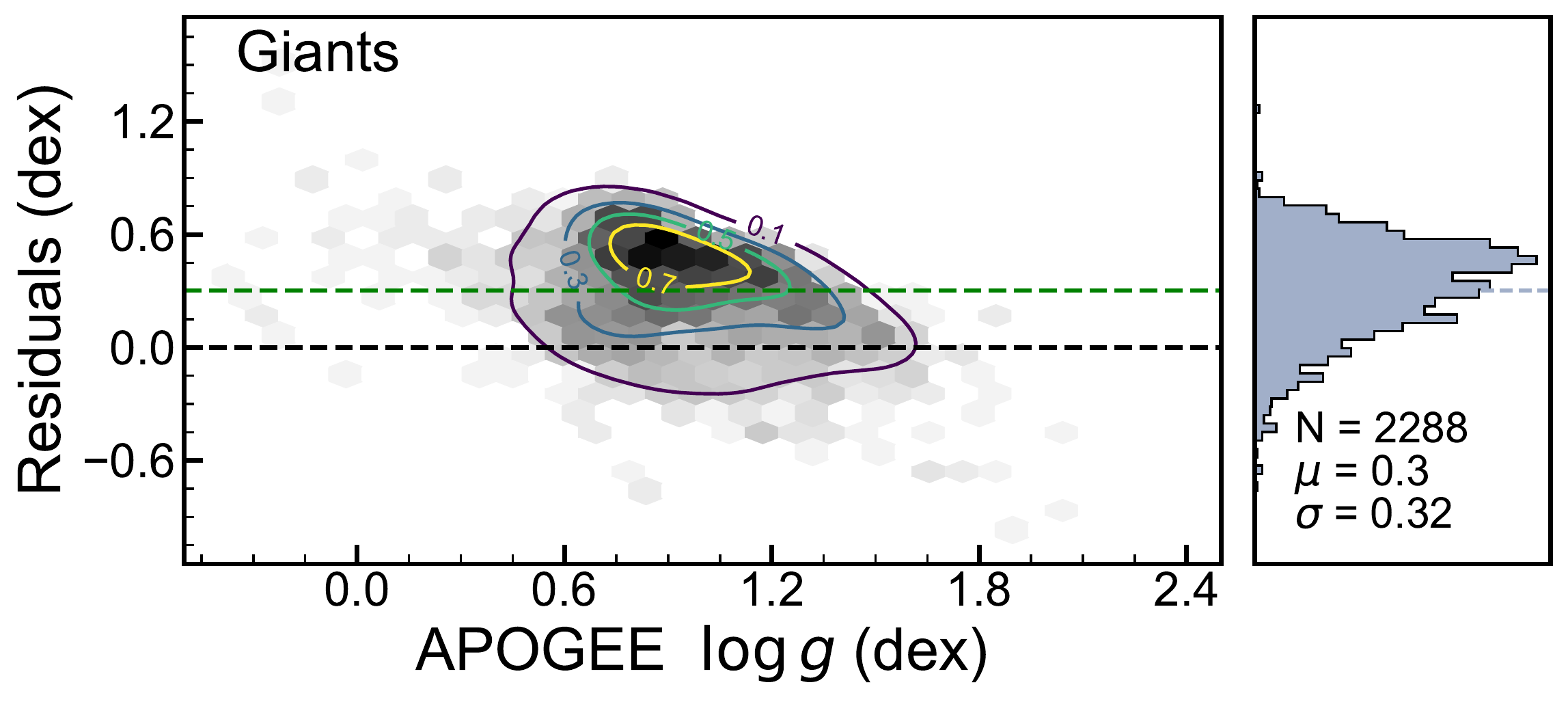}{0.45\textwidth}{}
    }  \vspace{-7mm}    
    
	\caption{Same as Figure~\ref{fig:TvsAPOGEE}, but for surface gravity.
	}
	\label{fig:GvsAPOGEE}
\end{figure*}

\begin{figure*}[!ht]
	\centering
    \gridline{
    \fig{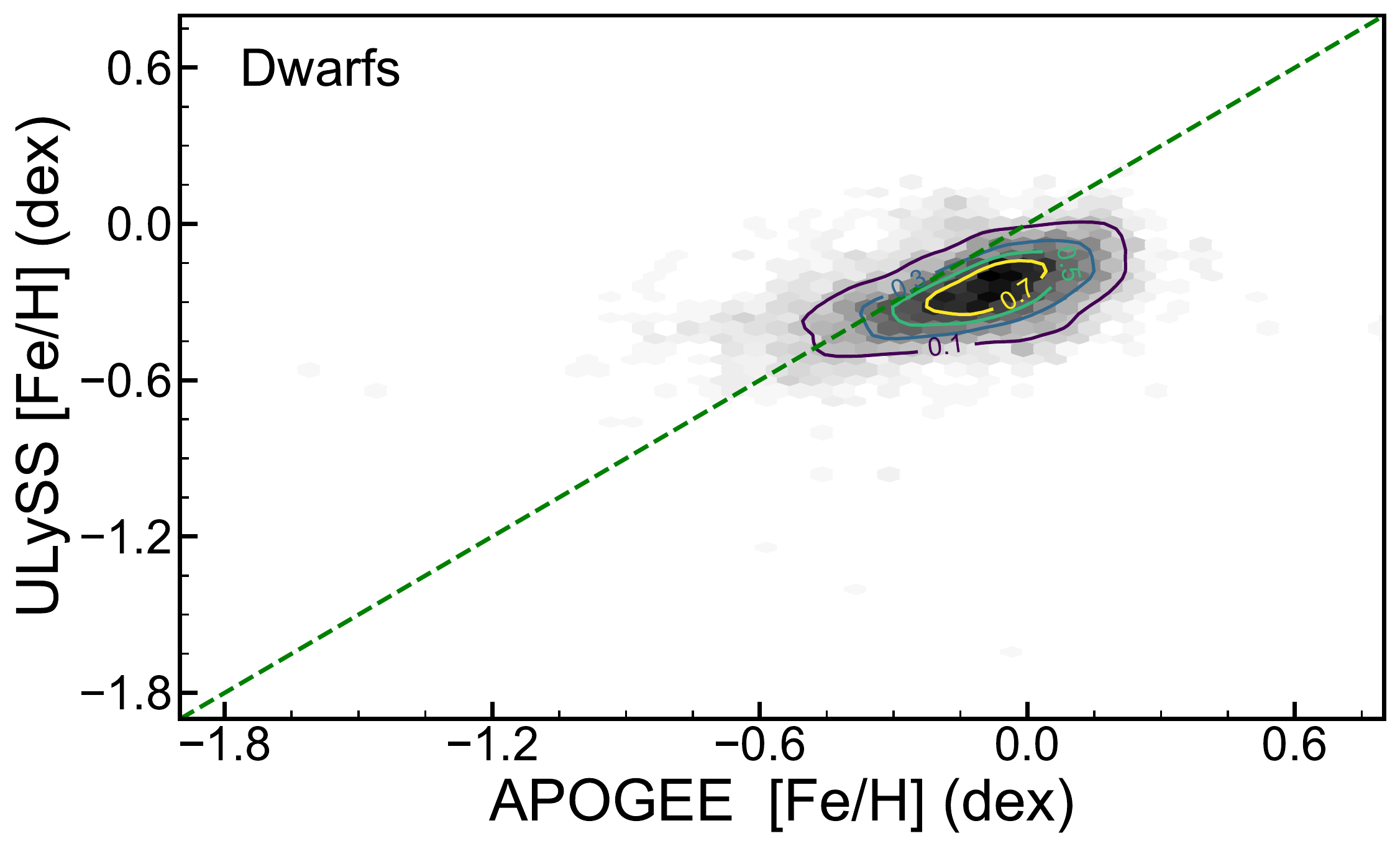}{0.45\textwidth}{}
    \fig{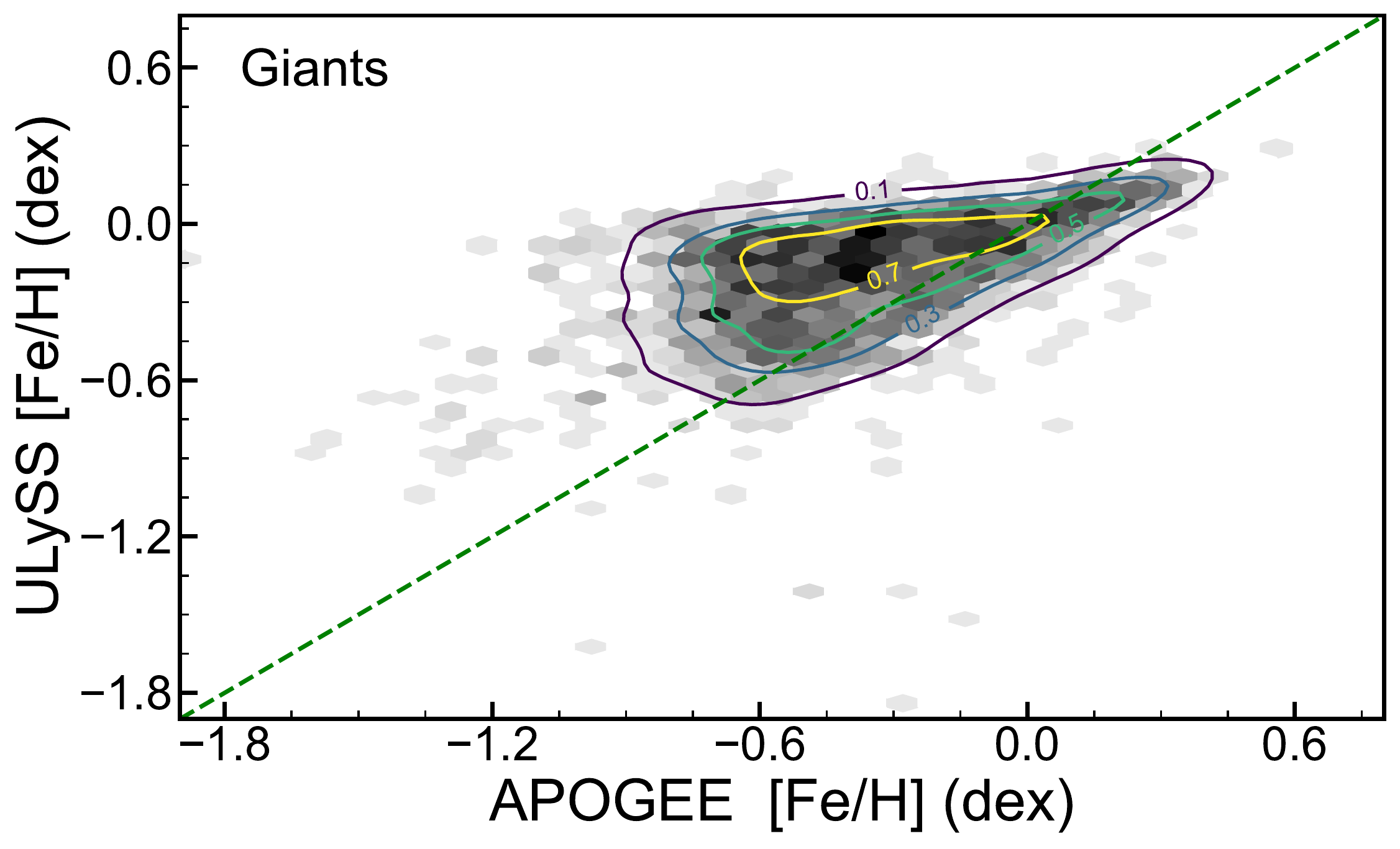}{0.45\textwidth}{}
	}  \vspace{-7mm}
        \gridline{
    \fig{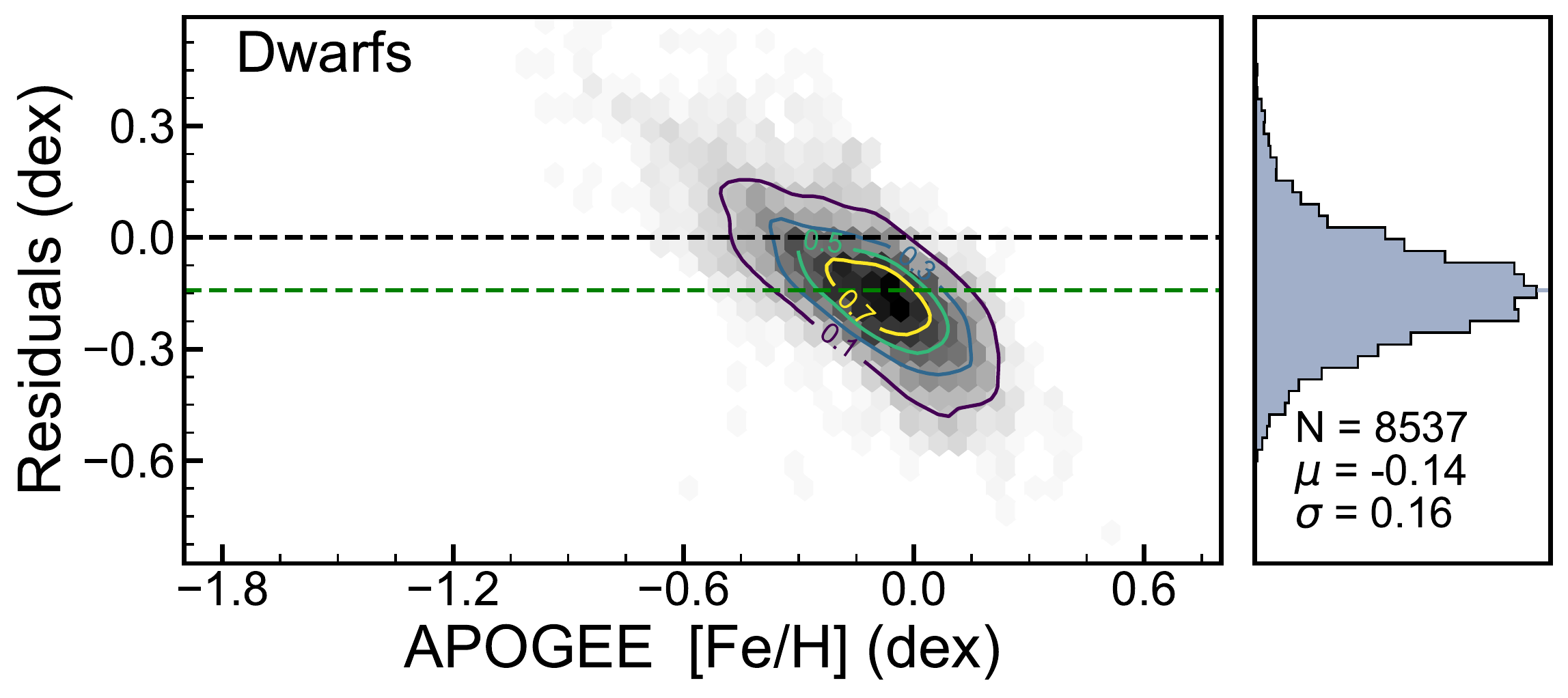}{0.45\textwidth}{}
	\fig{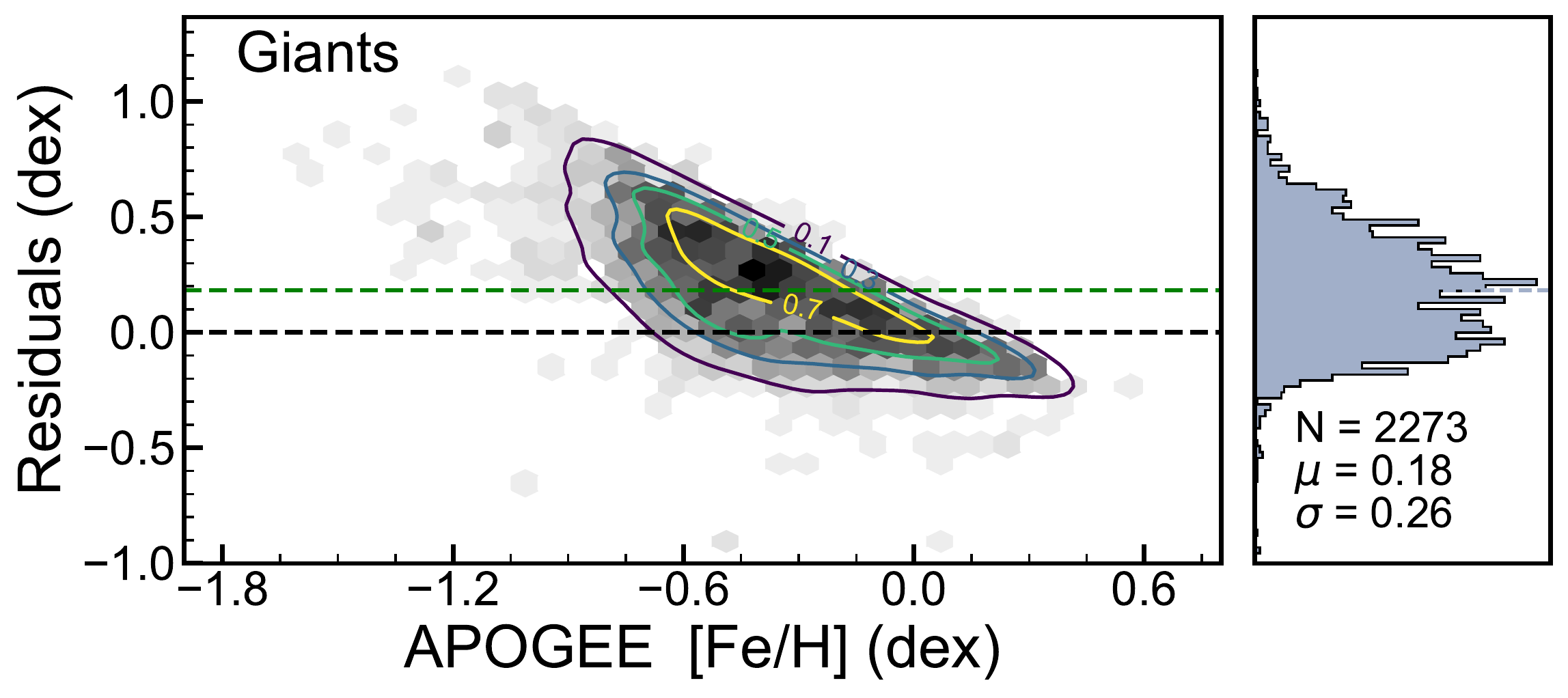}{0.45\textwidth}{}
    }  \vspace{-7mm}    
    
	\caption{Same as Figure~\ref{fig:TvsAPOGEE}, but for metallicity.
	}
	\label{fig:MvsAPOGEE}
\end{figure*}

As shown in Figures~\ref{fig:TvsAPOGEE}--\ref{fig:MvsAPOGEE}, the comparisons of the three stellar parameters show good agreements, although there are some systematic offsets. 
The bottom panels are residuals between our results and those from APOGEE, where the mean value of bias and the standard dispersion are also marked. 
The \teff\ comparison with those of APOGEE DR17 is plotted in Figure~\ref{fig:TvsAPOGEE}, and a consistent result is found despite systematic underestimations of \TDSA\ and \TGSA\ for dMs and gMs, respectively.
It needs to be pointed out that the APOGEE \teff\ used in Figure~\ref{fig:TvsAPOGEE} is calibrated by ASPCAP\footnote{https://dev.sdss.org/dr17/irspec/parameters/}.
To illustrate the difference, we also compare with those of APOGEE spectroscopic \teff\ in Figure~\ref{fig:TspecvsAPOGEE}, as a result, the systematic difference is significantly reduced to \TDSAs\ and \TGSAs\ for dMs and gMs, respectively, which suggests that there are systematic differences of \teff\ between APOGEE and other calibrations (also reported by \citealt{Birky.2020}).

The comparison of \logg\ is plotted in Figure~\ref{fig:GvsAPOGEE} for dMs and gMs.
The dMs have a larger sample size, and the comparison show good agreement for the vast majority of dMs with a small offset of \GDSA, while the surface gravities of gMs seem slightly overestimated with a systematic offset of \GGSA. 
The residuals generally exhibit a weak negative correlation comparing to those from APOGEE---stars with higher surface gravities are underestimated, while those with lower gravities are overestimated.
Similarly, the comparison with those of APOGEE spectroscopic \logg\ is shown in Figure~\ref{fig:GspecvsAPOGEE}. The dispersions of both dMs and gMs are worse than the calibrated \logg, and the systematic difference becomes larger for dMs (\GDSAs) and slightly smaller for gMs (\GGSAs).

When comparing our \feh\ and those from APOGEE DR17, we find a good consistency: the systematic offsets as well as dispersions are $0.14 \pm 0.16$\,dex and $0.18 \pm 0.26$\,dex for dMs and gMs, respectively, albeit with a negative correlation in residuals.

\subsection{Comparison with Other Literature}

While the comparison with APOGEE DR17 has displayed the good performance of our method, we do find some systematic deviations, so it is necessary to make comparisons with other literature.
Therefore, we collect stellar parameters from \RefNum\ other papers.
They both worked on M-type stars stellar parameter estimation using different methodologies.
We can take advantage of these results and make comparisons with well-determined stellar parameters.

\begin{figure}[htbp]
	\centering
	\includegraphics[width=0.5\textwidth]{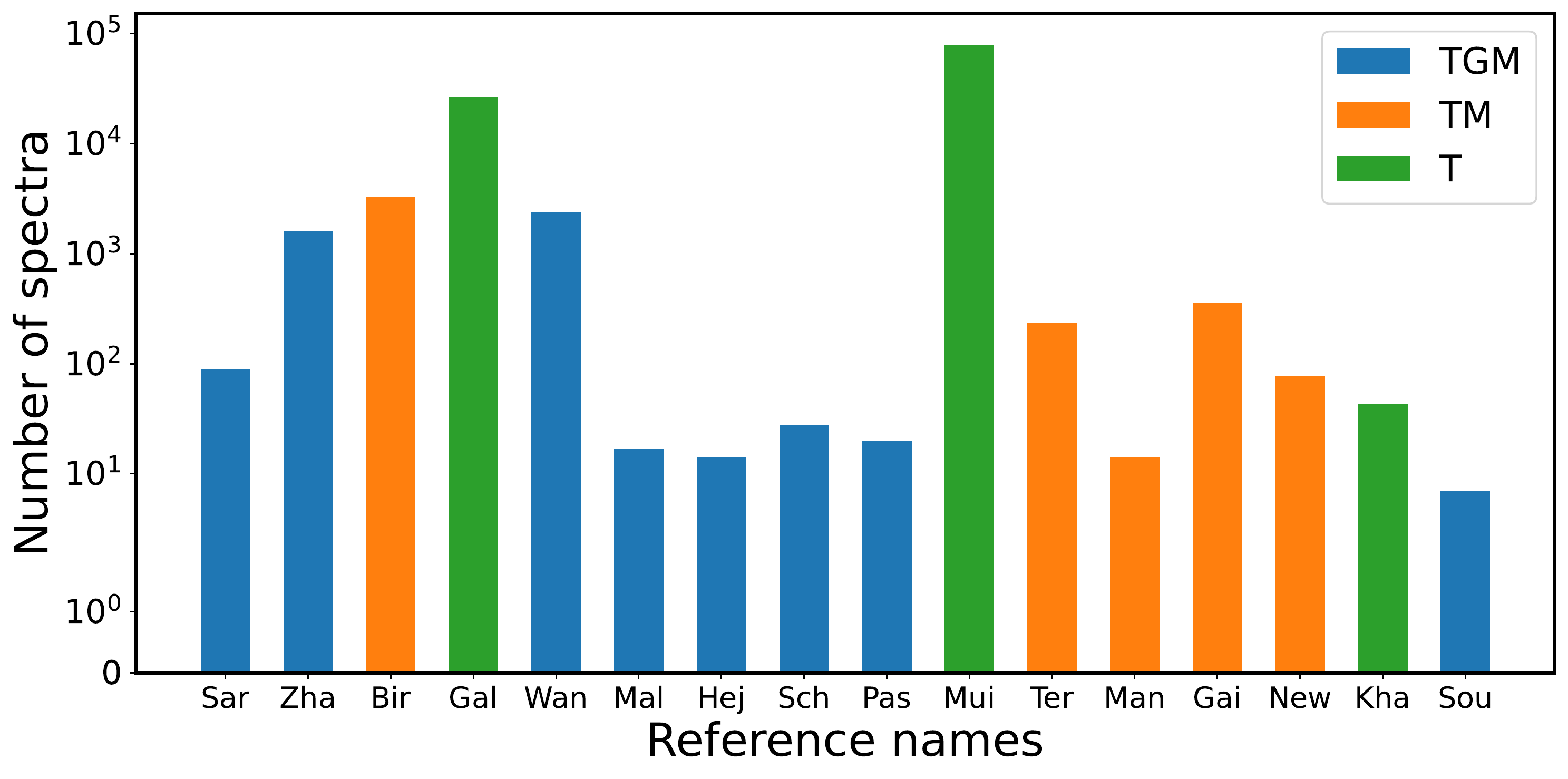}
	\caption{The number of valid stellar parameters provided by \RefNum\ papers from the literature. See Table~\ref{Tab:Ref_set} for detail.
	} 
	\label{fig:TGM_bar_plot}
\end{figure}

Figure~\ref{fig:TGM_bar_plot} displays the quantities of cross-matched stars for each literature's contribution in our reference set, detailed information can be found in Table~\ref{Tab:Ref_set}.

We assemble this reference set with 114,144 spectra cross-matched from those works, and exclude the binary and variable candidates, similar to Section~\ref{subsec:APOGG_valid}.
Finally, we have 49,698 common stars in this reference set.

\begin{figure}[!htbp]
	\centering
    \fig{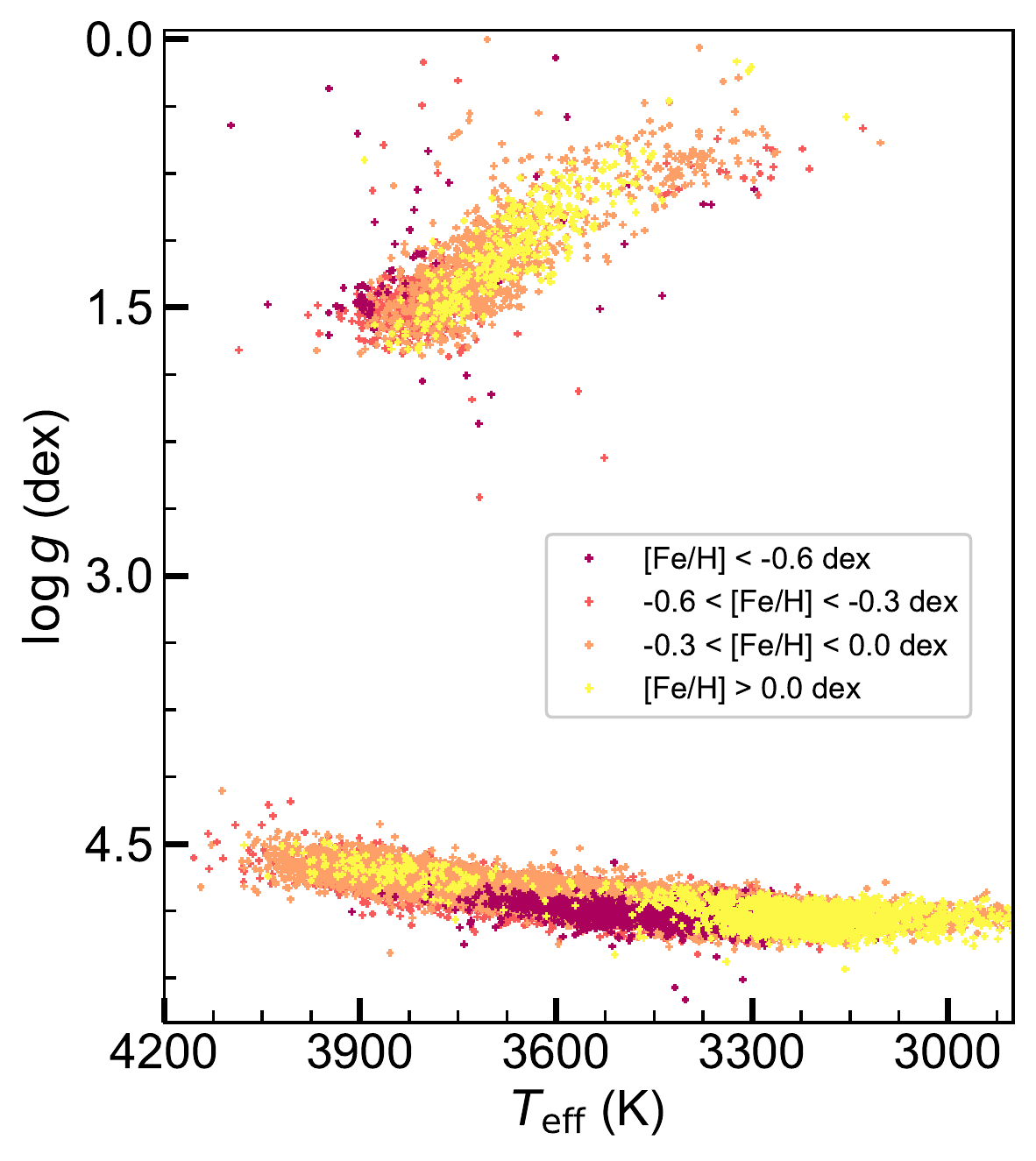}{0.4\textwidth}{}
	\caption{The distribution of \teff\ versus \logg\ for common stars from the reference set. Stars are color coded in different metallicity groups.}
	\label{fig:RS_para_dist}
\end{figure}

Figure~\ref{fig:RS_para_dist} shows the parameter distribution (\teff\ versus \logg) of common stars in this reference set,
 which displays a similar but broader parameter space comparing to Figure~\ref{fig:APG_para_dist}.

\begin{deluxetable}{lrccc}
\centering
\tablecaption{Literature sources for comparison.}
\tabletypesize{\footnotesize}
\label{Tab:Ref_set}
\tablehead{
\colhead{Ref names} & \colhead{Common stars} & \multicolumn{3}{c}{Parameters}}
\startdata
    \citet{sarmento.2021}   & 91            &\teff& \logg & \feh\  \\
	\citet{zhang.2021}      & 1579          &\teff& \logg & \feh\  \\
	\citet{Birky.2020} 		& 3047          &\teff& ---   &\feh\  \\
	\citet{galgano.2020} 	& 26502         &\teff& ---   & ---   \\
	\citet{Wang.2020} 		& 2284          &\teff& \logg & \feh\  \\
	\citet{maldonado.2020a} & 17            &\teff& \logg & \feh\  \\
	\citet{Hejazi.2020} 	& 14            &\teff& \logg & \feh\  \\
	\citet{Schweitzer.2019} & 28            &\teff& \logg & \feh\  \\
	\citet{Passegger.2018} 	& 20            &\teff& \logg & \feh\ \\
	\citet{muirhead.2018} 	& 79658         &\teff& ---   & ---   \\
	\citet{terrien.2015} 	& 237           &\teff& ---   & \feh\ \\
	\citet{Mann.2015}		& 14            &\teff& ---   &\feh\  \\
	\citet{gaidos.2014} 	& 355           &\teff& ---   & \feh\ \\
	\citet{Newton.2014} 	& 1192          &\teff& ---   & \feh\ \\
	\citet{khata.2021a} 	& 43			&\teff& ---   & ---  \\
	\citet{souto.2021} 		& 7			    &\teff& \logg & \feh\ \\
\enddata
\end{deluxetable}

\begin{figure*}[!ht]
	\centering
    \gridline{
    \fig{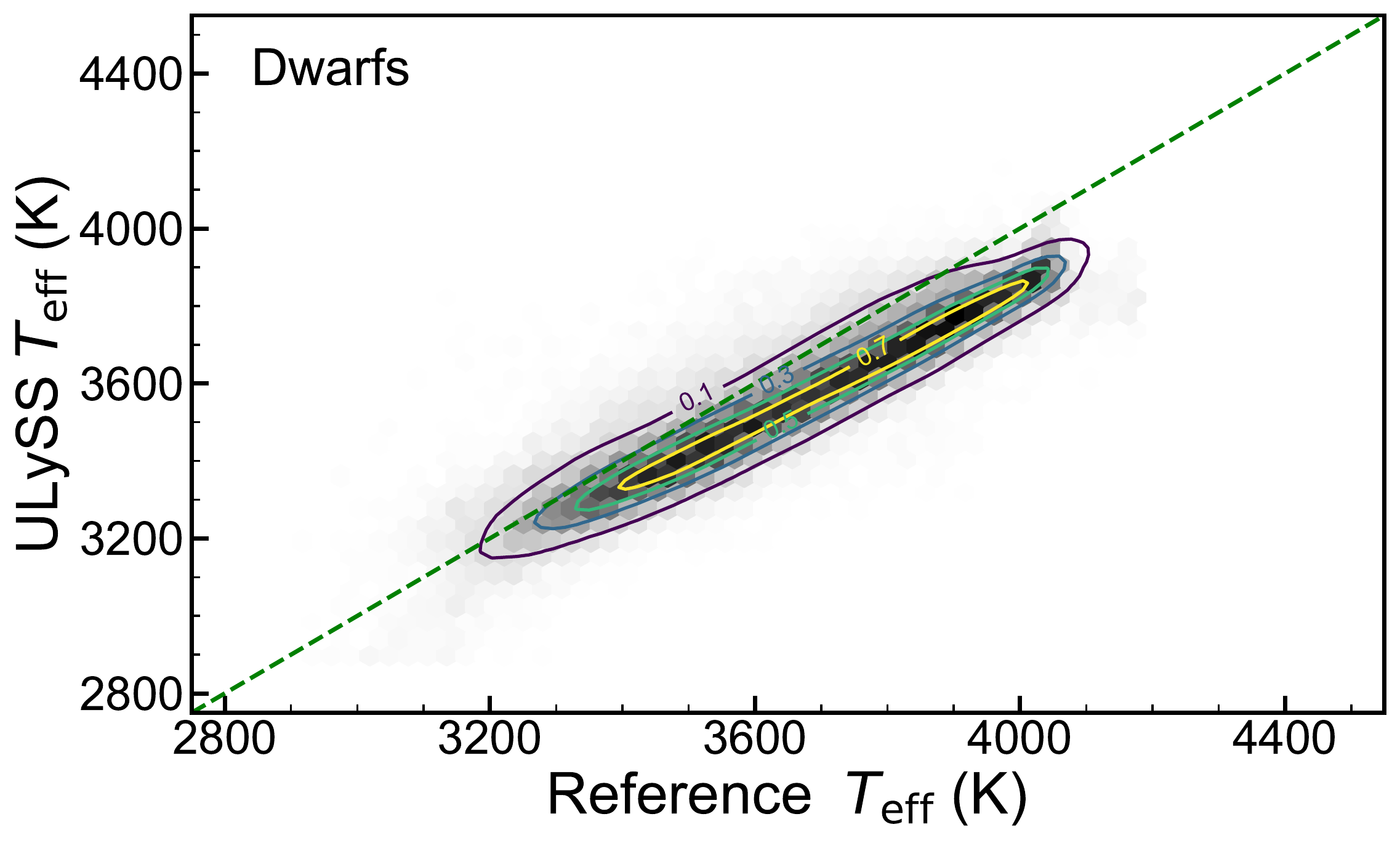}{0.45\textwidth}{}
    \fig{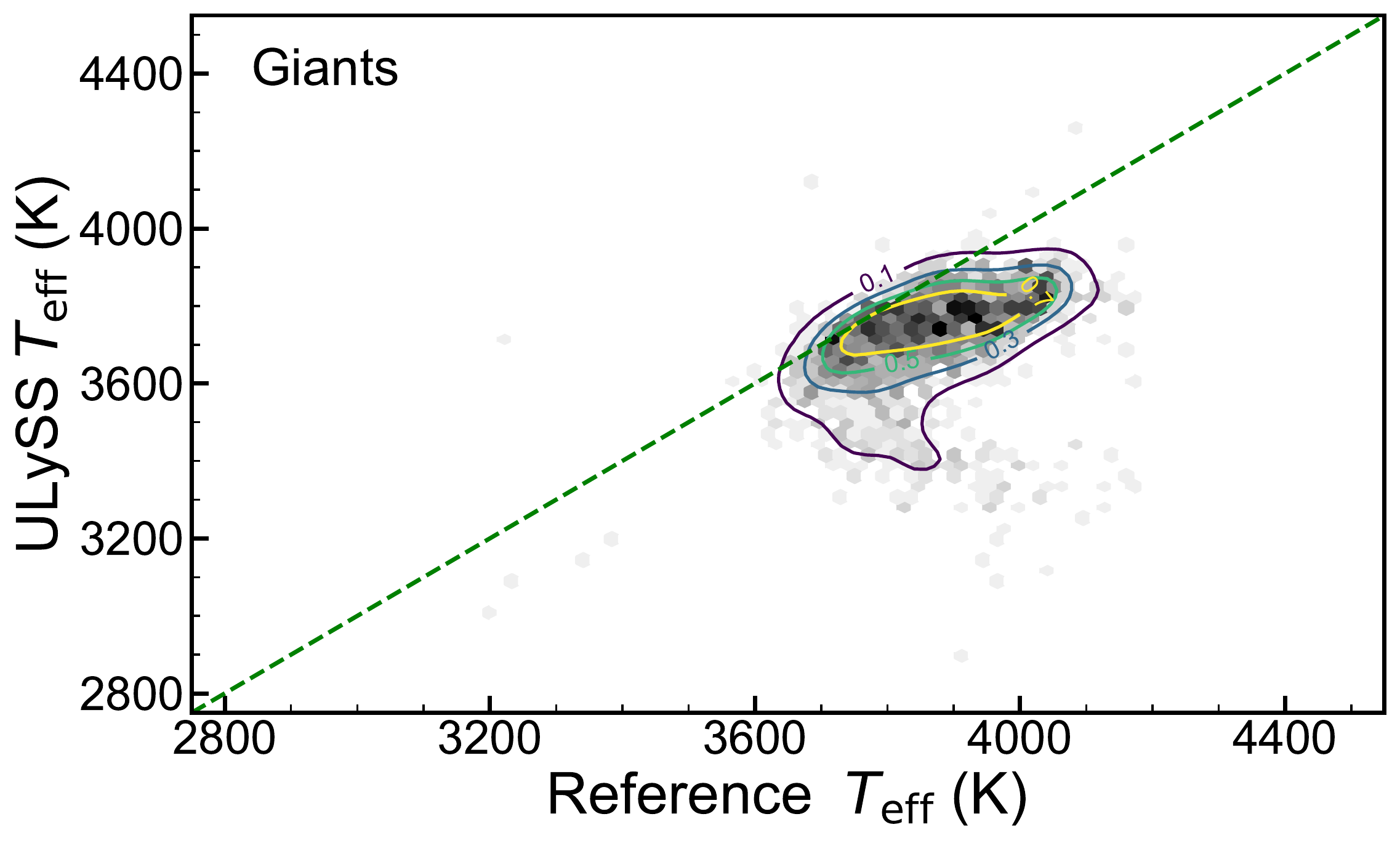}{0.45\textwidth}{}
	}  \vspace{-7mm}
        \gridline{
    \fig{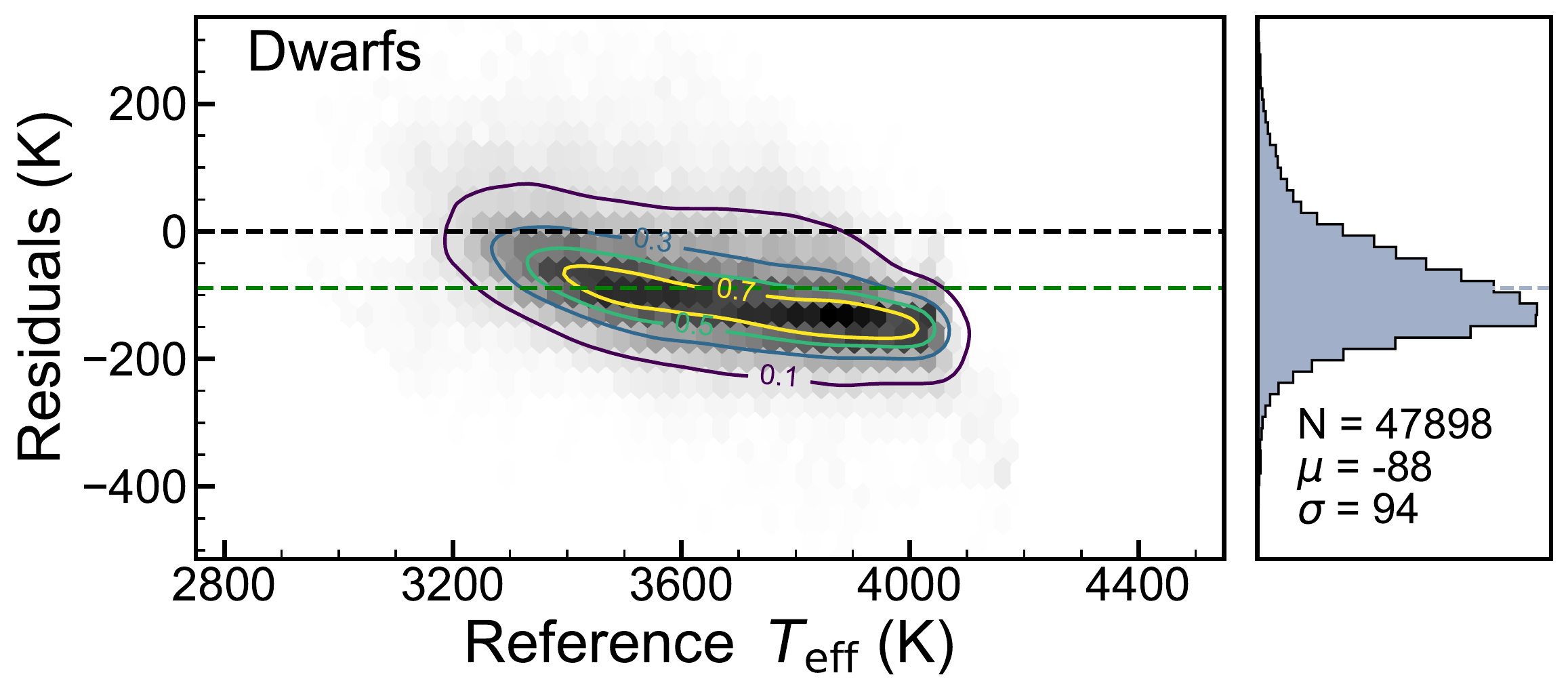}{0.45\textwidth}{}
    \fig{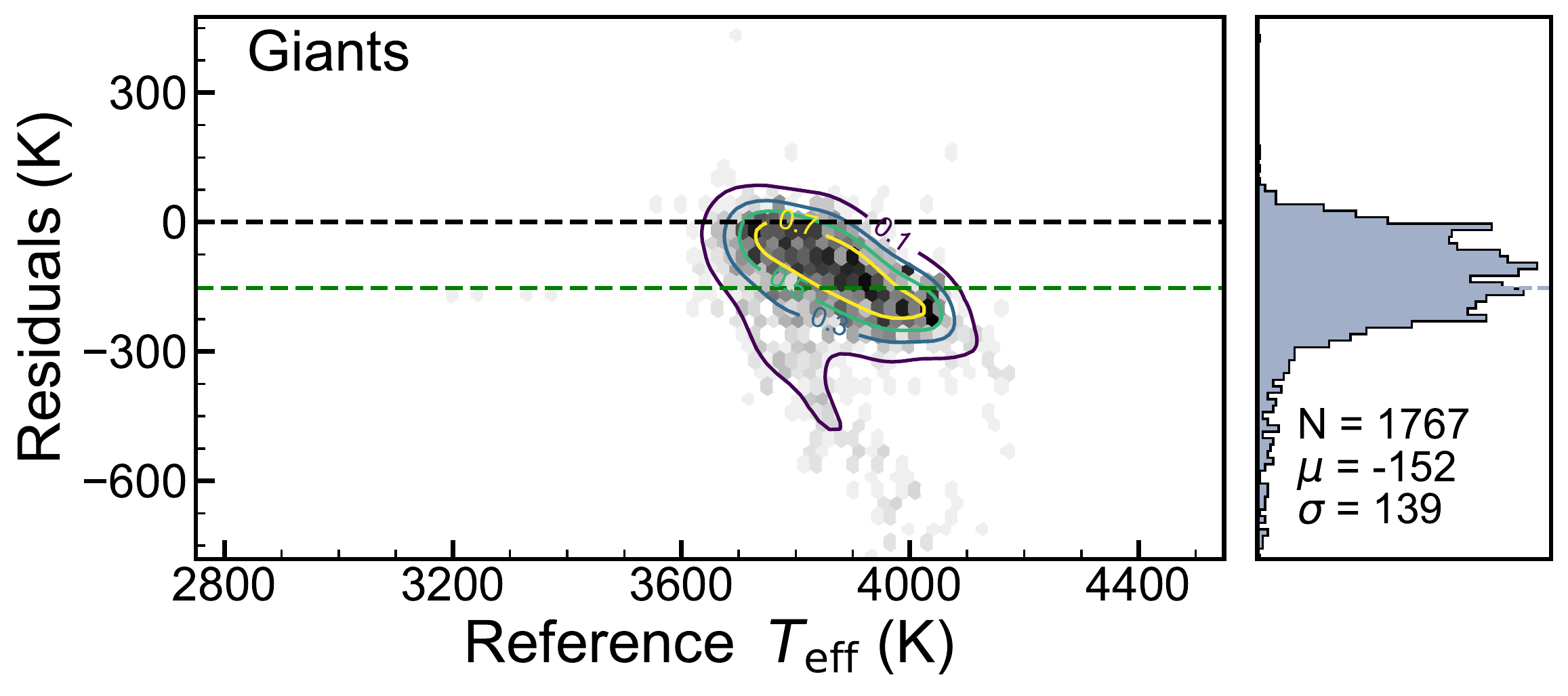}{0.45\textwidth}{}
    }  \vspace{-7mm}    
    
	\caption{The comparison of the effective temperatures derived from our method with the literature values. 
	The top-left panel shows the \teff\ comparison for dMs, while the top-right panel for gMs. 
	The bottom-left and bottom-right panels show the residuals' distributions for dMs and gMs, respectively. 
	The color scale as well as the contour lines indicate the number density of each region.
	}
	\label{fig:TvsRS}
\end{figure*}

\begin{figure*}[!ht]
	\centering
    \gridline{
    \fig{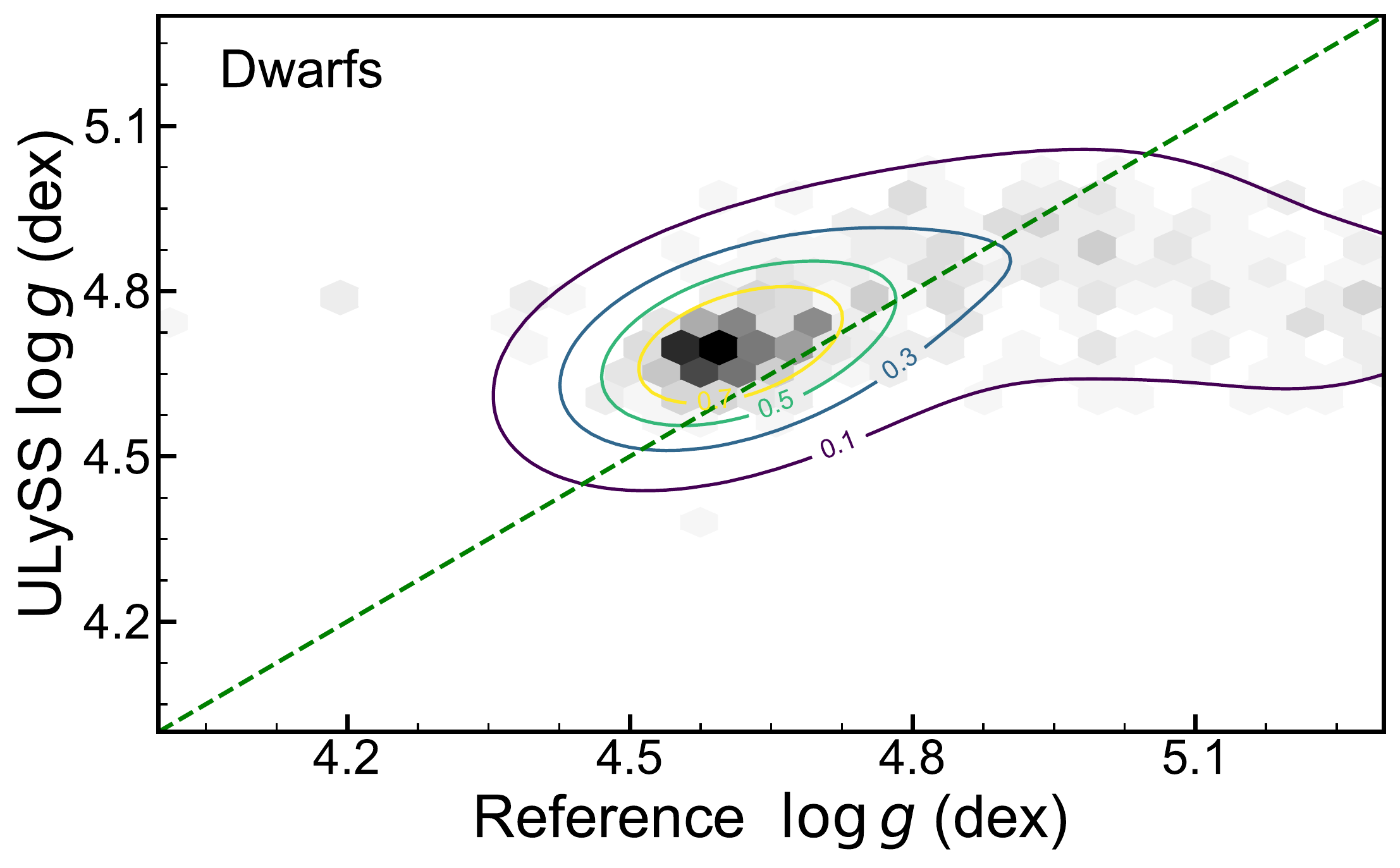}{0.45\textwidth}{}
    \fig{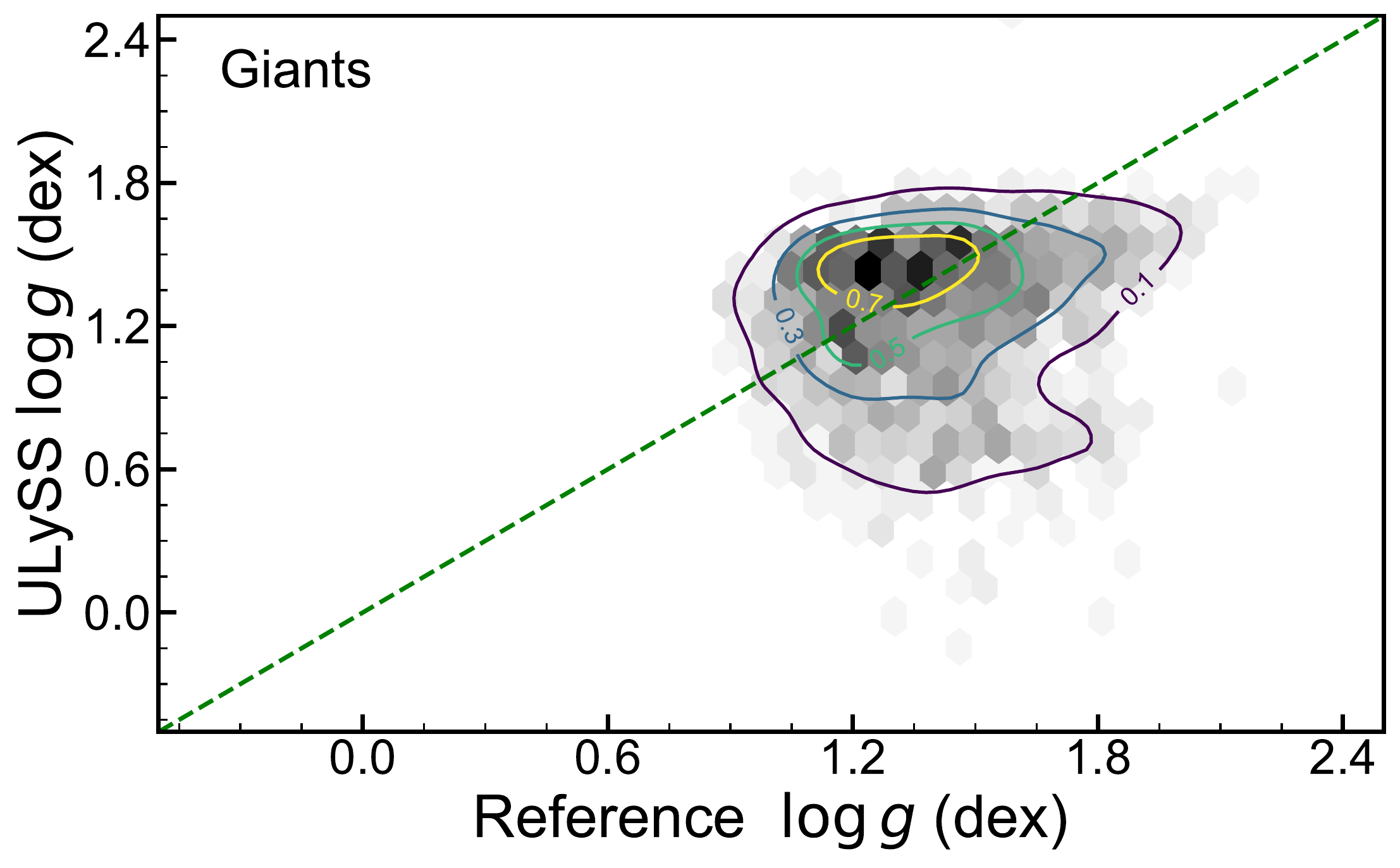}{0.45\textwidth}{}
	}  \vspace{-7mm}
        \gridline{
    \fig{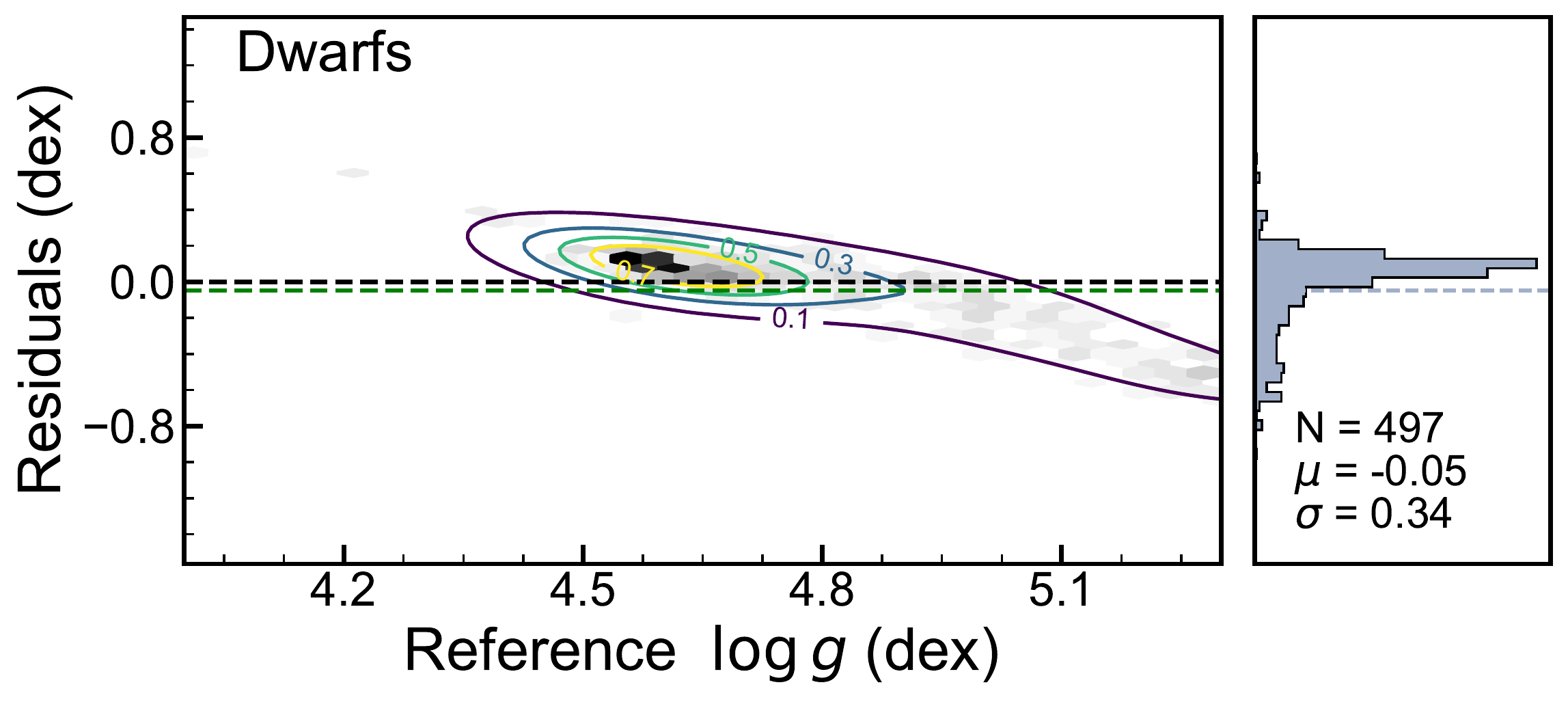}{0.45\textwidth}{}
    \fig{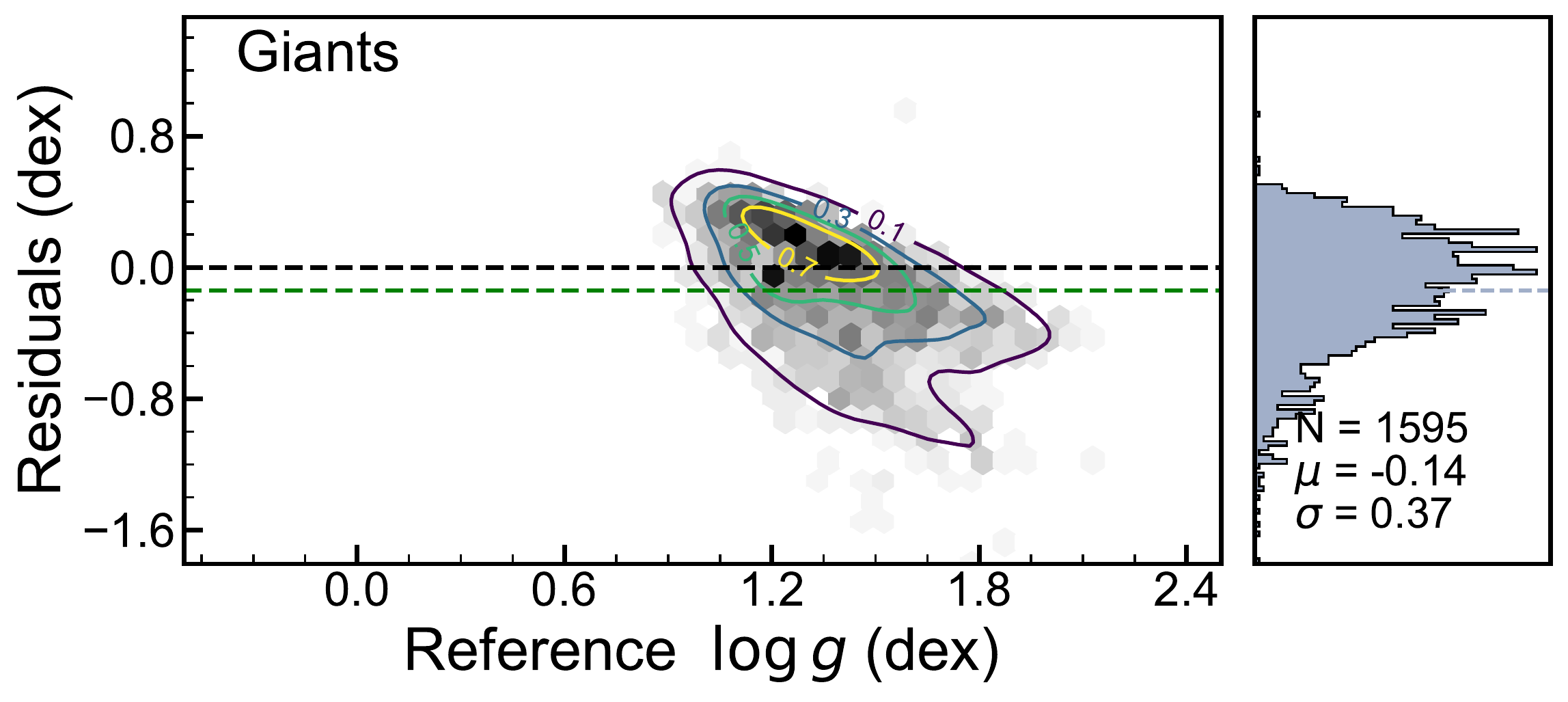}{0.45\textwidth}{}
    }  \vspace{-7mm}    
    
	\caption{Same as Figure~\ref{fig:TvsRS}, but for surface gravity.
	}
	\label{fig:GvsRS}
\end{figure*}

\begin{figure*}[!ht]
	\centering
    \gridline{
    \fig{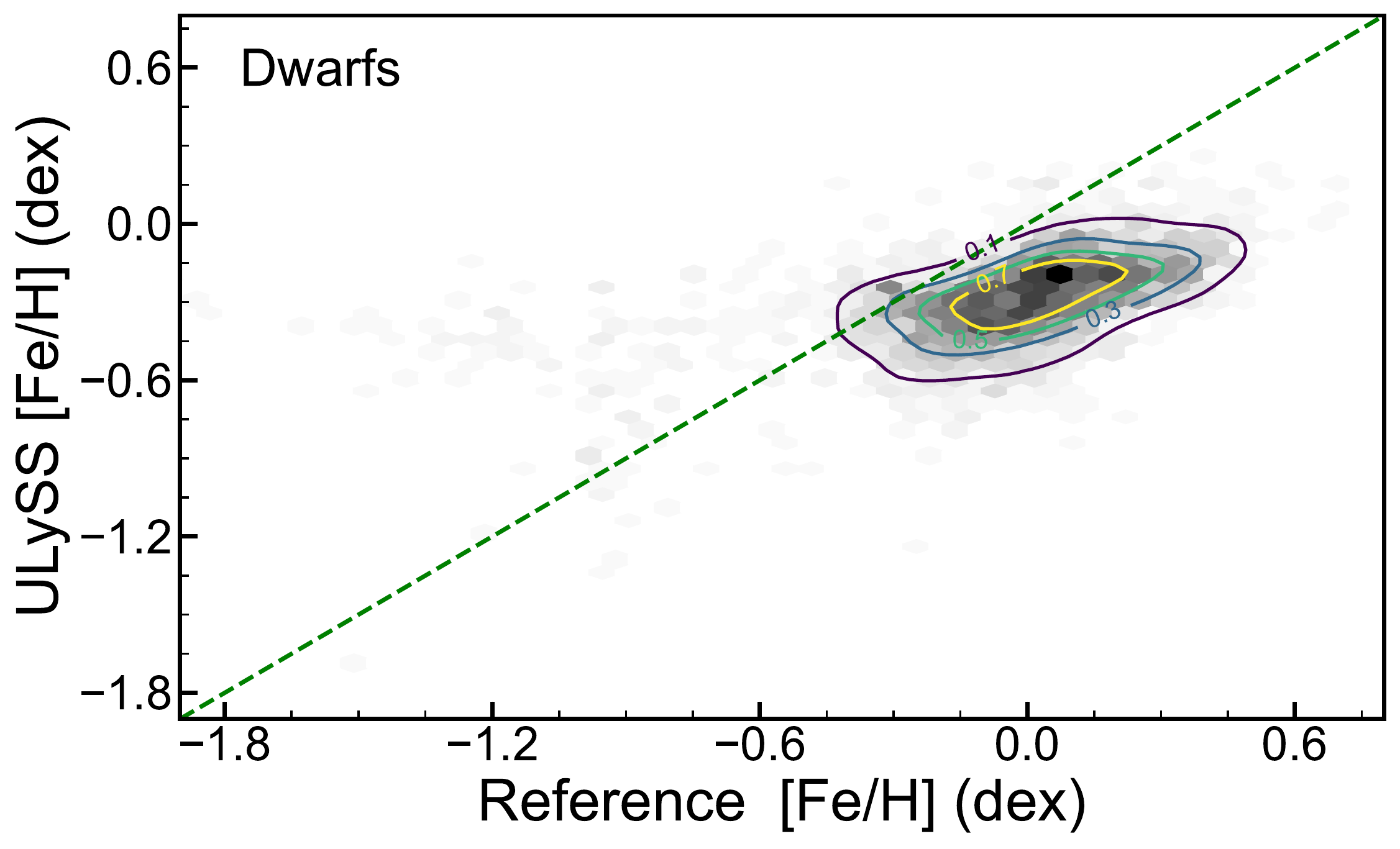}{0.45\textwidth}{}
    \fig{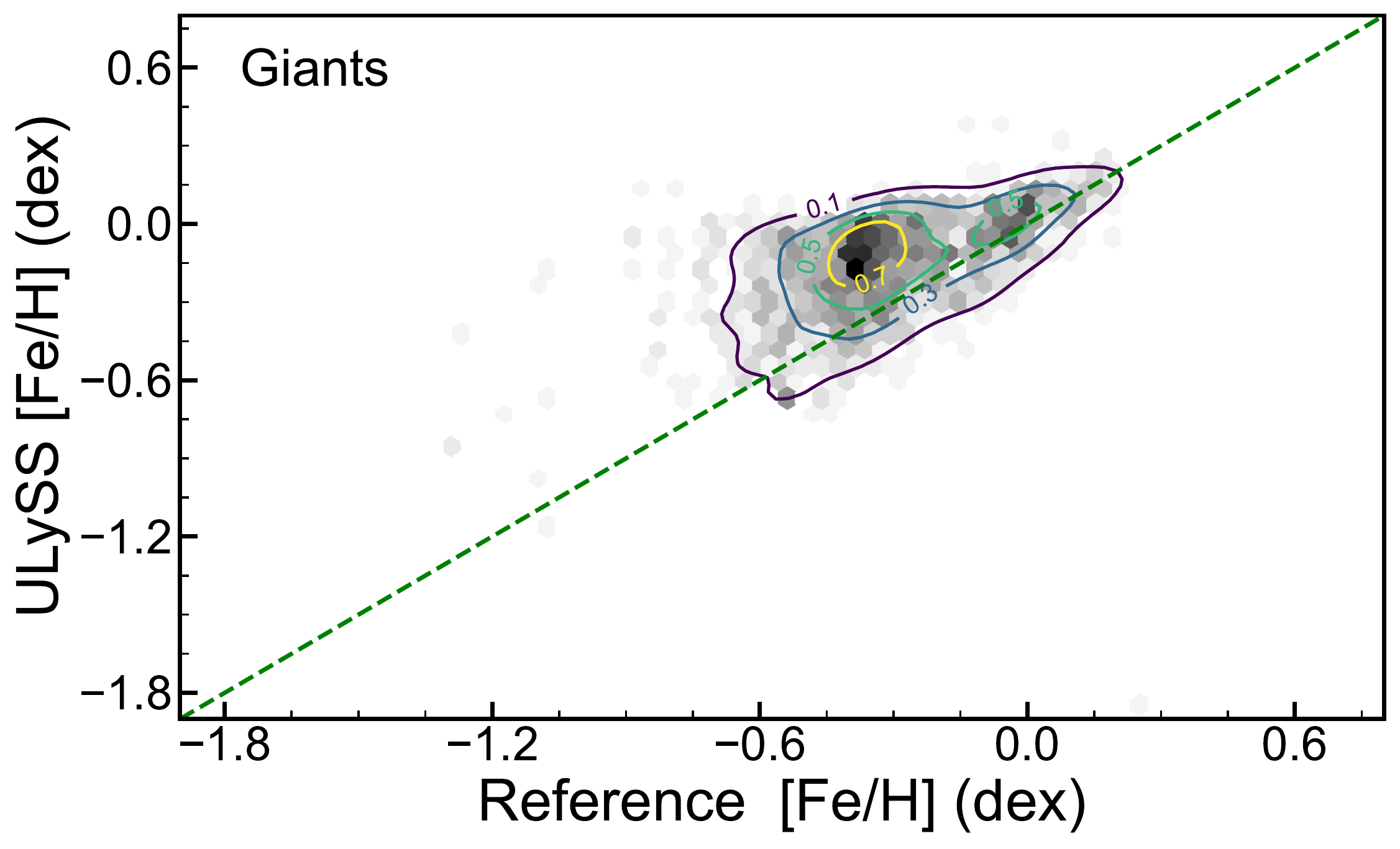}{0.45\textwidth}{}
	}  \vspace{-7mm}
        \gridline{
    \fig{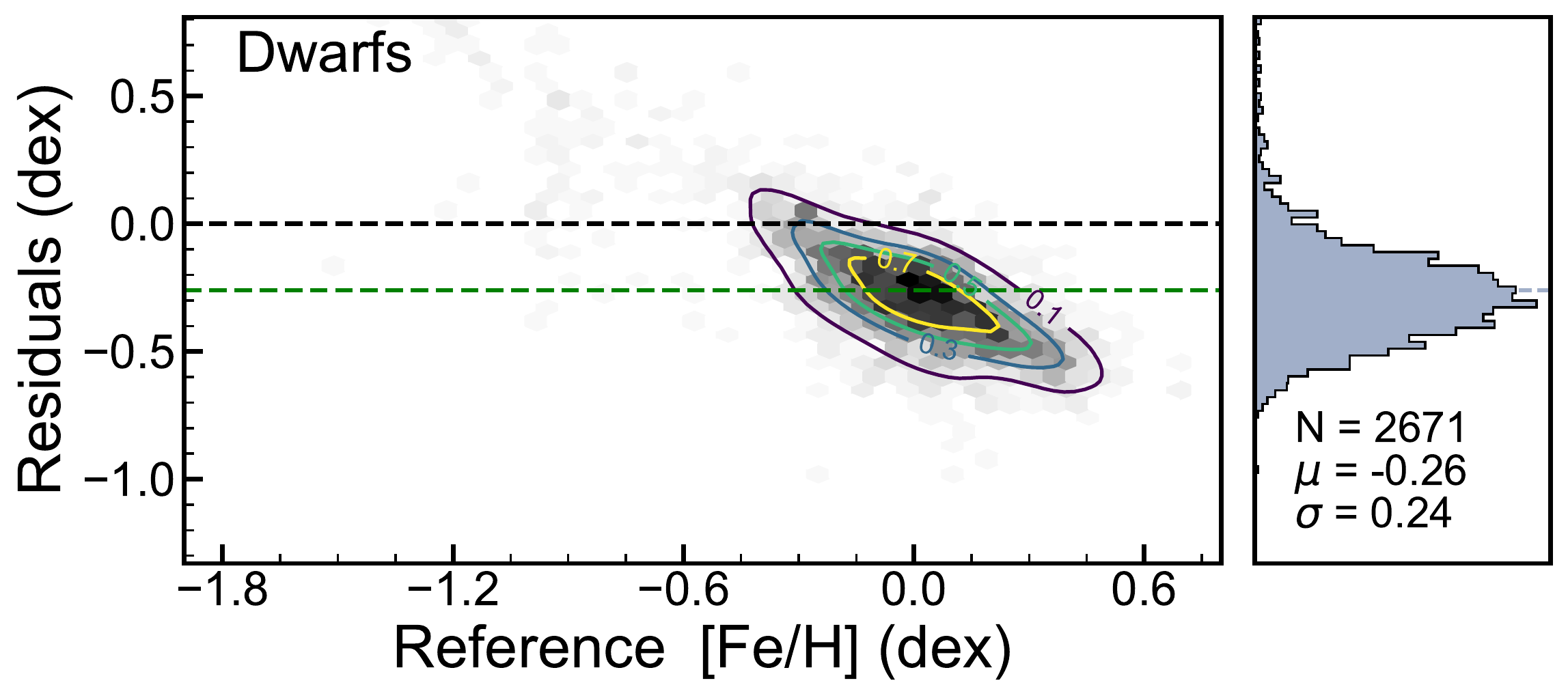}{0.45\textwidth}{}
    \fig{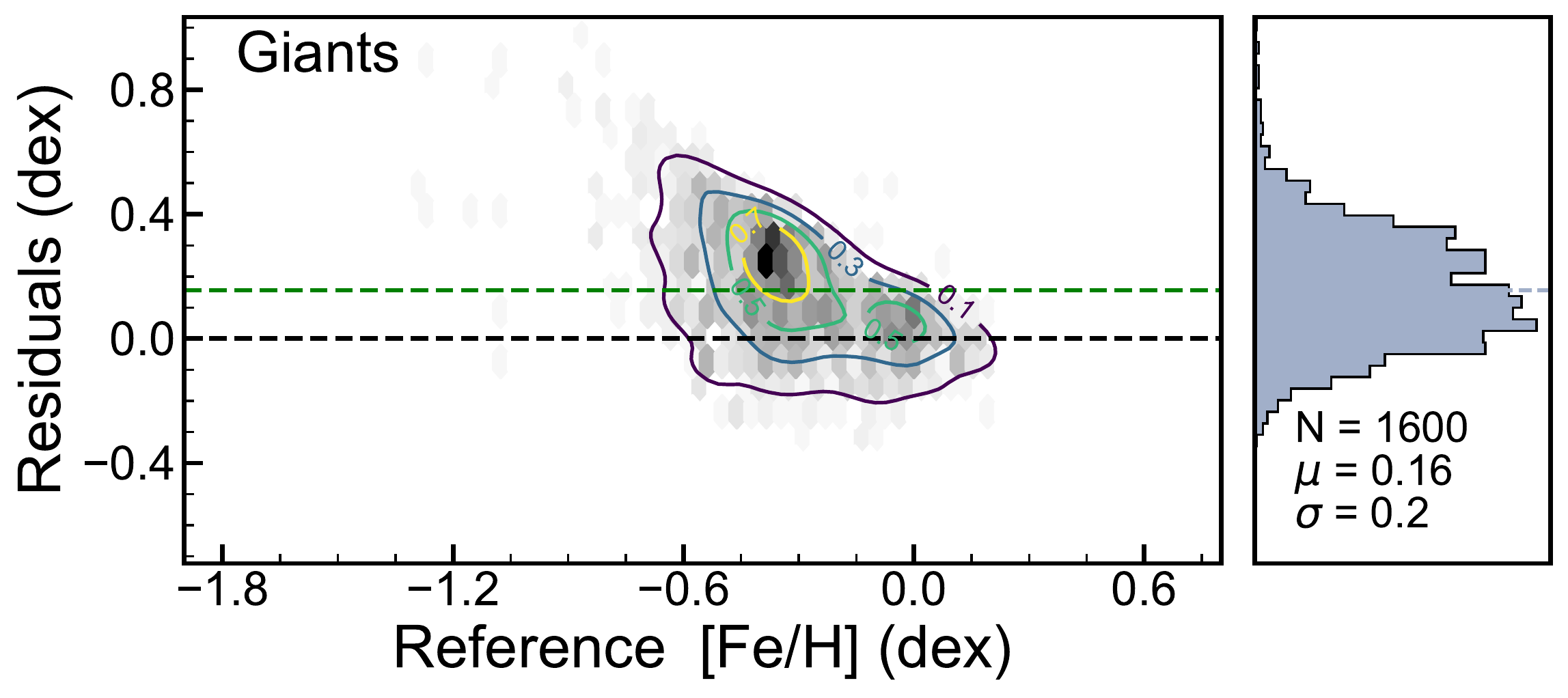}{0.45\textwidth}{}
    }  \vspace{-7mm}    
    
	\caption{Same as Figure~\ref{fig:TvsRS}, but for metallicity.
	}
	\label{fig:MvsRS}
\end{figure*}

The comparisons of these stellar parameters are presented in Figures~\ref{fig:TvsRS}--\ref{fig:MvsRS} with detailed offsets and standard deviations listed in the residual distributions.
In Figure~\ref{fig:TvsRS}, we have compared our \teff\ with those in the reference set, which benefits from a large effective temperature coverage from 2,900\,K to 4,200\,K. 
The \teff\ offset for dMs is $-$\TDSR, which is less than the systematic offset appearing in the APOGEE comparison.
For gMs, the systematic offset is $-$\TGSR\ with a larger dispersion of \TGDR.
The differences of residual offsets also indicate a systematic difference of $\sim$ 100\,K exists between \teff\ from APOGEE and those from other measurements.

In our reference set, the available \logg\ and \feh\ measurements are not as large as \teff, the comparisons are displayed in Figures~\ref{fig:GvsRS} and \ref{fig:MvsRS}.
The \logg\ for dMs derived from our method shows a modest systematic offset (\GDSR) and a good consistency comparing with those from different literature.
For the gMs, we find a systematic offset of $-$\GGSR, which is less than the offset found in APOGEE comparison. 

For \feh, the residuals between our results and other measurements show small systematic offset and dispersion, along with a weak overall decrease similar to the APOGEE comparison. 
The systematic offsets are $-$\MDSR\ and \MGSR\ for dMs and gMs, respectively, which suggests that our derived \feh\ is generally underestimated for dMs, while the derived \feh\ seems overestimated for gMs.

\section{Error Estimation} \label{sect:error}

The comparisons in Section~\ref{sect:valid} showed that the fitting errors of our results could be influenced by stellar parameters to some extent.
Besides, the quality of the observed spectra can also affect the precision of stellar atmospheric parameter measurement.
With the help of the overlapping observations of LAMOST survey, over 26\% stars have multiple visits in LAMOST DR8,
 which allows us to examine the error in stellar parameters, due to the SNR of spectra.

In this section, we estimate the errors, and investigate the precision of our measurements by comparing the stellar parameters derived from these stars with repeated observations.

\subsection{Random Errors due to the Quality of Spectra}
\label{subsect:random_err}
\begin{figure*}[ht!]
	\centering
    \gridline{
    \fig{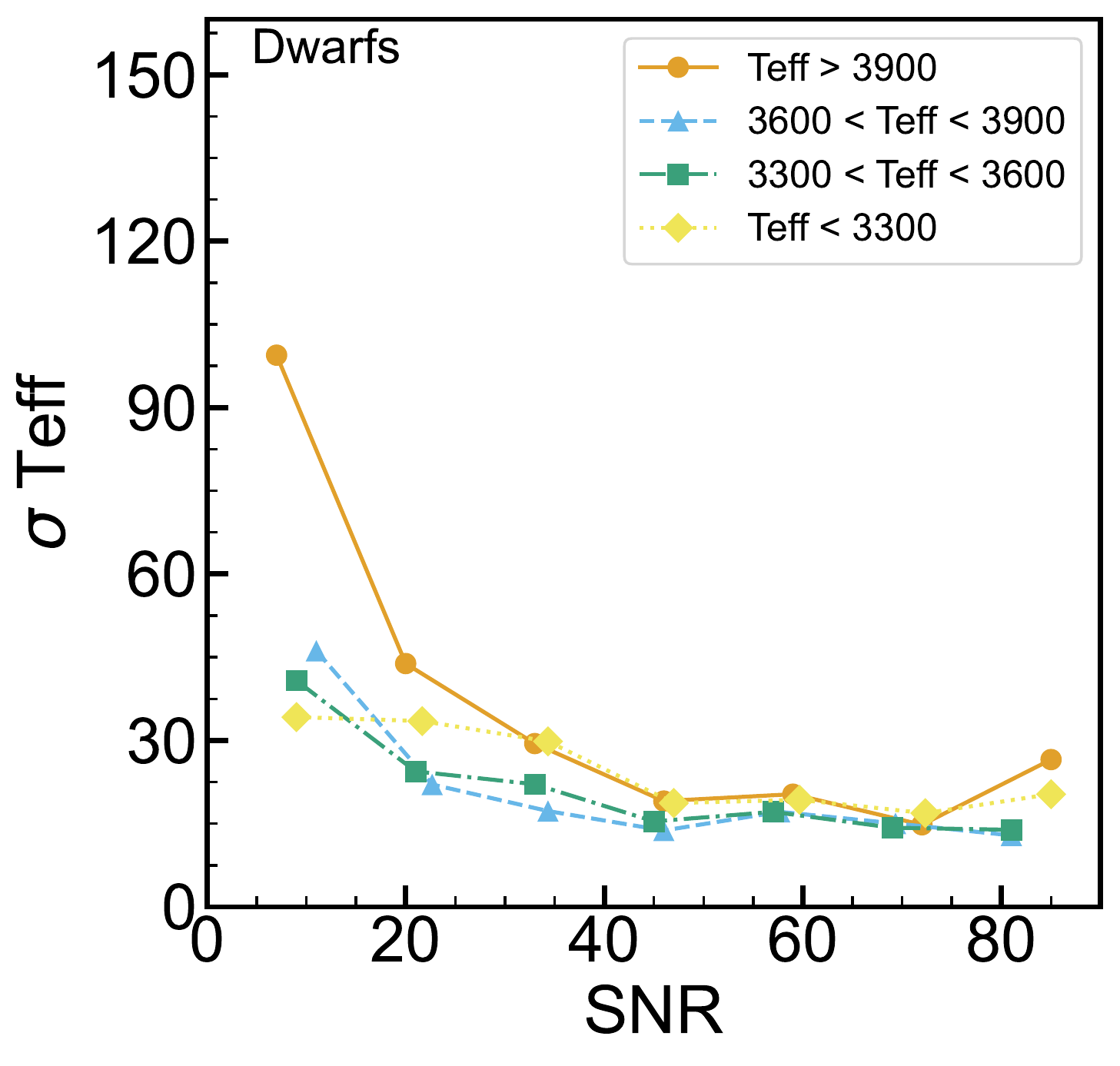}{0.32\textwidth}{}
    \fig{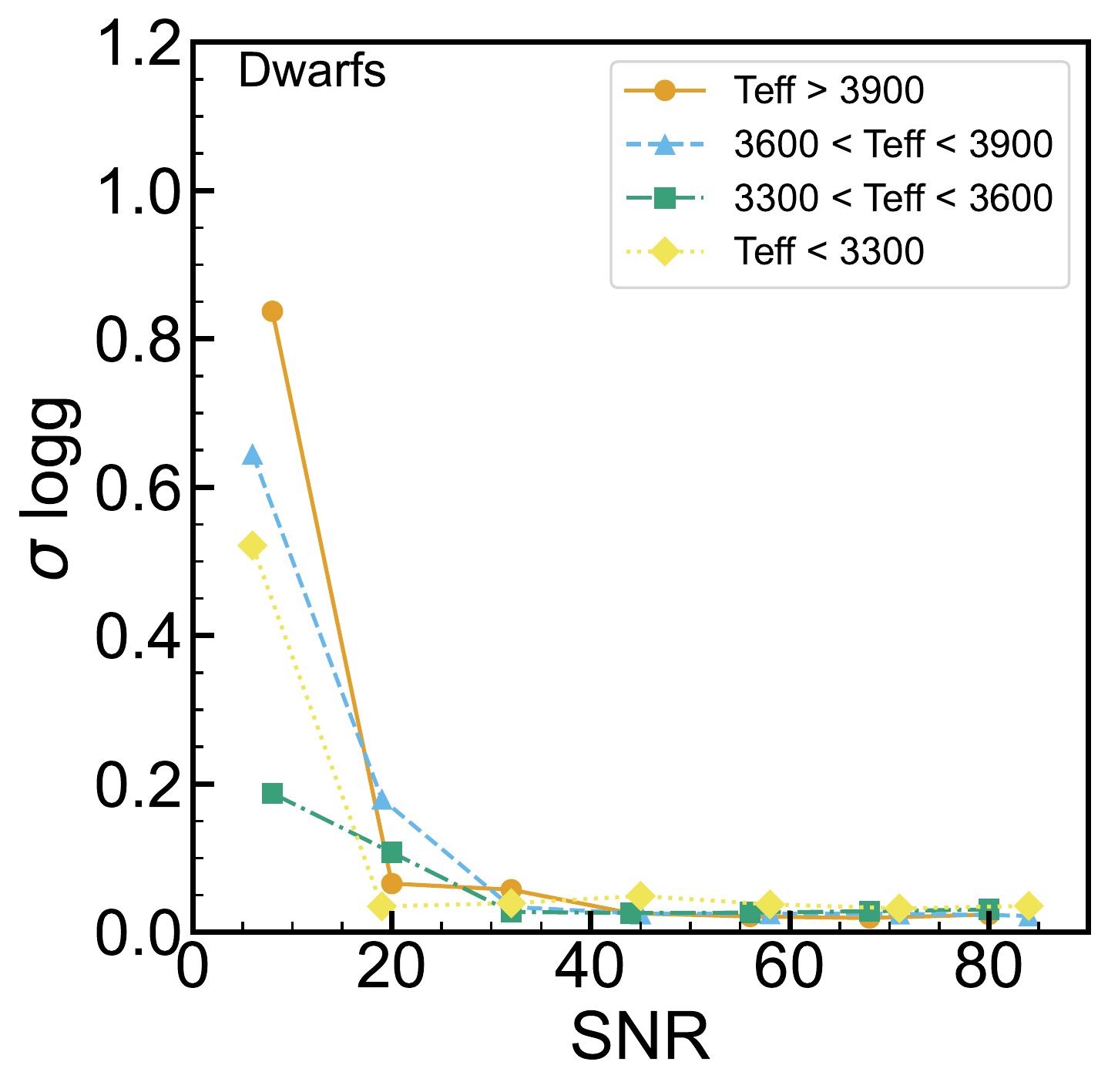}{0.32\textwidth}{}
    \fig{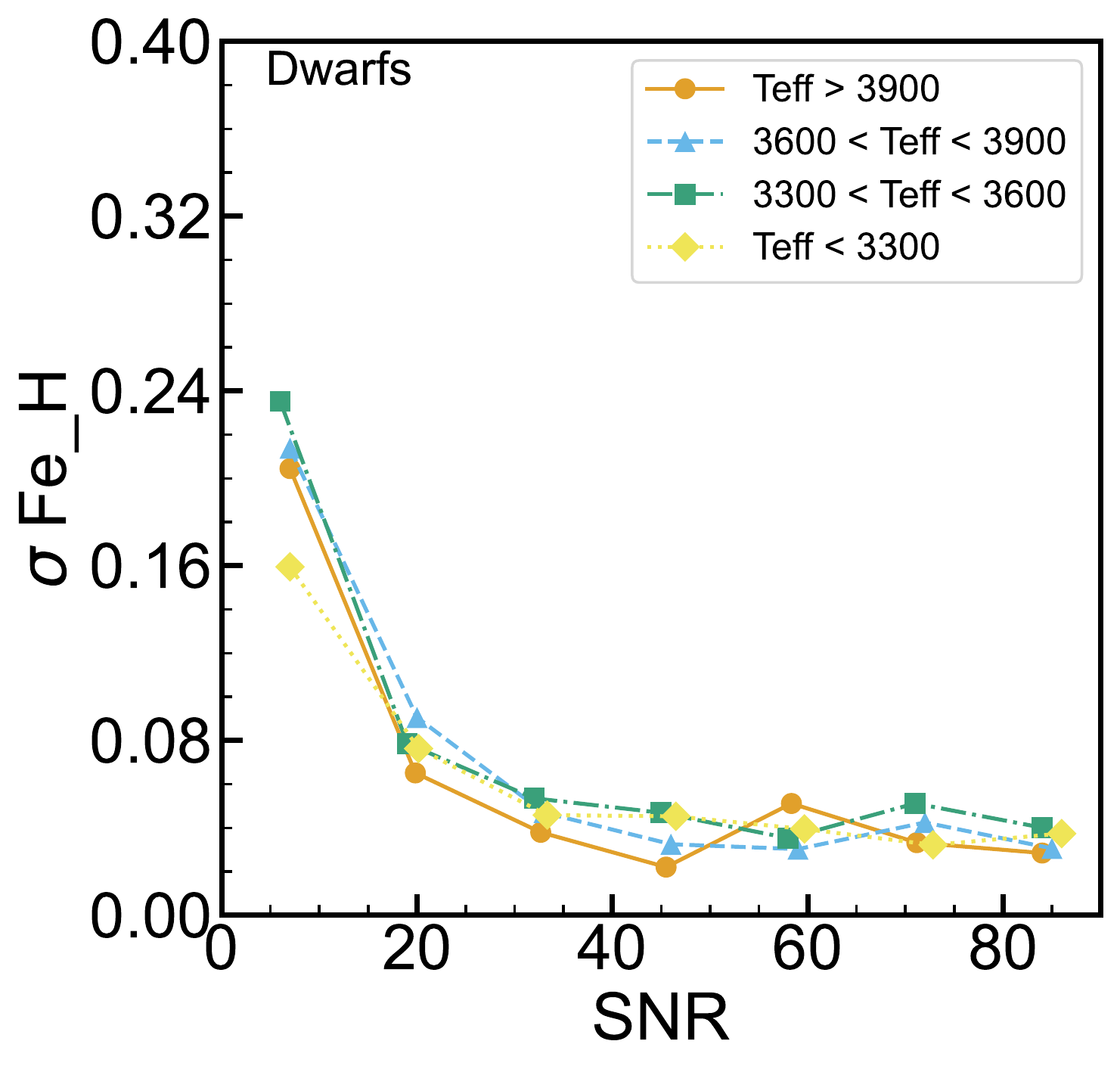}{0.32\textwidth}{}
    }  \vspace{-5mm}
        \gridline{
    \fig{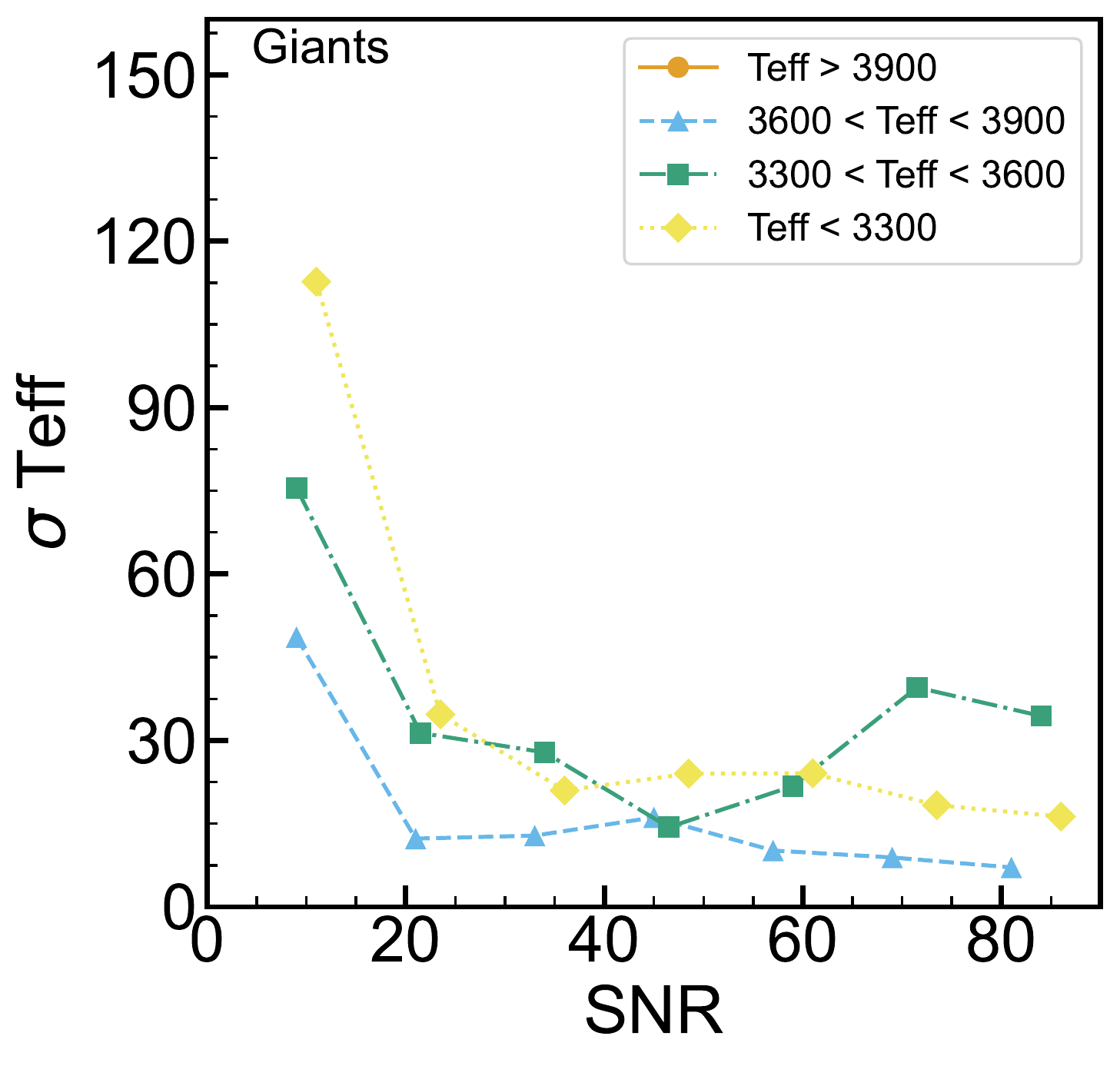}{0.32\textwidth}{}
    \fig{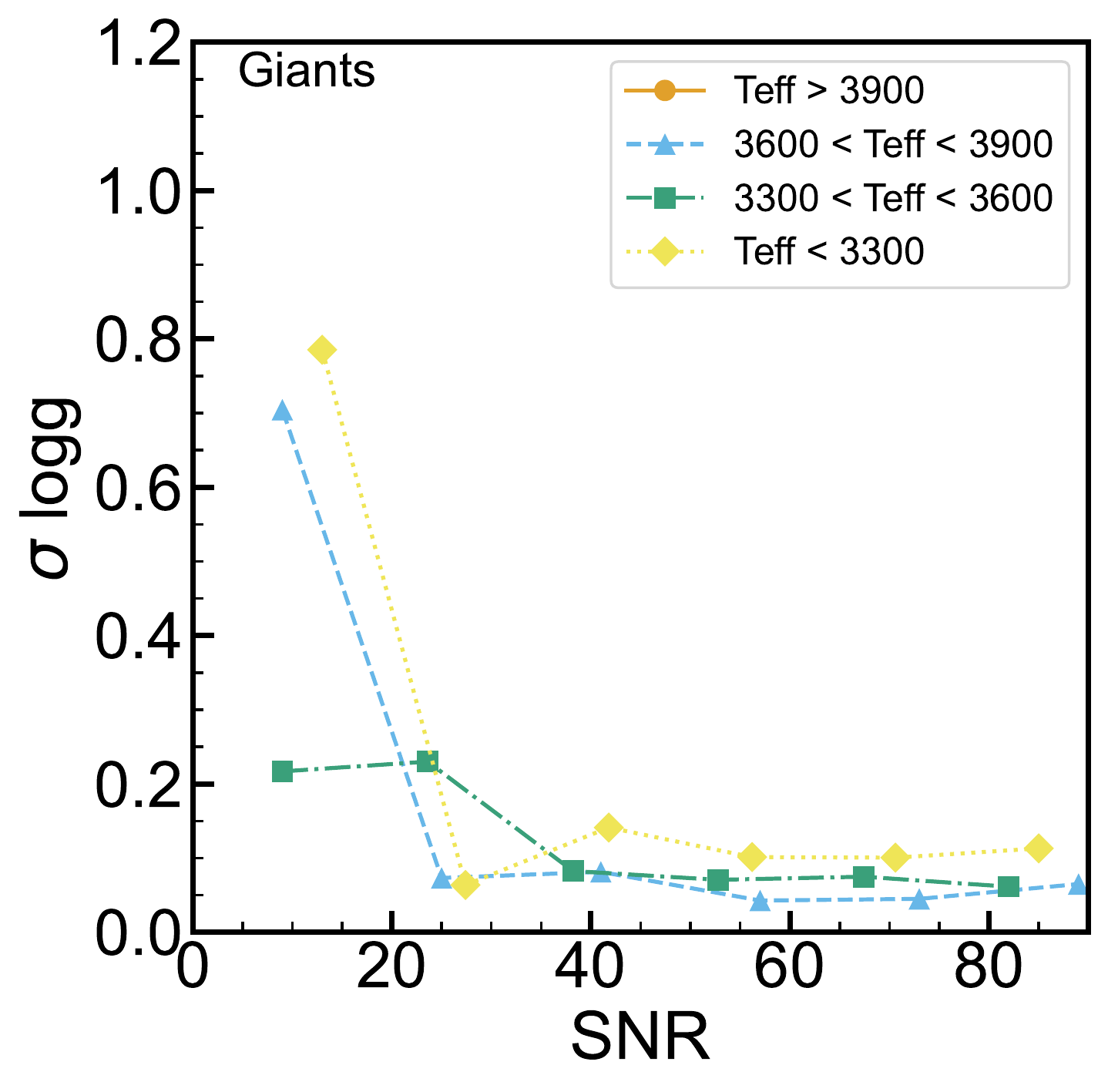}{0.32\textwidth}{}
    \fig{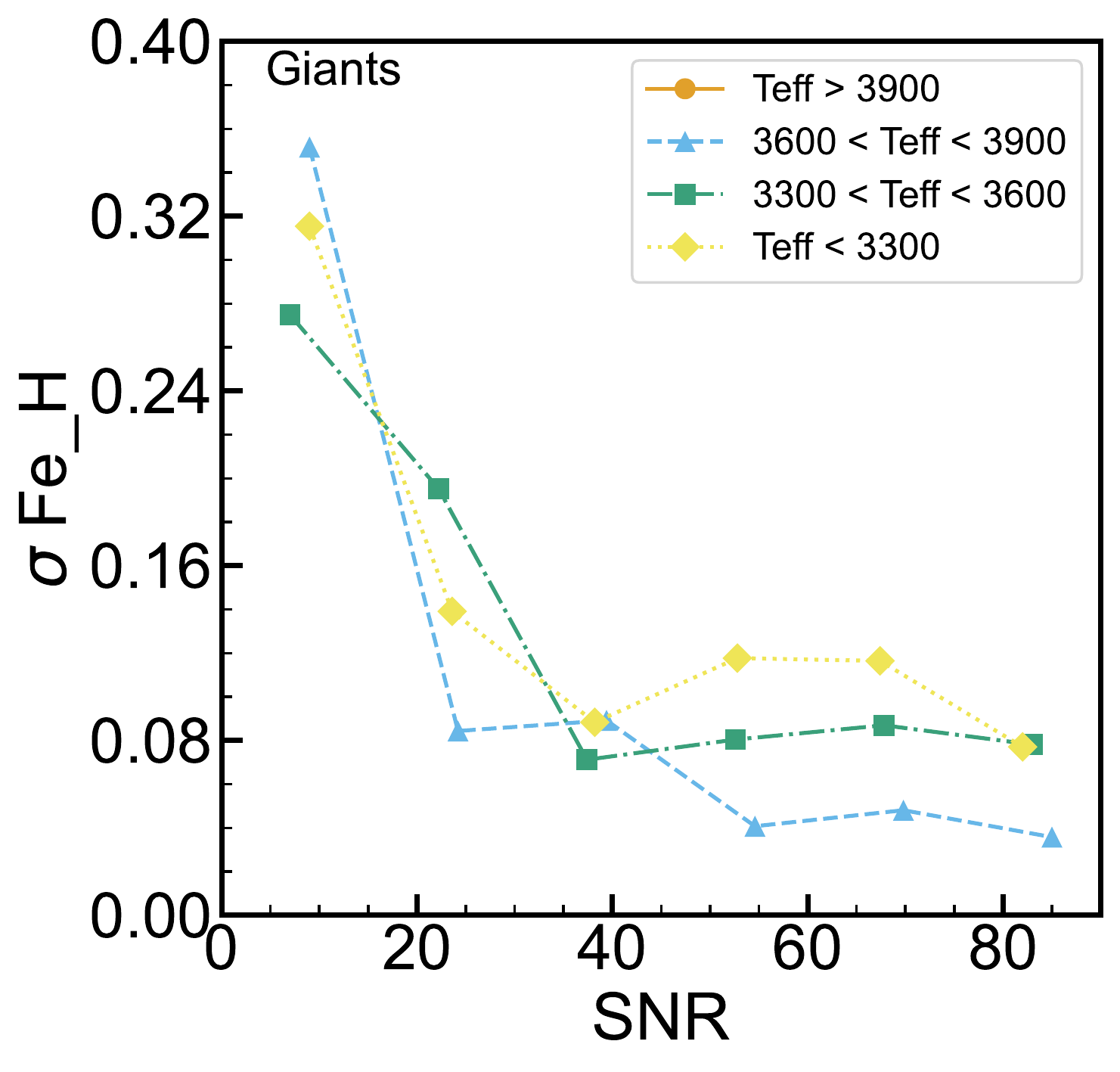}{0.32\textwidth}{}
    }  \vspace{-5mm}    
    
	\caption{The random errors of stellar parameters as functions of SNR. The dMs (top) and gMs (bottom) are grouped in different temperature bins. The dots represent the standard dispersion in each SNR bin.
	}
	\label{fig:Random_error_snrr}
\end{figure*}

To evaluate the effect of spectral quality on the stellar parameters, we first select spectra of repeated observations for the same stars, with a difference in SNR should be lower than 20\%.
Figure~\ref{fig:Random_error_snrr} shows the residuals of stellar parameters for these duplicate stars as a function of SNR.
We use the standard dispersion to represent the errors of stellar parameter determination, which present declining tendencies as the SNR of spectra increase.
We split our sample into dMs (top panel) and gMs (bottom panel), as well as group stars into different temperature bins.

The effective temperatures are less sensitive to SNR. 
Generally the cooler stars tend to have better precision, except for the coolest stars (\teff\ $<$ 3,000\,K), and there are not so many stars as other groups in our sample set.
For the spectra of SNR $\sim$ 5, the precision of \teff\ is better than 120\,K, and it immediately decays to 45\,K when SNR is higher than 20. 

However, the surface gravities are very sensitive to the SNR of spectra, which may be due to the molecular bands blurred by noise.
For both dMs and gMs, \logg\ can be determined with a precision better than 0.2\,dex for SNR $>$ 20. 
For the low SNR spectra, the measurement of \logg\ is visually poor, e.g. the dispersion of hot dwarfs (\teff\ $>$ 3,900\,K) can be as large as 0.85\,dex, which becomes a main reason to exclude them from our comparisons. 

A similar situation can be found for metallicities, which shows a significant dependency on SNR. 
In general, dMs can have more precise \feh\ than gMs. 
For dMs in each temperature group, the precision of \feh\ is about 0.24\,dex when SNR is about 5, and rapidly decreases to 0.08\,dex when SNR reaches 15. 
The determination of \feh\ for gMs is relatively poor as the precision reaches 0.36\,dex of SNR $\sim$ 5 and 0.2\,dex of SNR $>$ 25, respectively. 
In each temperature group, the \feh\ dispersion for gMs differs from the other groups, indicating a dependency of \feh\ determination on \teff.

\subsection{Systematic Errors due to SNR}
\label{subsect:systematic_snr}

\begin{figure*}[!htbp]
	\centering
    \gridline{
    \fig{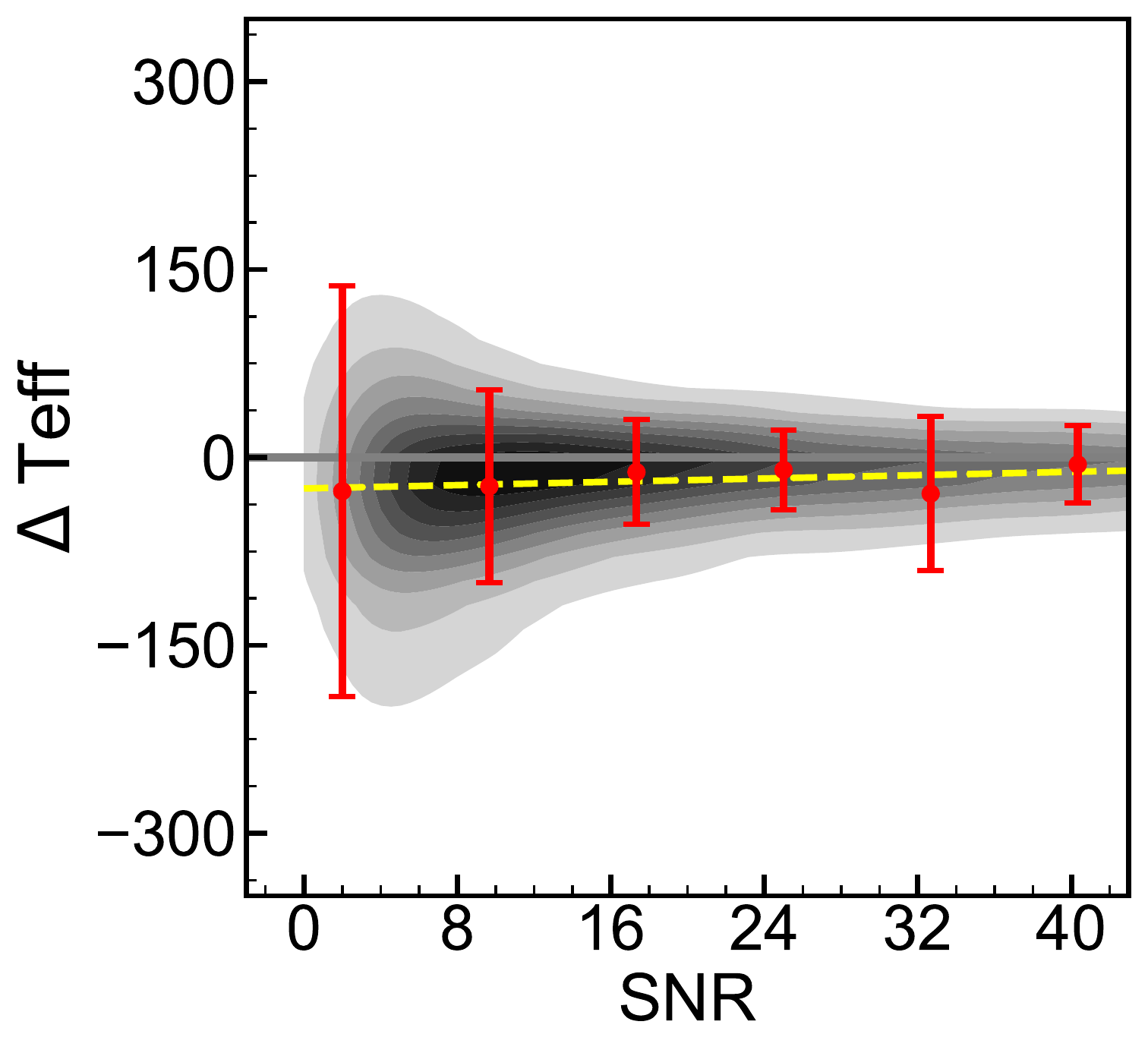}{0.32\textwidth}{}
    \fig{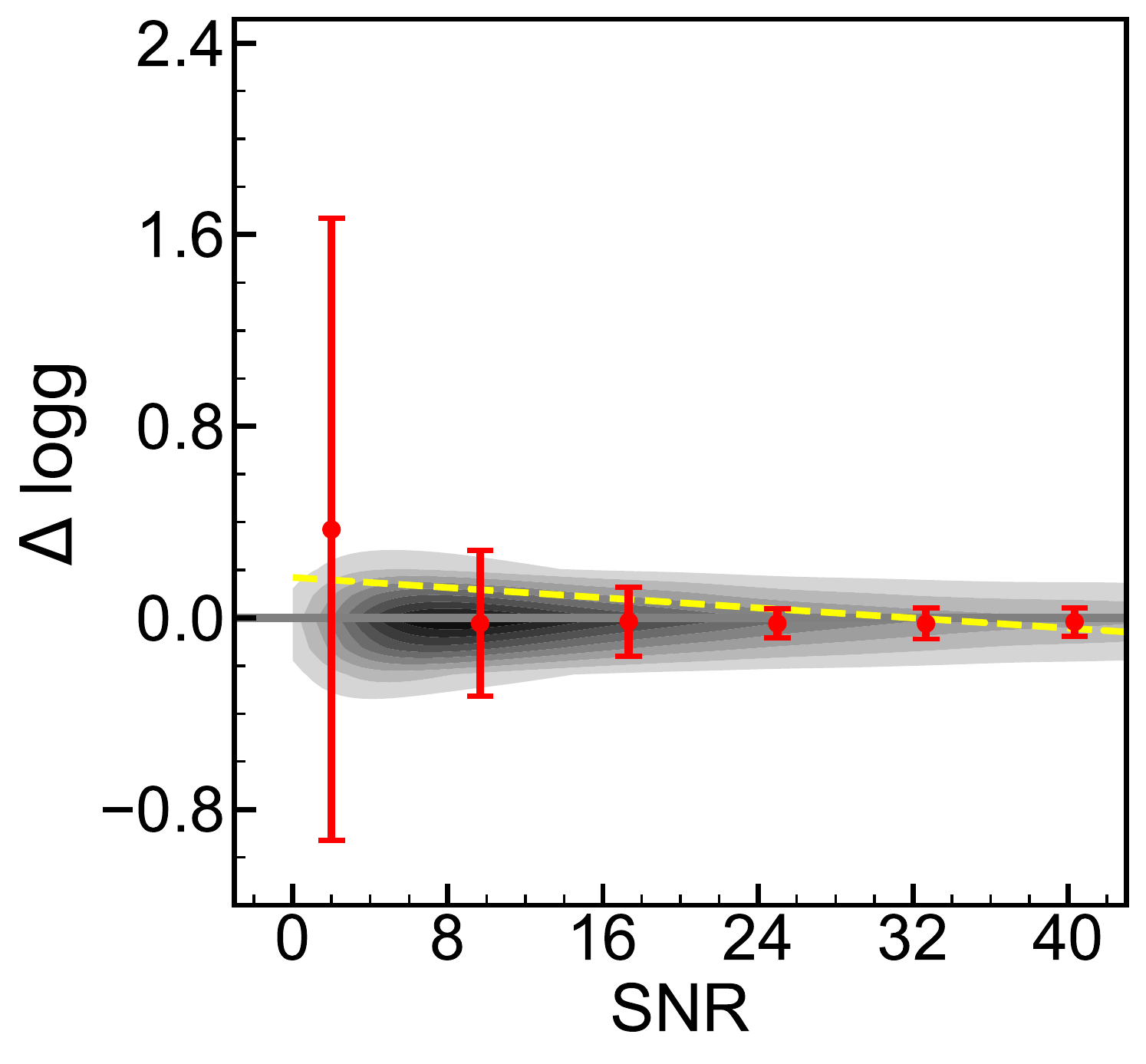}{0.32\textwidth}{}
    \fig{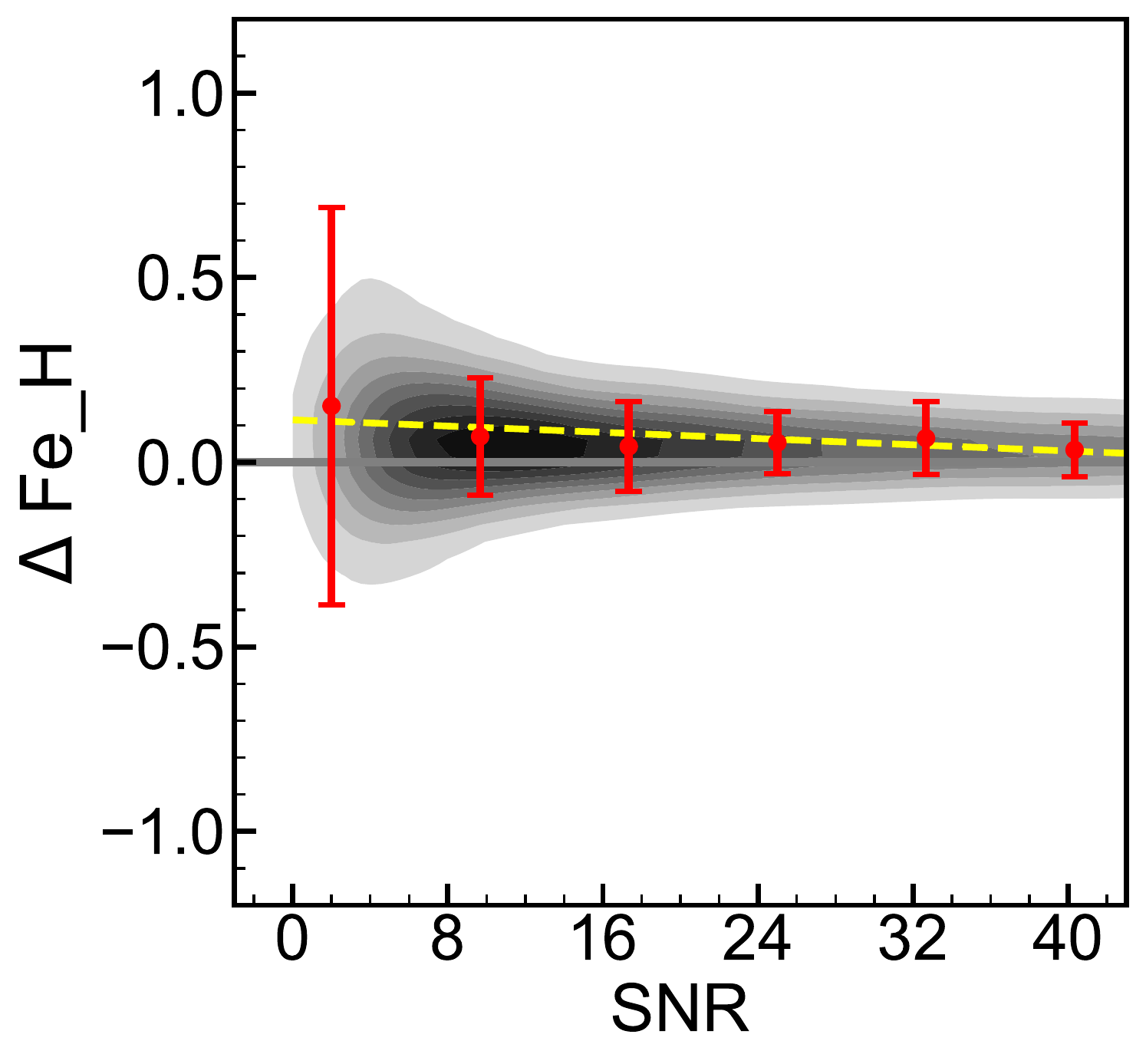}{0.32\textwidth}{}
    }  \vspace{-5mm}
	\caption{The systematic errors of stellar parameters as functions of SNR. Red error bars represent the standard deviations in each SNR bin, and the red dots in the center are the means. The yellow dashed lines are linear regressions of the means.}
	\label{fig:Systematic_error_snrr}
\end{figure*}

Furthermore, in order to investigate the possible systematic errors introduced by low spectral quality, we select spectra of repeated observations for the same targets with different SNR, of which the spectra with higher SNR should be 40 higher than the spectra of low SNR.
The residuals are represented by $\Delta P =  P_{\text{high}} - P_{\text{low}}$, where $P$ refers to the three stellar parameters, $P_{\text{high}}$ stands for the one of higher SNR, while $P_{\text{low}}$ is the one of lower SNR.  

Figure~\ref{fig:Systematic_error_snrr} exhibits the difference between stellar parameters deduced from spectra of repeated observations as a function of lower SNR. 
The asymmetry of residual distributions indicates that systematic deviations exist between high and low SNR results.

We use error bars to present the standard dispersion of each SNR bin and perform linear regressions of mean values in each bin (see the yellow dashed lines in Figure~\ref{fig:Systematic_error_snrr}) to generalize the gradients of systematic deviations for \teff, \logg\ and \feh. 
Similarly, \teff\ shows good precision and is less sensitive to SNR of lower spectral quality, and can be determined with a systematic difference better than 50\,K at SNR $\sim$ 10. 
The systematic errors of \logg\ are quite obvious in the low SNR region, which indicates that surface gravities deduced from low SNR spectra are systematically underestimated compared to those from high SNR.
It should be noted that, for some stars of SNR lower than 5, the differences between surface gravities derived from high and low SNR spectra can be as bad as $\sim$ 2.5\,dex, which suggests that our program tend to misclassify some dMs as gMs in the low SNR region.
For \feh, an obvious underestimation of $\sim$ 0.7\,dex appears at SNR $\sim$ 5, however, for sample stars with SNR higher than 20, the systematic errors are less than 0.2\,dex, which is less significant. 
 
\subsection{Internal Uncertainty Analysis}

\begin{figure*}[!htbp]
	\centering
	\fig{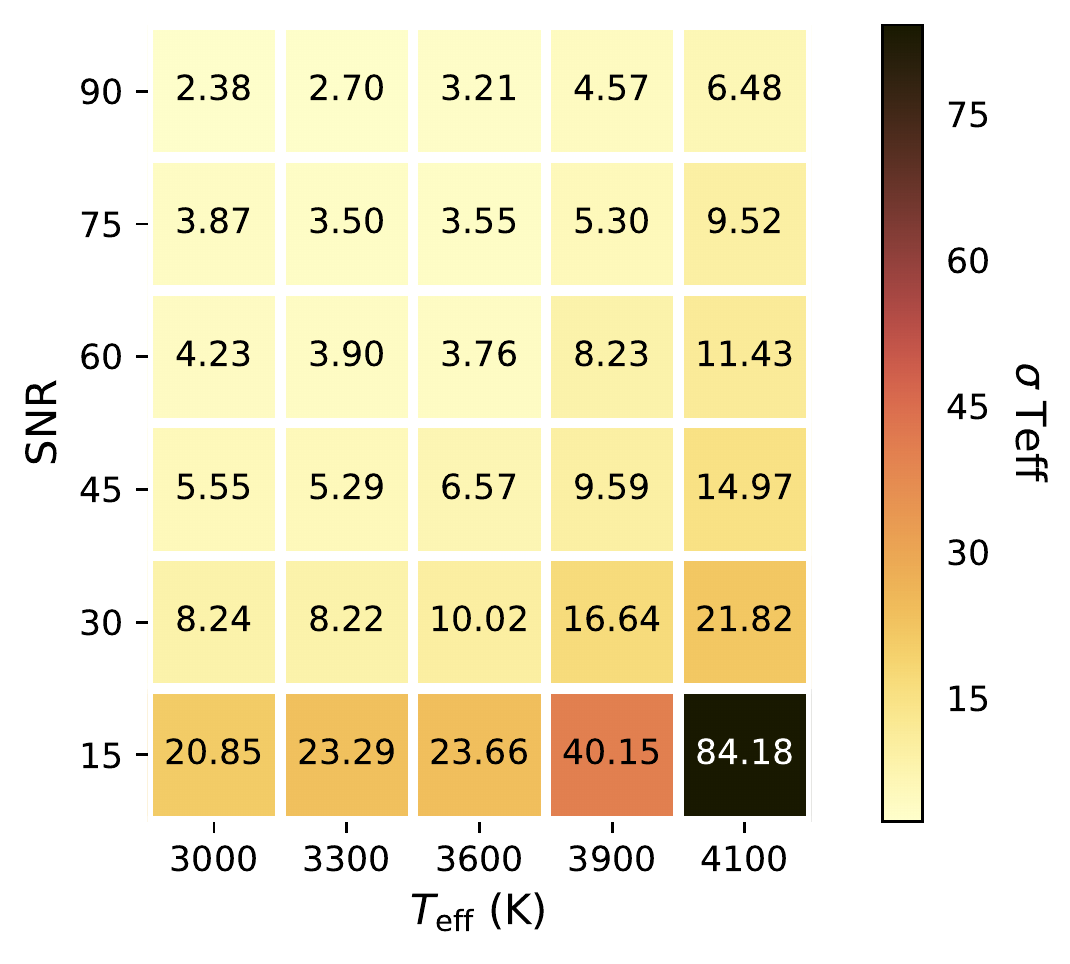}{0.28\textwidth}{}
	\fig{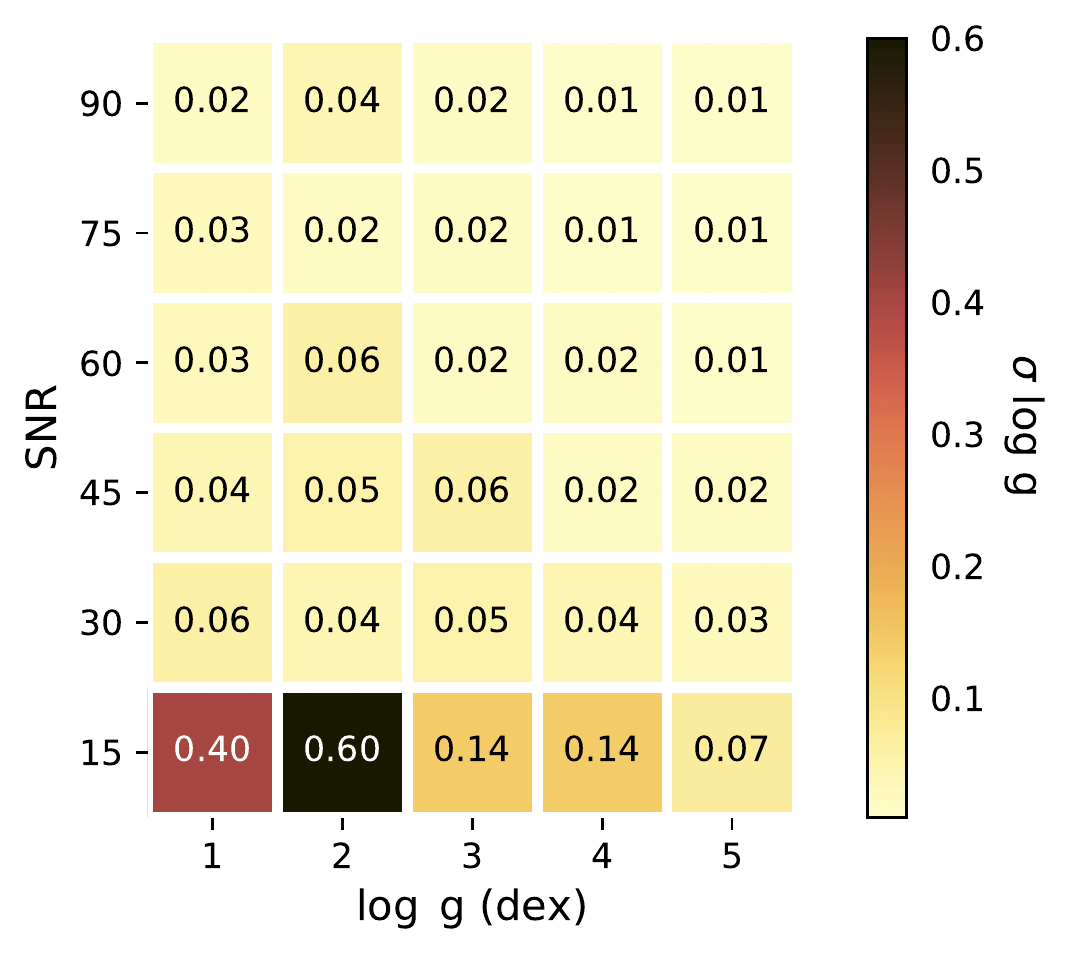}{0.28\textwidth}{}
	\fig{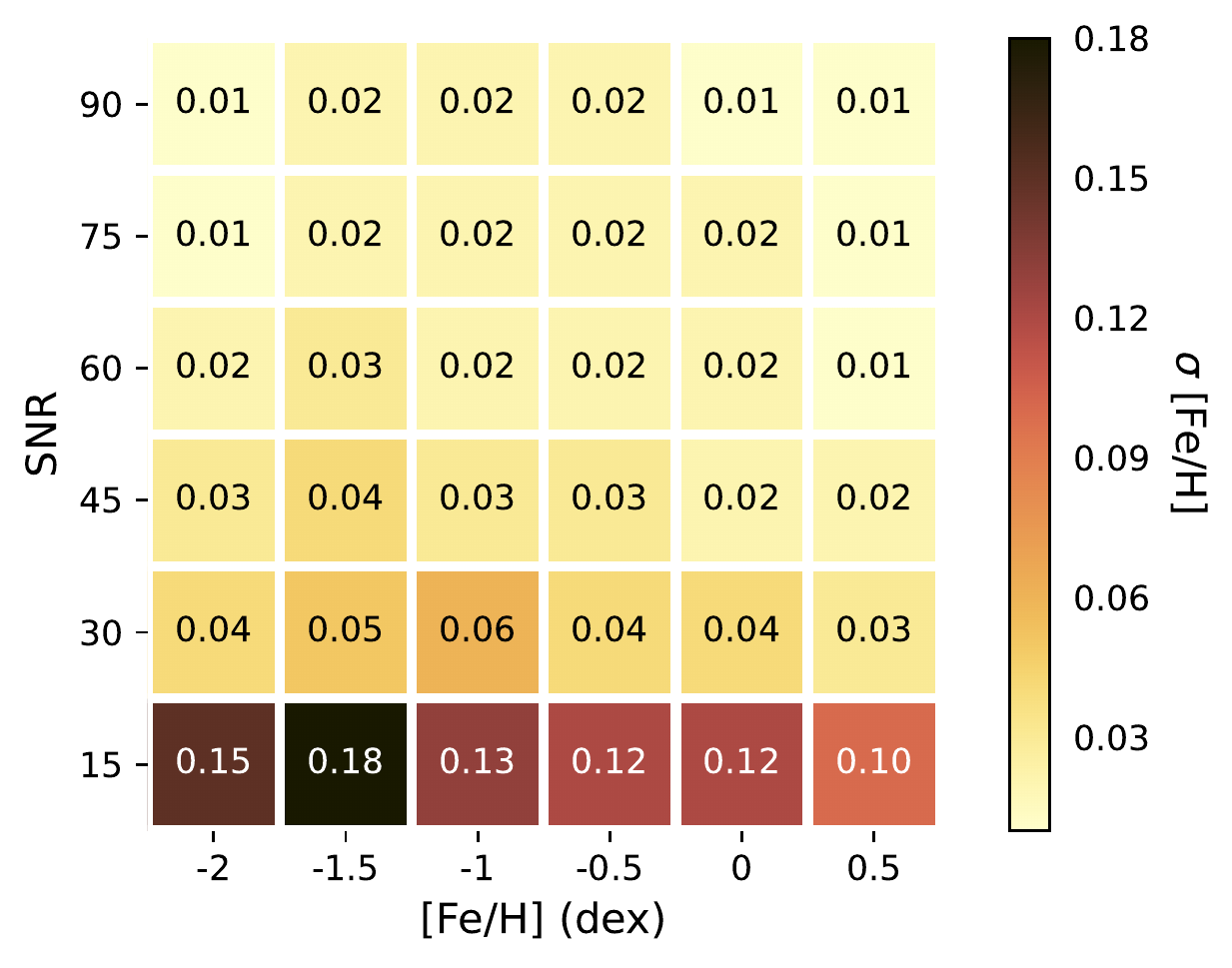}{0.32\textwidth}{}
	\caption{The internal uncertainties in different stellar parameters and SNR values. The values in each box stand for the standard deviation of residuals within each interval.}
	\label{fig:error_mcmc}
\end{figure*}

As we know, during the $\chi^2$ minimization, errors can be introduced due to the fitting procedure. 
We evaluate the internal uncertainties of our stellar parameter measurements in three steps. 
In the first step, we sample a set of stellar parameter (2900\,K--4500\,K for \teff, 0.0--6.0\,dex for \logg\ and $-2.5$ to $+1.0$\,dex for \feh) based on the distribution of stellar parameter space of our results. 
Second, we utilize these stellar parameters to obtain the corresponding spectra generated from V2 interpolator, and add Gaussian noises to simulate the distortion during observations. 
In the third step, we redetermine the stellar parameters for all these generated spectra from a single Markov Chain Monte Carlo (MCMC) simulation. 
We calculate the residuals between the input stellar parameters in step 1 and the outputs in step 3, and present the deviations of these residuals in Figure~\ref{fig:error_mcmc}. 

We notice that hot metal-poor giants generally have worse performance than the others, especially in the low SNR region. 
For spectra of SNR $>$ 20, stellar parameters can be determined with internal uncertainties better than 22\,K, 0.06\,dex and 0.06\,dex for \teff, \logg\ and \feh, respectively.
It should be noted that the error obtained in this simulation is lower than those in the external comparisons, because the theoretical spectrum with randomly added noise is more ideal than the observed spectrum.

\subsection{The Catalog of Stellar Atmospheric Parameters}

We present the stellar atmospheric parameter catalog of our results, which contains \specNum\ spectra of \starNum\ M-type stars from LAMOST LRS DR8.
In Table~\ref{Tab:example}, we show a sample set of our stellar parameter results and several important attributes,
 including the celestial coordinates, spectral quality information: SNR in u, g, r and i bands as well as the spectral identification information: the LAMOST designation (desig), LAMOST unique spectra ID (obsid) and source identifier (source\_id) from Gaia Early Data Release 3 \citep{gaiacollaboration.2021}.

\begin{figure}[!htbp]
	\centering
    \fig{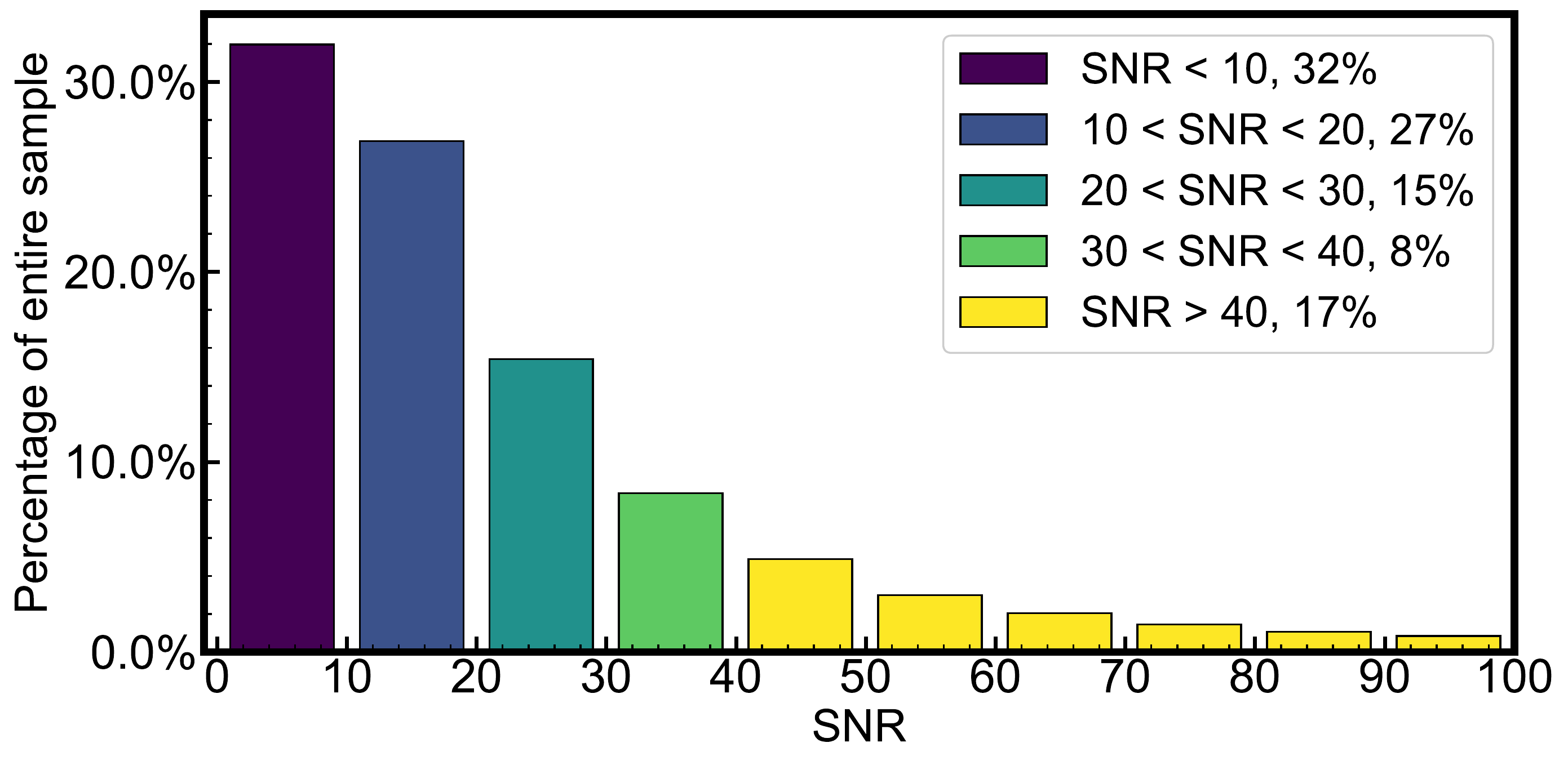}{0.45\textwidth}{}
	\caption{The percentage for the stars with SNR below 10, from 10 to 20, 20 to 30, 30 to 40 and above 40.}
	\label{fig:sample_SNR_hist}
\end{figure}

\begin{figure*}[htbp]
	\centering
    \fig{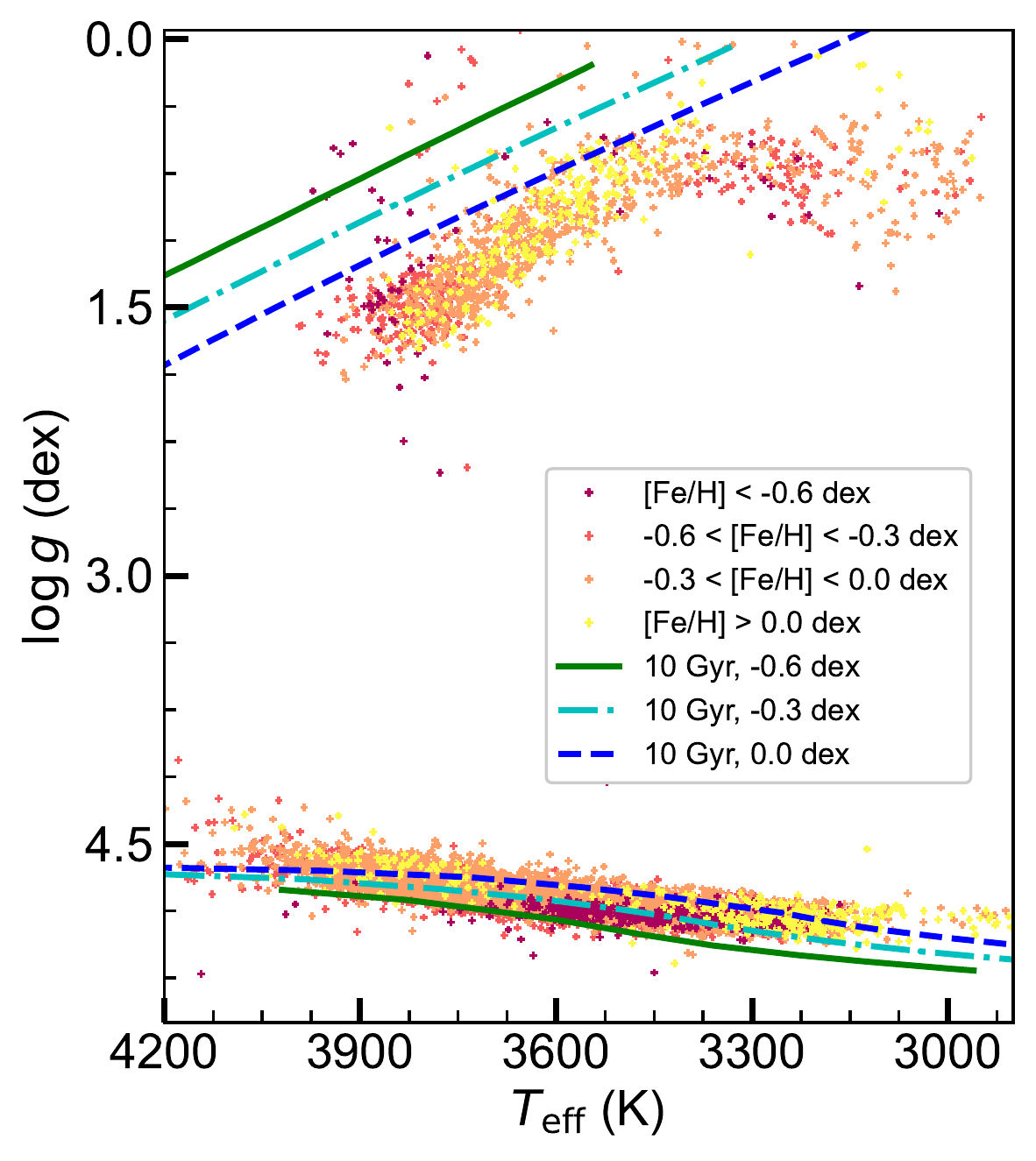}{0.4\textwidth}{}
    \fig{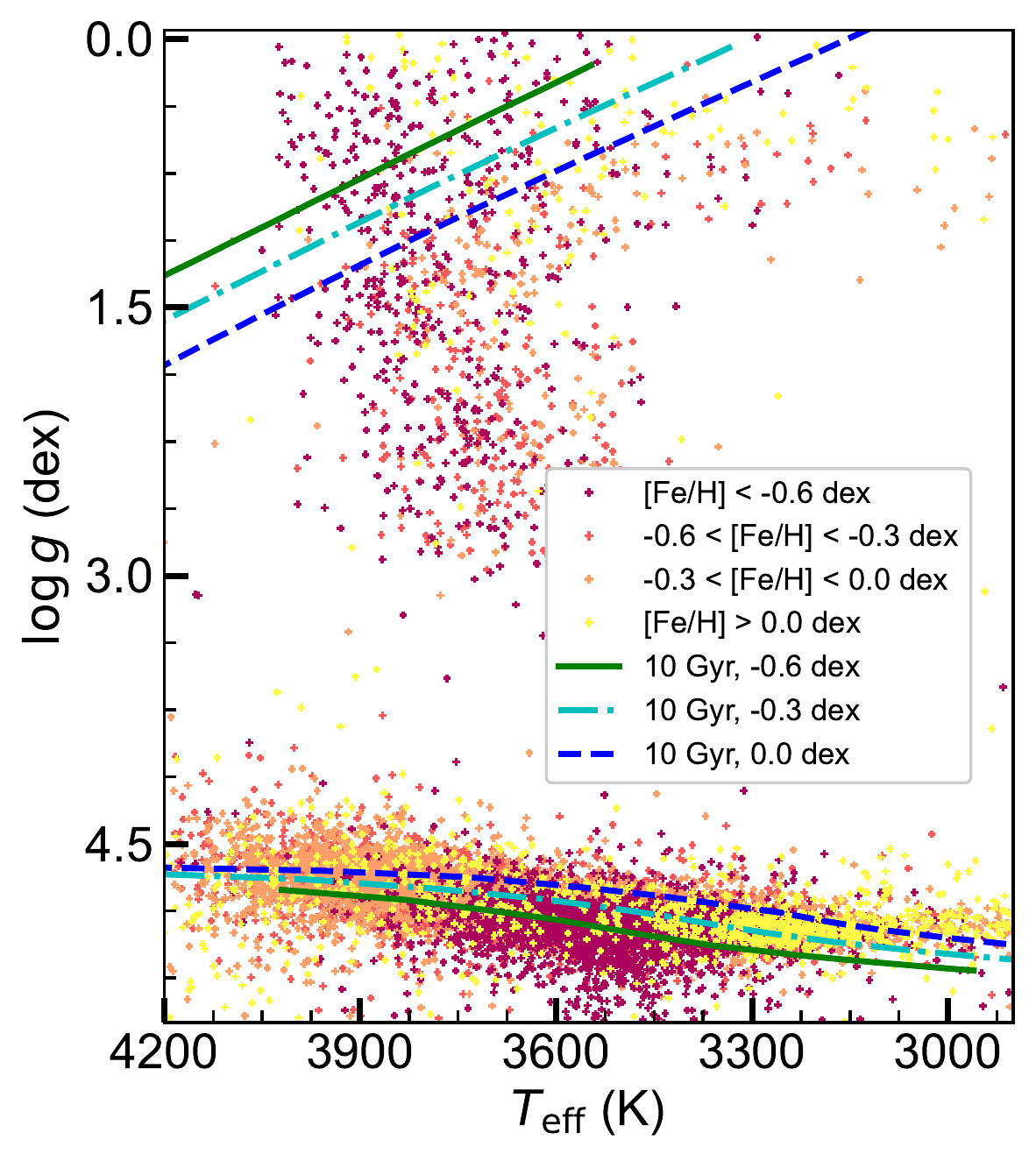}{0.4\textwidth}{}
	\caption{Left panel: the Kiel diagram colored by different metallicity groups of the samples with SNR $>$ 20. 
	Lines correspond to 10 Gyr PARSEC isochrones with \feh\ $=$ -0.6, -0.3 and 0.0 dex, respectively.
	Right panel: Same Kiel diagram but for stars of SNR $<$ 20.}
	\label{fig:HR_diagram_isoch}
\end{figure*}

In Figure~\ref{fig:sample_SNR_hist}, we present the distribution of SNR for the observational spectra. 
Spectra with SNR below 20 account for more than half of our samples, one should be careful using their stellar parameters. As discussed in Section~\ref{sect:error}, we recommend using the stellar parameters derived from spectra with SNR higher than 20 for reliability and good consistency.
Additionally, in Figure~\ref{fig:HR_diagram_isoch}, we plot the results derived from our method along with theoretical isochrones from the PAdova and TRieste Stellar Evolution Code \citep[PARSEC,][]{bressan.2012} for stars with SNR above 20 and below 20. 
As illustrated in the top panel, the stars of metallicity above the solar value (\feh\ $>$ 0.0 dex) can be clearly distinguished via the isochrone. 
Moreover, the comparison between the top and bottom panels show the improvement of SNR cut discussed above. Conspicuous outliers are rare for stars of SNR $>$ 20.

\begin{figure}[htbp]
	\centering
    \fig{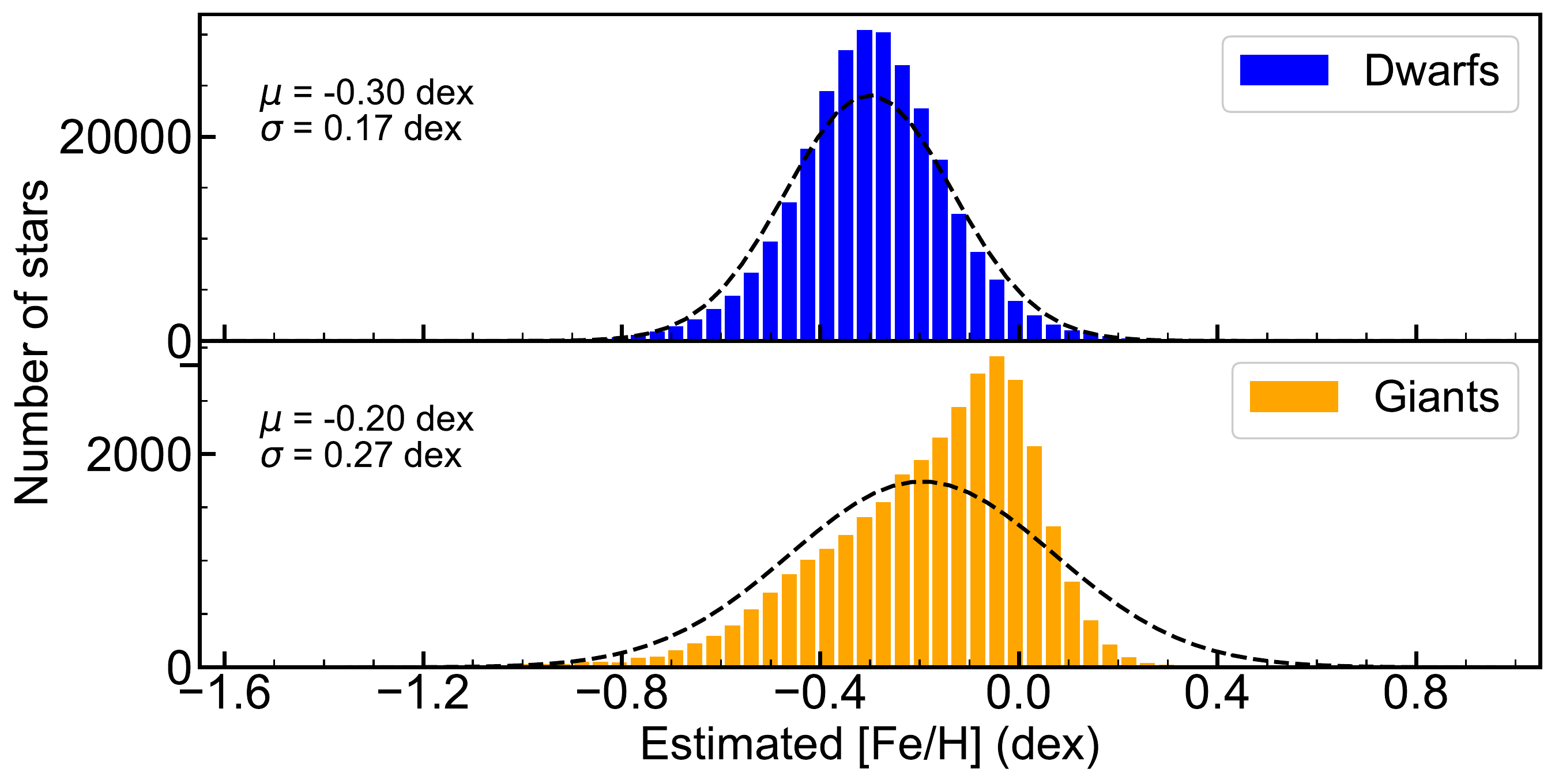}{0.45\textwidth}{}
	\caption{The histograms of \feh\ for dMs (top panel) and gMs (bottom panel). 
	The histograms only display \feh\ range from -1.6\,dex to 1.0\,dex, as there are only a few metal-poor stars (\feh\ $<$ -1.6\,dex).}
	\label{fig:sample_Fe_H_hist}
\end{figure}

We note that the lower limit of derived \teff\ is as low as 2,900\,K. 
For gMs, the lower limit of \logg\ can reach -0.24\,dex, and for dMs the upper limit is 5.9\,dex. 
Figure~\ref{fig:sample_Fe_H_hist} displays the distributions of \feh\ values for dMs and gMs, separately. 
The dMs show a more symmetrical distribution with a peak \feh\ around $-$0.30\,dex. 
For gMs, however, the peak is more skewed towards the solar abundance.

\subsection{Potential Future Uses}

This work builds on a number of other parameter determination studies for cool stars. 
Our study provides a significantly larger sample than the current literature sources that we have presented detailed comparisons with in Table~\ref{Tab:Ref_set}. 
This allows our catalog to provide well-characterized properties which might be useful for a wide range of purposes. 
Objects that are found to be time-variable in some way greatly benefit from having known properties. 
These properties of cool stars may help with the immediate interpretation or prove valuable for further follow-up studies. 
These might be individual systems that are for example found to host a planet or perhaps stars where an asteroseismic analysis enables the identification of interesting oscillation modes. 
These two use cases are also appropriate for the case of larger samples. 
It might be interesting to look at planetary formation as a function of metallicity \citep[e.g.,][]{lu.2020} or metallicity might be related to asteroseismic evolutionary information for large numbers of stars \citep[e.g.,][]{bellinger.2019}. 

Our catalog can also be valuable as a means to recognize important outliers among the two key populations of cool stars: the most numerous stars (dMs) and most luminous stars (gMs). 
Such studies can provide valuable insights for understanding and modelling the evolution of both stellar properties as well as interferences about our galaxy and beyond. 
The properties of cool stars with age can provide a range of different information useful in a range of contexts. 
The slow evolution of dMs provides the most pristine view of elements unchanged by their lack of evolution through only a small fraction of their main sequence lifetime \citep[e.g.,][]{laughlin.1997}. On the other hand gMs are useful probes of the mixing of evolved stars as well as dust. It is not clear that these are fully consistent with the predictions of stellar evolutionary models \citep[e.g.,][]{lancon.2018}. 
Such efforts together underpin systematic empirical grids of stellar types which are in turn a vital component of galaxy evolution models. It has long been known that the detailed properties of gMs on the horizontal branch as a function of stellar properties are vital \citep[e.g.,][]{worthey.1994, cabrera-ziri.2022} though dMs can also dominate the energy output of galaxies \citep[e.g.,][]{conroy.2012a, vandokkum.2021}. 
In the era of Gaia parameters, large scale spectroscopic properties from surveys such as LAMOST can be key in a detailed modern understanding of our galaxy. 
Our future work includes repeat observations, examination and determinations of our sample to further enhance our robust characterisation for the properties of M stars.

\section{Conclusion} \label{sect:conclusion}

Utilizing the MILES V2 interpolator and ULySS, we derive the stellar atmospheric parameters for \specNum\ spectra of \starNum\ M-type stars from LAMOST DR8. 
We compare our stellar parameters with external references including APOGEE DR17 and other literature, and good consistencies can be found. 
To evaluate the precision of our results, we analyze the stars with repeated observations. 
The precision of derived stellar parameters is mainly dependent on the SNR and effective temperature. 
For the SNR higher than 20, the typical precisions are better than \TrandSNR, \GrandSNR\ and \MrandSNR\ for \teff, \logg\ and \feh, respectively. 
A Monte-Carlo simulation is applied to check the performance of our method, and the internal uncertainty of our method is $\sigma_{T_{\rm eff}} <$ 22\,K, $\sigma_{\log{g}} <$ 0.06\,dex and $\sigma_{\rm [Fe/H]} <$ 0.06\,dex, for the spectra of SNR better than 20.

Our results supply LAMOST LRS DR8 with large numbers of well-determined M-type star stellar atmospheric parameters, and our method could be applied to the works of stellar parameter determination for M-type star in future low resolution surveys, for example, LAMOST LRS DR9.
Moreover, it is important to further investigate the metallicity distributions and kinematic properties of different Galactic populations.

\begin{acknowledgments}
We thank the referee for the helpful comments which have helped us to improve the manuscript.
Our research is supported by National Key R\&D Program of China No.2019YFA0405502, the National Natural Science Foundation of China
under grant Nos. 12090040, 12090044, 11833006, 12022304, 11973052, 11973042 and U1931102. We acknowledge the science research grants from the China Manned Space Project with NO.CMS-CSST-2021-B05.
This work is supported by Chinese Academy of Sciences President's International Fellowship Initiative. Grant No. 2020VMA0033.
H.-L.Y. acknowledges the supports from Youth Innovation Promotion Association, Chinese Academy of Sciences (id. 2019060).
Guoshoujing Telescope (the Large Sky Area Multi-Object Fiber Spectroscopic Telescope LAMOST) is a National Major Scientific Project built by the Chinese Academy of Sciences.
Funding for the project has been provided by the National Development and Reform Commission.
LAMOST is operated and managed by the National Astronomical Observatories, Chinese Academy of Sciences.

\end{acknowledgments}

\restartappendixnumbering
\appendix

\section{Appendix Figure} \label{sect:appendixFigure}
\begin{figure*}[!ht]
	\centering
    \gridline{
    \fig{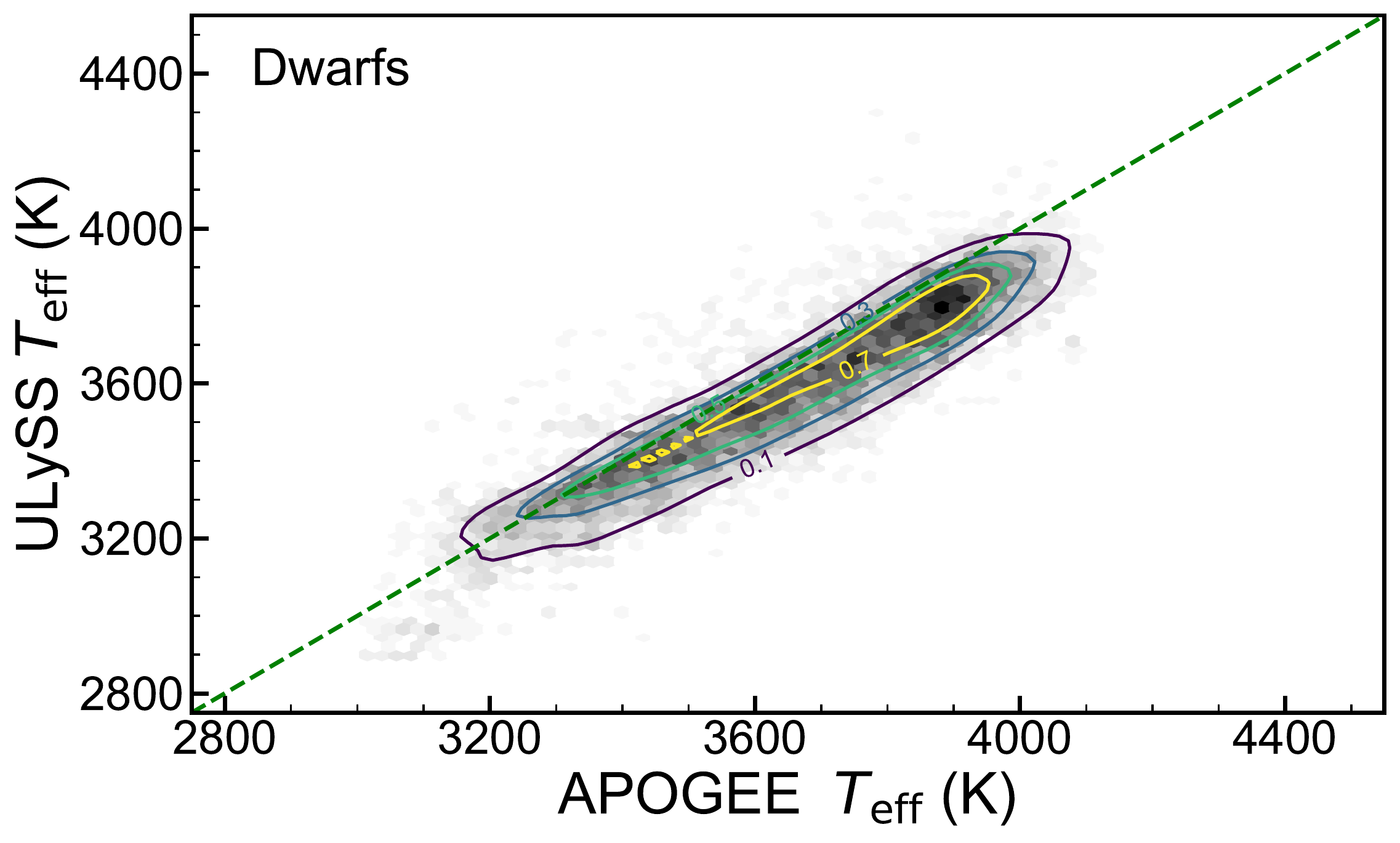}{0.45\textwidth}{}
    \fig{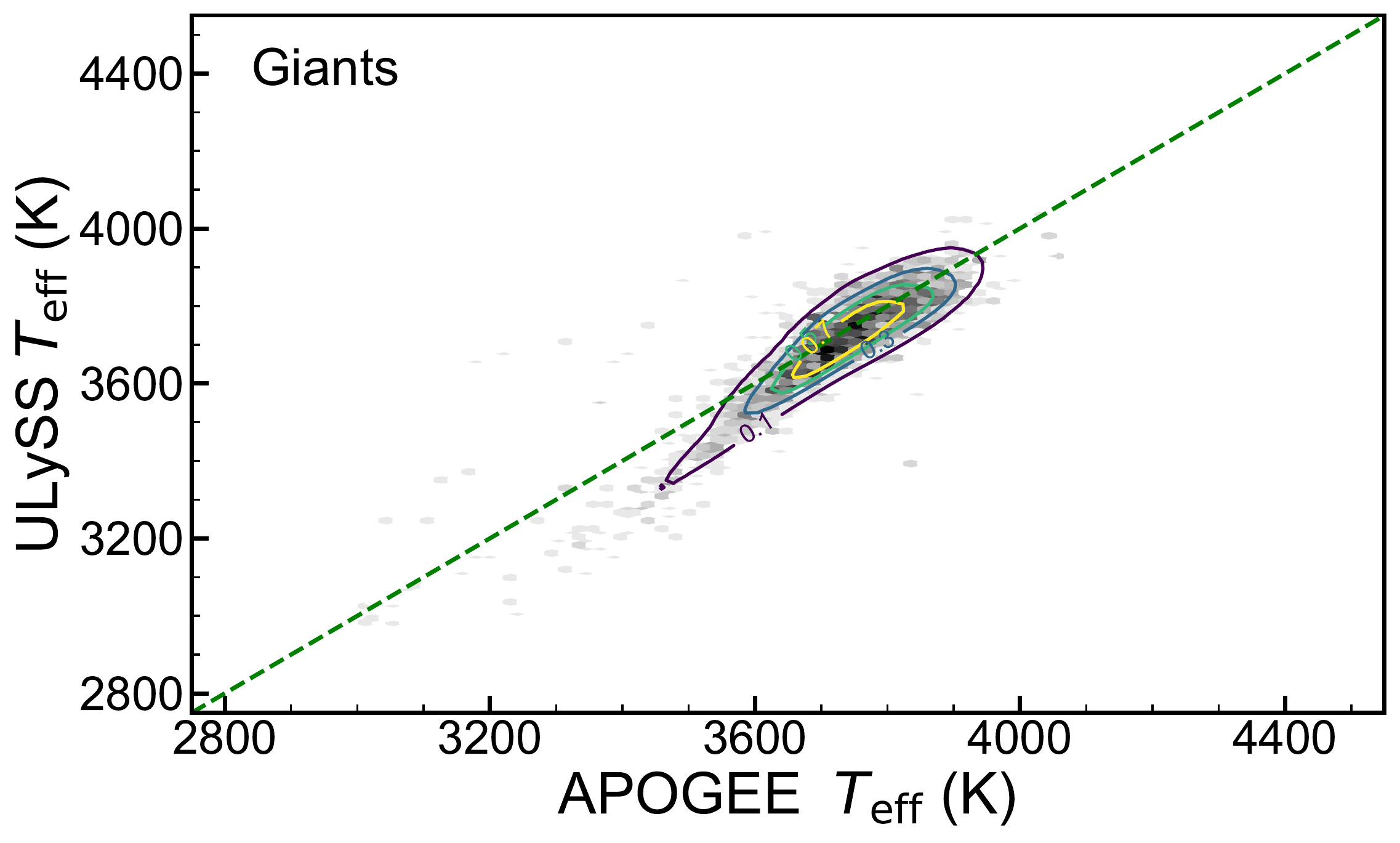}{0.45\textwidth}{}
	}  \vspace{-7mm}
        \gridline{
    \fig{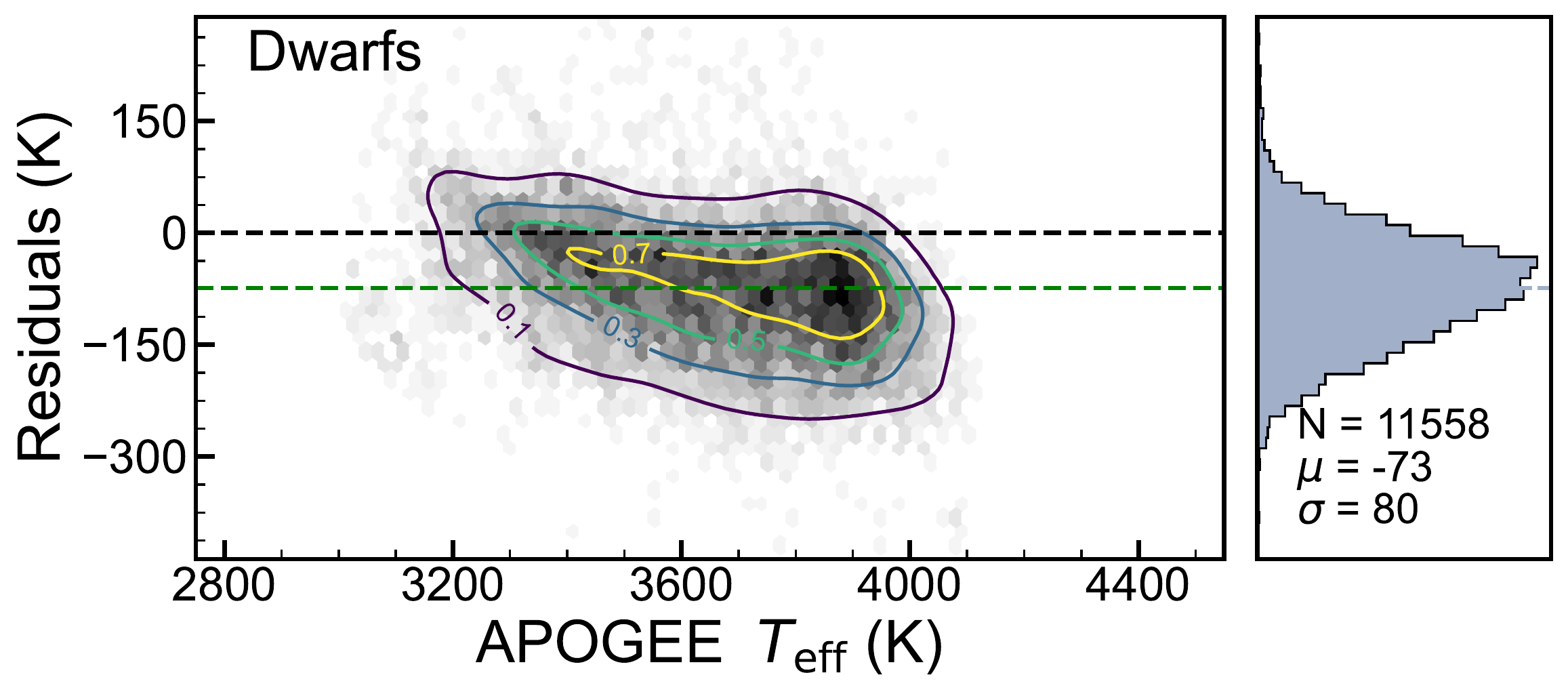}{0.45\textwidth}{}
	\fig{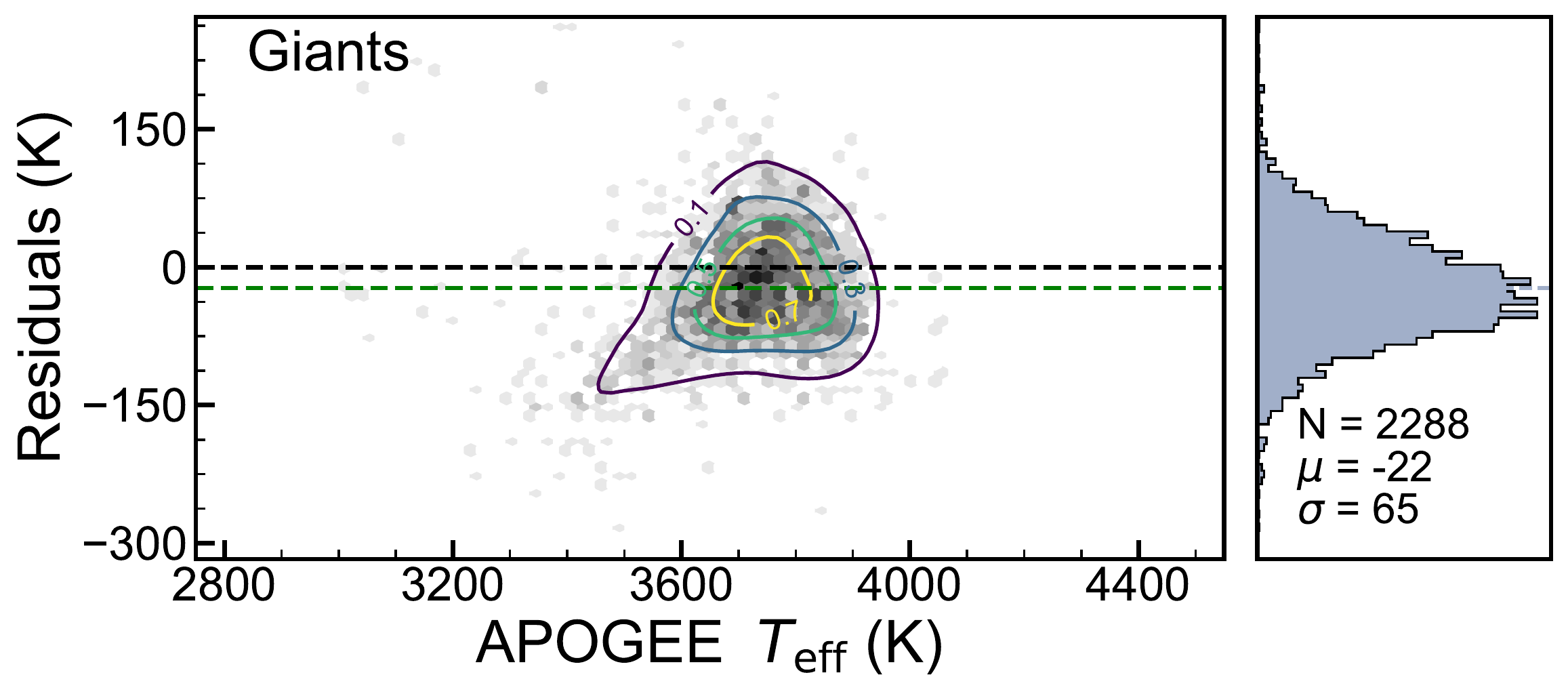}{0.45\textwidth}{}
    }  \vspace{-7mm}
    
	\caption{The comparison of the effective temperatures derived from our method with the APOGEE DR17 spectroscopic values. 
	The top-left panel shows the \teff\ comparison for dMs, while the top-right panel for gMs. 
	The bottom-left and bottom-right panels show the residuals' distributions for dMs and gMs, respectively. 
	The color scale as well as the contour lines indicate the number density of each region.
	}
	\label{fig:TspecvsAPOGEE}
\end{figure*}

\begin{figure*}[!ht]
	\centering
    \gridline{
    \fig{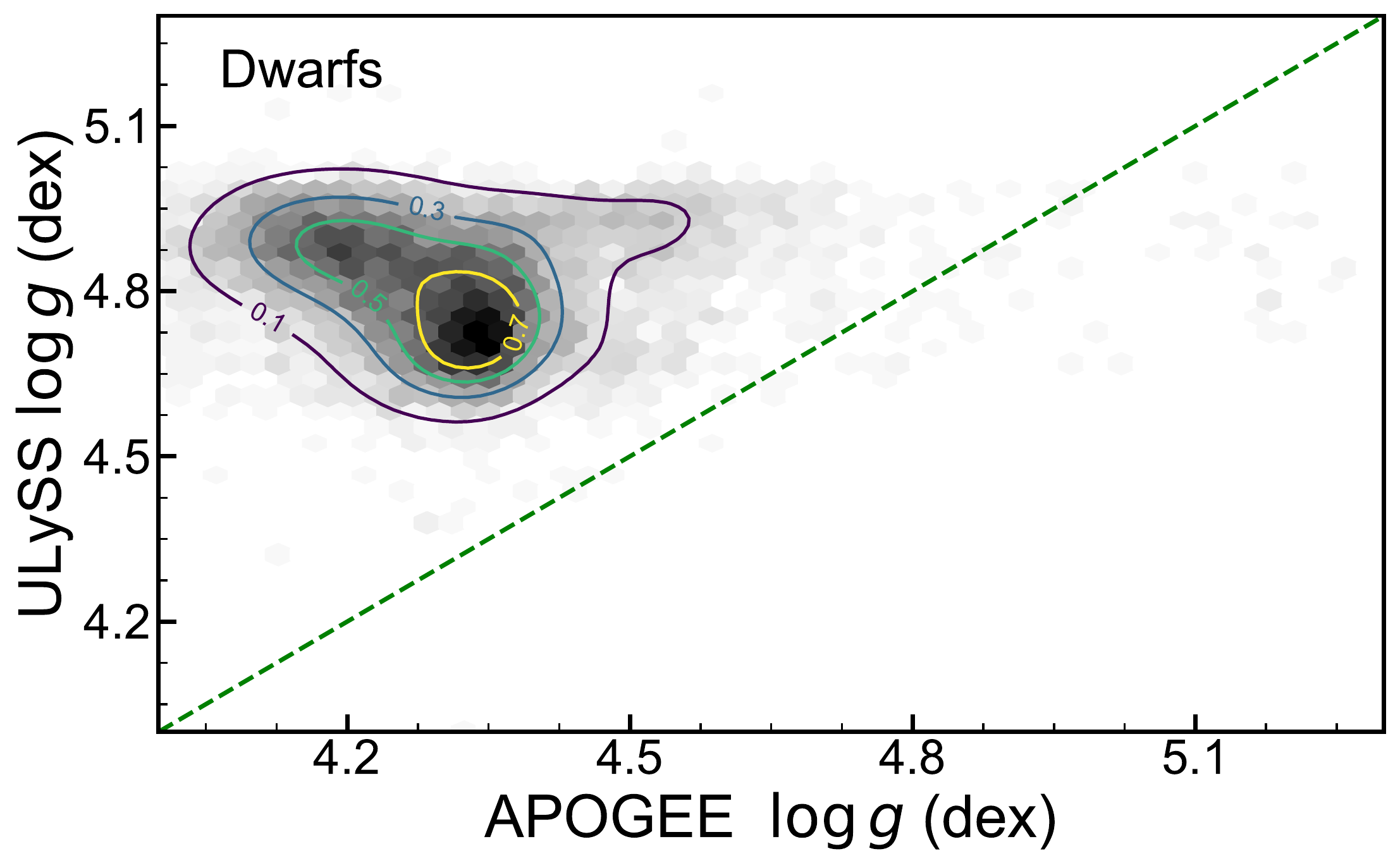}{0.45\textwidth}{}
    \fig{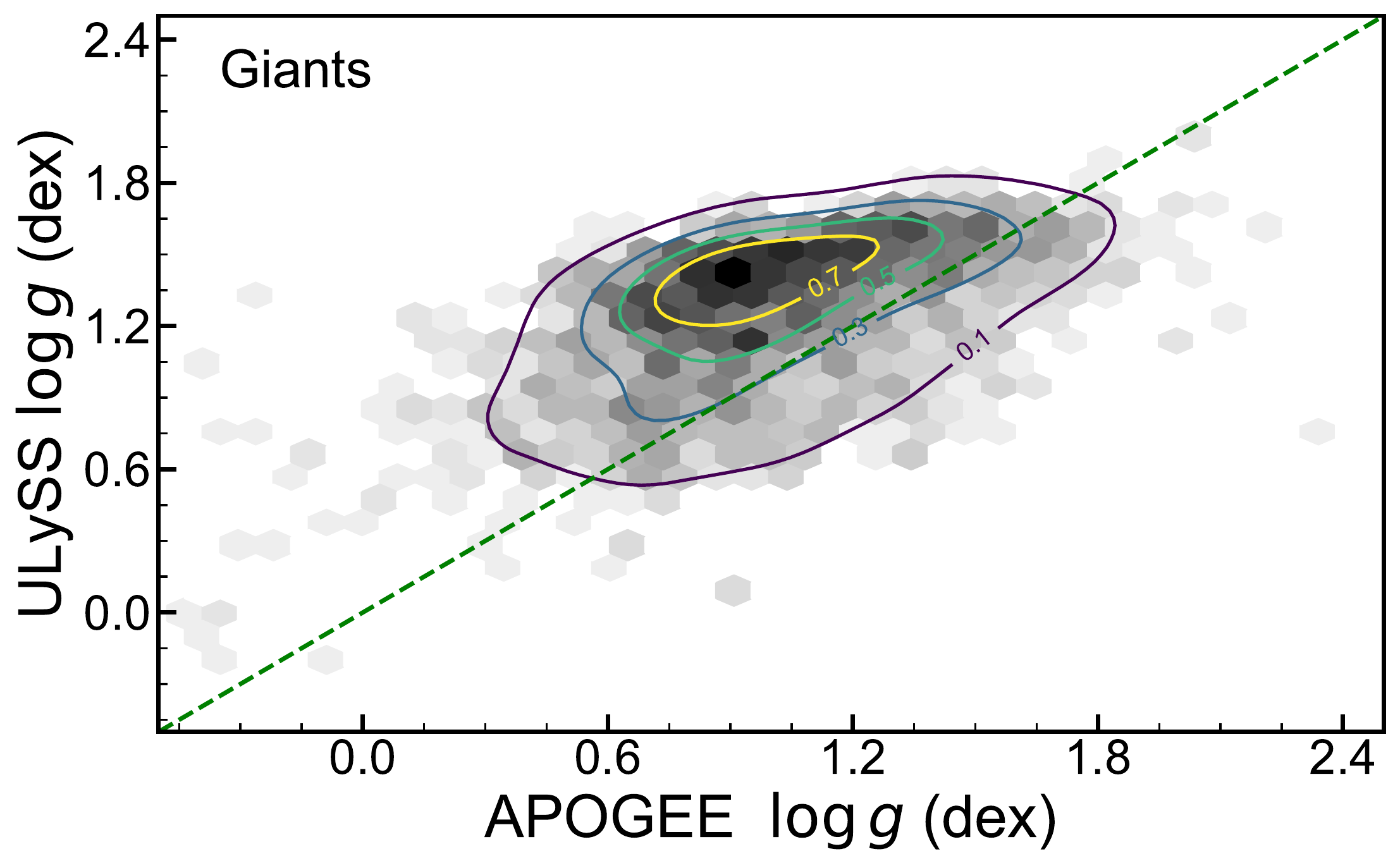}{0.45\textwidth}{}
	}  \vspace{-7mm}
        \gridline{
    \fig{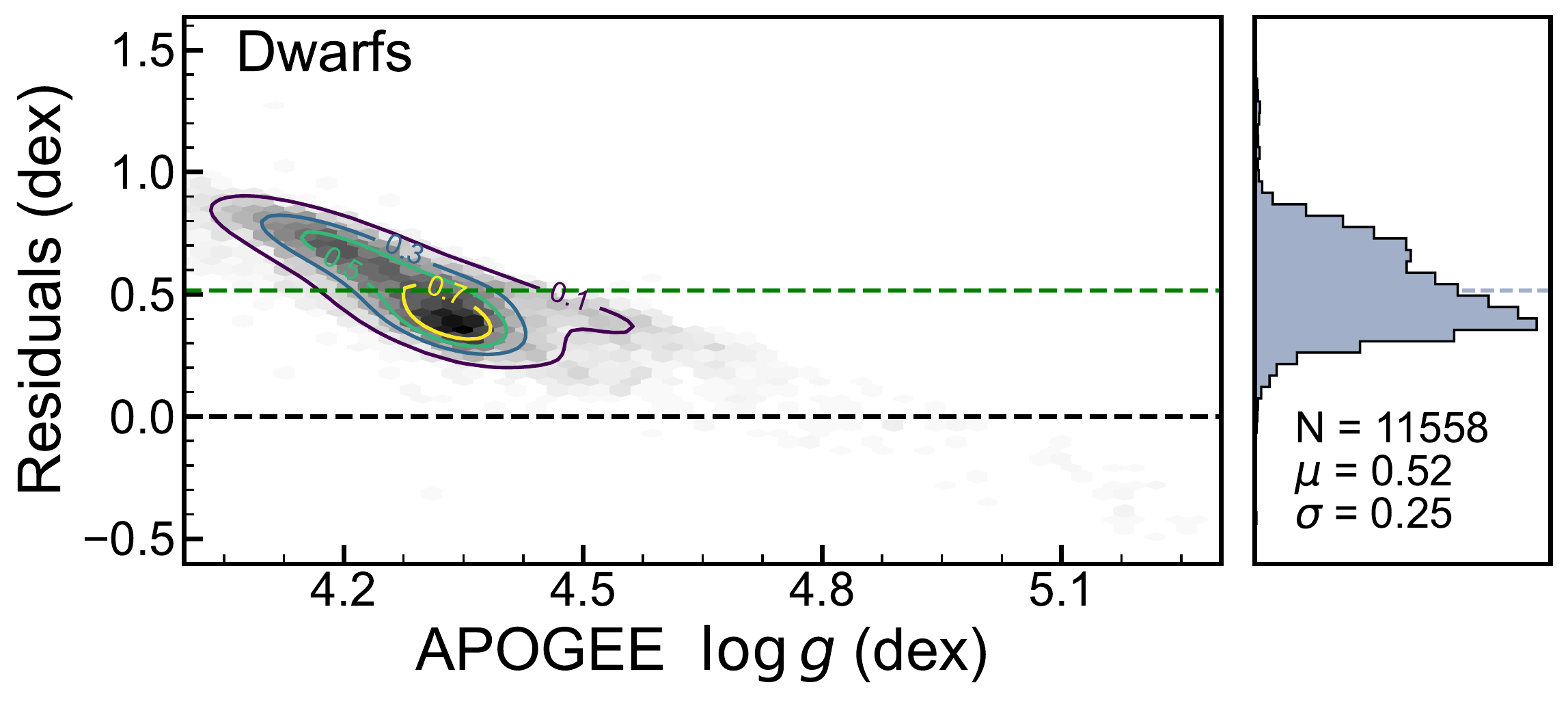}{0.45\textwidth}{}
    \fig{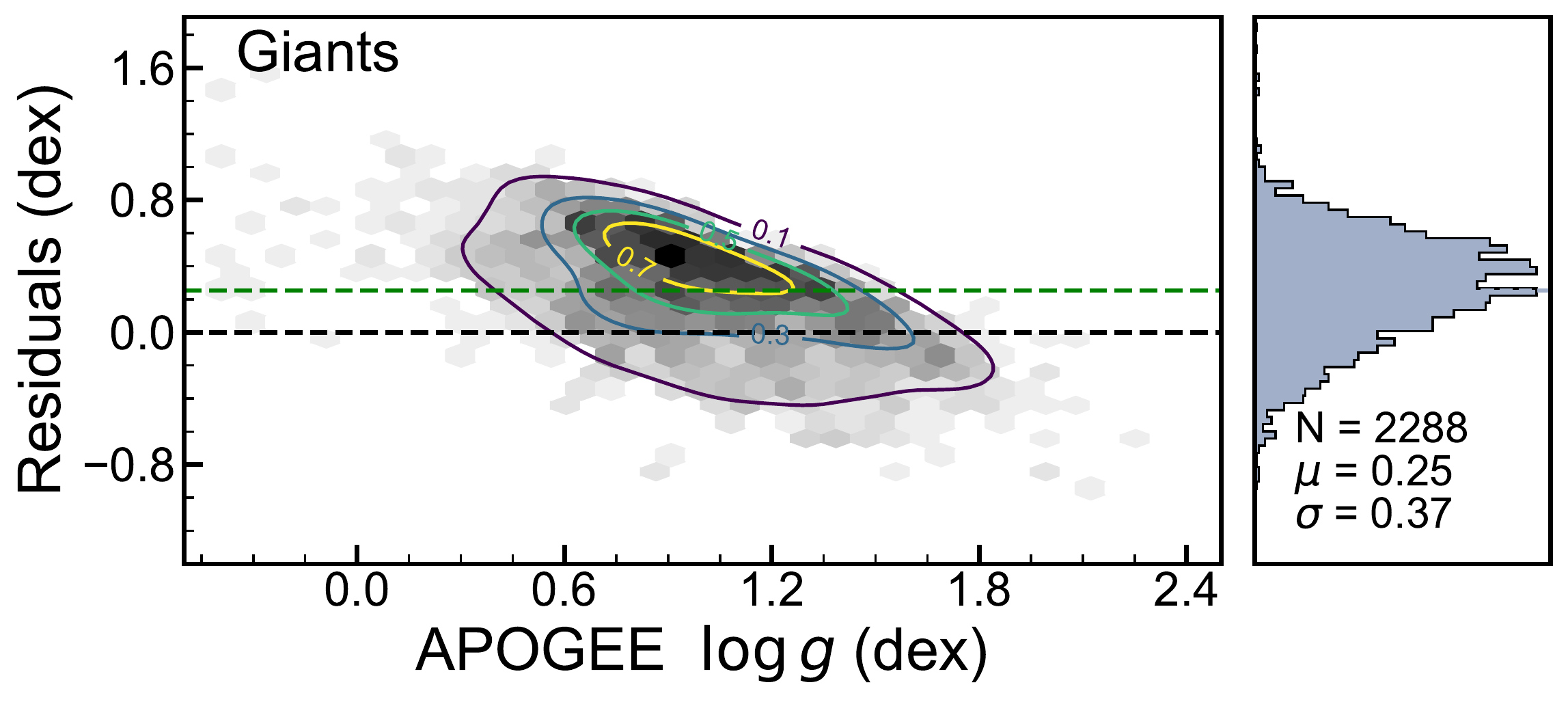}{0.45\textwidth}{}
    }  \vspace{-7mm}    
    
	\caption{Same as Figure~\ref{fig:TspecvsAPOGEE}, but for surface gravity.
	}
	\label{fig:GspecvsAPOGEE}
\end{figure*}

\section{Appendix Table} \label{sect:appendixTable}

\begin{longrotatetable}
\label{Tab:example}
\begin{deluxetable*}{rrrrrrrrrr}
\tablewidth{700pt}
\tablecaption{Stellar parameters of randomly selected objects.}
\centering
\tablehead{\colhead{desig} & \colhead{obsid} & \colhead{source\_id} & \colhead{RA} & \colhead{DEC} & \colhead{SNR\_r} & \colhead{RV} & \colhead{T\_eff} & \colhead{log g} & \colhead{[Fe/H]} \\ 
\colhead{} & \colhead{} & \colhead{} & \colhead{hms (J2000)} & \colhead{dms (J2000)} & \colhead{} & \colhead{(km$\cdot s^{-1}$)} & \colhead{(K)} & \colhead{(dex)}  & \colhead{(dex)}  } 
    \startdata
		 J$003233.56+023235.4$ & $182907177$ & $2547554358360573184$ & $00:32:33.57$ & $+02:32:35.4$ & $21    $ & $-0.6 $ & $3865$ & $4.66$ & $-0.23$ \\
		 J$005018.74+383849.1$ & $407239   $ & $367934386366947456 $ & $00:50:18.75$ & $+38:38:49.1$ & $12    $ & $-38.7$ & $3699$ & $4.73$ & $-0.11$ \\
		 J$005137.86+382236.2$ & $612710127$ & $367725002416518272 $ & $00:51:37.87$ & $+38:22:36.3$ & $35    $ & $-20.0$ & $3739$ & $4.73$ & $-0.40$ \\
		 J$062058.46+210949.0$ & $437312190$ & $3375942048114941696$ & $04:11:37.63$ & $+33:59:16.6$ & $144   $ & $-26.4$ & $3409$ & $0.52$ & $-0.11$ \\
		 J$041137.62+335916.6$ & $200512017$ & $170751686193676288 $ & $06:14:31.81$ & $+08:40:37.2$ & $46    $ & $-63.6$ & $3755$ & $2.39$ & $-0.64$ \\
		 J$061431.80+084037.2$ & $503710146$ & $3328420434007985280$ & $06:20:58.46$ & $+21:09:49.0$ & $22    $ & $1.9  $ & $3506$ & $1.58$ & $-0.52$ \\
		 J$063953.53+534915.5$ & $546315051$ & $994279332683432320 $ & $06:39:53.54$ & $+53:49:15.5$ & $14    $ & $-23.7$ & $3782$ & $1.46$ & $-0.47$ \\
		 J$065018.79+264542.8$ & $46204100 $ & $3385128472059867904$ & $06:50:18.80$ & $+26:45:42.8$ & $20    $ & $-33.3$ & $3907$ & $4.64$ & $-0.21$ \\
		 J$071141.54+195445.8$ & $175006180$ & $3363557565751137152$ & $07:11:41.54$ & $+19:54:45.9$ & $42    $ & $-32.5$ & $3813$ & $4.63$ & $-0.16$ \\
		 J$082845.70+104545.1$ & $337405103$ & $600889912104424192 $ & $08:28:45.71$ & $+10:45:45.1$ & $49    $ & $15.5 $ & $3915$ & $4.65$ & $-0.15$ \\
		 J$085752.81+155051.6$ & $629901215$ & $610580732712396032 $ & $08:57:52.82$ & $+15:50:51.6$ & $67    $ & $27.7 $ & $3657$ & $4.89$ & $-0.45$ \\
		 J$103456.90-054946.3$ & $422001160$ & $3777400028612544640$ & $10:34:56.91$ & $-05:49:46.3$ & $17    $ & $-52.6$ & $3947$ & $4.67$ & $-0.38$ \\
		 J$104029.59+040235.2$ & $546101111$ & $3857828655644765440$ & $10:40:29.60$ & $+04:02:35.2$ & $15    $ & $53.0 $ & $3970$ & $4.50$ & $-0.11$ \\
		 J$143657.78+285648.1$ & $566705048$ & $1281154251515481344$ & $14:36:57.78$ & $+28:56:48.2$ & $29    $ & $-8.8 $ & $3792$ & $4.75$ & $-0.37$ \\
		 J$143700.28+313008.0$ & $319815214$ & $1286110746854188288$ & $14:37:00.29$ & $+31:30:08.0$ & $32    $ & $14.3 $ & $3765$ & $4.73$ & $-0.29$ \\
		 J$144136.81+162250.9$ & $450707153$ & $1234646382833797632$ & $14:41:36.81$ & $+16:22:50.9$ & $17    $ & $-52.4$ & $3574$ & $4.85$ & $-0.40$ \\
		 J$144852.72-001936.8$ & $649007156$ & $3650703754716377344$ & $14:48:52.46$ & $-00:19:36.9$ & $209   $ & $50.1 $ & $3045$ & $0.68$ & $-0.09$ \\
		 J$235515.89+042523.8$ & $505908050$ & $2743370468663782528$ & $23:55:15.90$ & $+04:25:23.9$ & $20    $ & $23.5 $ & $3791$ & $4.58$ & $-0.25$ \\
		 ... & ...  &  ... &  ... & ... & ... & ... & ... &... & ... \\
	\enddata
	 \tablecomments{The attribute column is not fully presented here. The complete catalog is accessible in its entirety in a machine-readable form.}
\end{deluxetable*}
\end{longrotatetable}

\bibliography{sample631}{}
\bibliographystyle{aasjournal}

\end{document}